\documentclass[a4paper,11pt]{article}
\usepackage{jheppub} 

\usepackage[T1]{fontenc} 

\title{\boldmath Universality of Loop Corrected Soft Theorems in 4d}

\author[a]{Hare Krishna}
\author{and}
\author[b]{Biswajit Sahoo}

\affiliation[a]{C. N. Yang Institute for Theoretical Physics, Stony Brook University,\\ Stony Brook, NY 11794, USA}
\affiliation[b]{Fields and Strings Laboratory, Institute of Physics,\\ Ecole Polytechnique Federale de Lausanne (EPFL),\\ CH-1015 Lausanne, Switzerland}

\emailAdd{harekrishna.harekrishna@stonybrook.edu, biswajit.sahoo@epfl.ch}

\abstract{In \cite{1808.03288}, logarithmic correction to subleading soft photon and soft graviton theorems have been derived in four spacetime dimensions from the ratio of IR-finite S-matrices. This has been achieved after factoring out IR-divergent components from the traditional electromagnetic and gravitational S-matrices using Grammer-Yennie prescription. Although the loop corrected subleading soft theorems are derived from one-loop scattering amplitudes involving scalar particles in a minimally coupled theory with scalar contact interaction, it has been conjectured that the soft factors are universal (theory independent) and one-loop exact (don't receive corrections from higher loops).

This paper extends the analysis conducted in \cite{1808.03288} to encompass general spinning particle scattering with non-minimal couplings permitted by gauge invariance and general coordinate invariance. By re-deriving the $\ln\omega$ soft factors in this generic setup, we establish their universal nature. Furthermore, we summarize the results of loop corrected soft photon and graviton theorems up to sub-subleading order, which follows from the analysis of one and two loop QED and quantum gravity S-matrices. While the classical versions of these soft factors have already been derived in the literature, we put forth conjectures regarding the quantum soft factors and outline potential strategies for their derivation.}

\begin{document} 
\hfill YITP-SB-2023-26

\maketitle
\flushbottom

\section{Introduction and Result}
\label{S:intro_res}
The soft theorem examines the infrared properties of a scattering amplitude involving a low (soft) momentum photon or graviton, in addition to other asymptotic particles. It establishes a relationship between this amplitude and the one without the low momentum photon or graviton. In a series of papers \cite{1703.00024,1706.00759,1707.06803,1809.01675,1802.03148}, it has been established that tree level soft photon and soft graviton theorems in four spacetime dimensions ($D=4$) are just the manifestation of gauge invariance and general coordinate invariance at the scattering amplitude level. Soft factorisation alone does not provide profound insights into the ultraviolet completion of QED or quantum gravity theory, nor does it impose additional constraints on the quantum theory beyond what has already been achieved by gauge invariance and general coordinate invariance.\footnote{We would like to emphasize that the Ward identities relating these soft theorems to asymptotic symmetries also do not provide any additional constraints on the quantum theory of electromagnetism or gravity beyond what has already been achieved by gauge invariance and general coordinate invariance \cite{1703.05448}. Instead, the Ward identities are simply the manifestation of the equations of motion for low-frequency photons or graviton fields at the level of scattering amplitudes. Where the soft charge is represented as an integral over the radiation mode of gauge or graviton fields and the hard charge is an integral over the inverse propagator operating on the current or stress tensor associated with finite energy scattered particles. The asymptotic symmetry parameters (large gauge transformation and asymptotic radial modes of bulk diffeomorphism) on the celestial sphere are just smearing functions that appears in both the integrands of the soft and hard charge expressions. The conservation of asymptotic charges about spatial infinity manifests as crossing symmetry relation between the two QED or gravitational S-matrices involving the soft photon/graviton in the ingoing and outgoing states.} Instead, given an effective field theory (EFT) action with potential non-minimal interactions permitted by gauge invariance or general coordinate invariance, one can systematically compute the non-universal soft factors up to a certain order in the soft momentum expansion \cite{1706.00759,1809.01675,1611.07534}. In the past, there were challenges in obtaining loop corrections to the subleading soft photon and graviton theorems in $D=4$ due to the presence of infrared divergence in traditional scattering amplitudes \cite{1405.1410,1405.1015}. However, this issue has been successfully resolved in \cite{1808.03288} by directly working in $D=4$ and carefully analyzing the possible non-analytic structures around $\omega=0$. In this work, the subleading soft photon and soft graviton theorems have been derived at the one-loop level, taking into account both electromagnetic and gravitational interactions. Interestingly, it has been observed that the subleading soft factors emerge at an order $\ln\omega$, where $\omega$ represents the energy of the soft photon or graviton. The loop corrected subleading soft factor is dominant compare the tree level subleading soft factor, which is of order $\omega^0$, as the energy approaches zero ($\omega\rightarrow 0$). The existence of the $\ln\omega$ soft graviton theorem has been confirmed in \cite{1901.10986, 1812.08137} with perfect agreement with the result of \cite{1808.03288} in the massless limit.

The soft graviton theorem results offer an intriguing application in deriving low-frequency gravitational wave forms and gravitational memory for astrophysical scattering events from their classical limit \cite{1801.07719,1804.09193,1906.08288,1912.06413,1411.5745,1712.01204,1502.06120,1806.01872,2008.04376,2105.08739,2106.10741}. In a typical classical gravitational scattering scenario, one provides initial scattering data such as masses, velocities, sizes, intrinsic angular momenta, and impact parameters of the scattered objects, along with the specified interaction among them. The goal is to determine the gravitational waveform as an output. However, the classical limit of the universal soft graviton theorem directly provides the low-frequency gravitational waveform in terms of both the initial and final scattering data, regardless of the knowledge of the interaction involved in the scattering process. This suggests a novel approach for deriving low-frequency and late-time gravitational waveforms by directly studying classical gravitational scattering processes with both initial and final scattering data, known as the classical soft graviton theorem. This approach has been successfully pursued in \cite{1906.08288,1912.06413,2008.04376,2106.10741}. The derivation of the classical soft graviton theorem readily extends to higher orders in the low-frequency expansion of the gravitational waveform, and numerous higher-order terms have been derived. However, deriving their quantum counterparts from the analysis of scattering amplitudes proves to be challenging in general, as discussed in the main body of the paper.

Consider a gravitational scattering amplitude involving $N$ number of finite energy particles (hard particles) with momenta, spins and polarizations  $\lbrace p_i , \Sigma_i,\epsilon_i \rbrace $ for $i=1,2,\cdots ,N$ ($N\geq 4$) and one low-energy (soft) graviton\footnote{The graviton is the particle created by operating metric fluctuation field $h_{\mu\nu}(x)$ on vacuum, where the classical metric fluctuation is defined by $h_{\mu\nu}(x)=\f{1}{2\sqrt{8\pi G}}\left(g_{\mu\nu}(x)-\eta_{\mu\nu}\right)$.} with momentum and polarization $k,\varepsilon$, and denote this scattering amplitude by $\mathcal{A}^{(N+1)}\big(\lbrace \epsilon_i,p_i ,\Sigma_i\rbrace ,\varepsilon ,k\big)$. We are following the convention that all the particles are ingoing, so if some of the particles are outgoing then we have to flip the sign of the four momenta for those particles. We will treat the soft graviton as outgoing with energy $\omega$, so according to our convention $k^\mu=-\omega \mathbf{n}^\mu$ with $\mathbf{n}^\mu$ being the null vector whose spatial part denotes the direction of soft graviton emission. Now the soft expansion of this $(N+1)$ particle amplitude takes the following form
\begin{align}
\mathcal{A}^{(N+1)}\big(\lbrace \epsilon_i,p_i ,\Sigma_i\rbrace ,\varepsilon ,k\big)=\sqrt{8\pi G}&\sum_{i=1}^{N}\epsilon_{i,\alpha} \left[(\mathbb{S}^{gr}_{\text{tree}})^{\ \alpha}_{\beta}+(\mathbb{S}^{gr}_{\text{1-loop}})^{\ \alpha}_{\beta}+(\mathbb{S}^{gr}_{\text{2-loop}})^{\ \alpha}_{\beta}+\cdots\right]\nn\\
&\times\mathcal{A}_{(i)}^{(N)\beta}\left(p_i\right)\ , \label{Soft_theorem}
\end{align}
where $\mathcal{A}_{(i)}^{(N)\beta}(p_i)$ represents the $i$'th particle polarisation $(\epsilon_{i\beta})$ stripped $N$-particle amplitude $\mathcal{A}^{(N)}\big(\lbrace \epsilon_i,p_i ,\Sigma_i\rbrace\big)$,  which is defined by the following relation
\be
\mathcal{A}^{(N)}\big(\lbrace \epsilon_i,p_i ,\Sigma_i\rbrace\big)\equiv \epsilon_{i\beta}\ \mathcal{A}_{(i)}^{(N)\beta}\left(p_i\right)\ .\label{eq:stripped_amp_def}
\ee
In \eqref{Soft_theorem}, the expression of tree level ``soft factor''\footnote{The actual tree level soft factor should be think of the  expression \eqref{eq:S_tree_gr} with a sum over hard particles from $i=1,2,\cdots, N$. Throughout the whole paper, we follow the same terminology ``soft factor'' referring to the soft factor expression without hard particle sum.} for single soft graviton emission reads \cite{weinberg1,weinberg2,jackiw1,jackiw2,1103.2981,1401.7026,1404.4091,1405.3533,1406.6987,1706.00759,1406.6574,1703.00024,1611.07534}
\be
(\mathbb{S}^{gr}_{\text{tree}})^{\ \alpha}_{\beta}&=& \f{\varepsilon_{\mu\nu}p_i^\mu p_i^\nu}{p_i\cdot k}\delta^\alpha_\beta +\f{\varepsilon_{\mu\nu}p_i^\mu k_\rho }{p_i\cdot k}\Bigg(\Big\lbrace p_i^\nu \f{\p }{\p p_{i\rho}}-p_i^\rho \f{\p}{\p p_{i\nu}}\Big\rbrace \delta^\alpha_\beta +\big(\Sigma_i^{\rho\nu}\big)_\beta^{\ \alpha}\Bigg)\nn\\
&&+\mathcal{O}(\omega^n, n\geq 1)\ .\label{eq:S_tree_gr}
\ee
In $D=4$ analyzing tree level scattering amplitudes for effective field theory the non-universal sub-subleading soft graviton factor at order $\omega$ has also been derived in \cite{1706.00759,1611.07534}. Reference \cite{1706.00759} explicitly evaluated the non-universal contribution to the sub-subleading soft factor in terms of the non-minimal coupling of two finite energy fields to a soft graviton field through the Riemann tensor, and  the general structure of the three-point 1PI (one-particle irreducible) vertex involving two hard particles and a soft graviton. By extending the analysis of  \cite{1706.00759}, it becomes evident that a complete soft factorization is not achievable at order $\omega^n$ for $n\geq 2$ in a generic theory of quantum gravity involving all possible higher derivative corrections allowed by general covariance. However, a partial soft factorization has been accomplished by enforcing linearized gauge invariance of the $(N+1)$-particle amplitude in \cite{1801.05528,1802.03148}. The generalization of the tree-level soft factor $\mathbb{S}^{gr}_{\text{tree}}$ for multiple soft graviton emissions up to subleading order can be found in \cite{1503.04816,1504.05558,1504.05559,1507.00938,1604.02834,1607.02700,1702.02350,
1705.06175, 1707.06803, 1709.07883, 1809.01675}.

The one-loop contribution to the ``soft factor'' for single soft graviton emission in \eqref{Soft_theorem} reads
\be
(\mathbb{S}^{gr}_{\text{1-loop}})^{\ \alpha}_{\beta}&=& \ K_{phase}^{reg}\ \f{\varepsilon_{\mu\nu}p_i^\mu p_i^\nu}{p_i\cdot k}\delta^\alpha_\beta +\f{\varepsilon_{\mu\nu}p_i^\mu k_\rho }{p_i\cdot k} \left\lbrace p_i^\nu \f{\p K_{gr}^{reg}}{\p p_{i\rho}}-p_i^\rho \f{\p K_{gr}^{reg}}{\p p_{i\nu}}\right\rbrace \delta^\alpha_\beta \non\\
&& +\ K_{phase}^{reg}\ \f{\varepsilon_{\mu\nu}p_i^\mu k_\rho }{p_i\cdot k}\left(\left\lbrace p_i^\nu \f{\p }{\p p_{i\rho}}-p_i^\rho \f{\p}{\p p_{i\nu}}\right\rbrace \delta^\alpha_\beta +\big(\Sigma_i^{\rho\nu}\big)_\beta^{\ \alpha}\right)\non\\
&&+\f{\varepsilon_{\mu\nu} k_\rho k_\sigma }{p_i\cdot k}\left\lbrace p_i^\mu \f{\p K_{gr}^{reg}}{\p p_{i\rho}}-p_i^\rho \f{\p K_{gr}^{reg}}{\p p_{i\mu}}\right\rbrace \left(\left\lbrace p_i^\nu \f{\p }{\p p_{i\sigma}}-p_i^\sigma \f{\p}{\p p_{i\nu}}\right\rbrace \delta^\alpha_\beta +\big(\Sigma_i^{\sigma\nu}\big)_\beta^{\ \alpha}\right)\nn\\
&&\ +\ \mathcal{O}(\omega^n,n\geq 0)\  +\ \mathcal{O}(\omega^n\ln\omega,n\geq 2)\ , \label{S1_loop_gr}
\ee
where
\begingroup
\allowdisplaybreaks
\begin{align}
K^{reg}_{gr}\ =\ \f{i}{2}\ (8\pi G)\sum_{\ell=1}^{N}\  \sum_{\substack{j=1\\j\neq \ell}}^{N}\Big{\lbrace} (p_{\ell}.p_{j})^{2}-\f{1}{2}p_{\ell}^{2}p_{j}^{2}\Big{\rbrace}\ \int_{\omega}^{\Lambda} \f{d^{4}\ell}{(2\pi)^{4}} \ \f{1}{\ell^{2}-i\epsilon}\ \f{1}{(p_{\ell}\cdot \ell +i\epsilon)\ (p_{j}\cdot \ell - i\epsilon)}\nn\\
\simeq  -iG\ (\ln\omega)\sum_{\ell=1}^{N} \sum_{\substack{j=1\\j\neq \ell}}^{N} \ \f{\Big{\lbrace} (p_{\ell}.p_{j})^{2}-\f{1}{2}p_{\ell}^{2}p_{j}^{2}\Big{\rbrace}}{\sqrt{(p_{\ell}.p_{j})^{2}-p_{\ell}^{2}p_{j}^{2}}} \Bigg{\lbrace}\delta_{\eta_{\ell}\eta_{j},1}
-\f{i}{2\pi}\ln\Bigg(\f{p_{\ell}.p_{j}+\sqrt{(p_{\ell}.p_{j})^{2}-p_{\ell}^{2}p_{j}^{2}}}{p_{\ell}.p_{j}-\sqrt{(p_{\ell}.p_{j})^{2}-p_{\ell}^{2}p_{j}^{2}}}\Bigg)\Bigg{\rbrace}\ , \label{K_gr_reg}
\end{align}
\endgroup 
and
\be
K_{phase}^{reg}\ &=&\ i\ (8\pi G)\ \sum_{j=1}^{N}(p_{j}.k)^{2}\int_{R^{-1}}^{\omega}\f{d^{4}\ell}{(2\pi)^{4}}\ \f{1}{\ell^{2}-i\epsilon}\f{1}{k.\ell+i\epsilon}\f{1}{p_{j}.\ell -i\epsilon}\non\\
&\simeq &\ -2iG\ (\ln\omega) \ \Bigg[ \sum_{\substack{j=1\\ \eta_{j}=-1}}^{N}p_{j}.k\ -\ \f{i}{2\pi}\sum_{\substack{j=1}}^{N}p_{j}.k\ \ln\Bigg(\f{p_{j}^{2}}{(p_{j}.\mathbf{n})^{2}}\Bigg)\Bigg]\ .\label{K_phase_reg}
\ee
In the expressions \eqref{K_gr_reg} and \eqref{K_phase_reg}, $\eta_j=+1$ if $j$'th particle is ingoing and $\eta_j=-1$ if $j$'th particle is outgoing. Under the sign $\simeq $ we only keep the logarithmic contributions after performing the integrations following \cite{1808.03288}. The upper limit of the integration $\Lambda$ in \eqref{K_gr_reg} represents the order of the energy of hard particles and  the lower limit of the integration $R^{-1}$ in \eqref{K_phase_reg} represents the order of the energy resolution of the detector. The first line of \eqref{S1_loop_gr} contains the $\mathcal{O}(\ln\omega)$ soft factor which has been derived in  \cite{1808.03288} as an one-loop exact result, analyzing one-loop gravitational S-matrices in the theory of scalar coupled to gravity. In \cite{1808.03288} a correction to $\mathcal{O}(\ln\omega)$ soft graviton factor due to electromagnetic  interaction has also been derived when the scattering particles carry some electric charges as well. In this article we re-derive the $\mathcal{O}(\ln\omega)$ soft factor for single soft graviton emission in a generic theory of quantum gravity for scattering of particles with arbitrary mass and spin. This investigation will demonstrate the universal (independent of theory) nature of the $\mathcal{O}(\ln\omega)$ soft factor, while also extending the infrared divergence factorization prescription proposed in \cite{1808.03288} to encompass a broad range of quantum gravity theories. In this paper we also conjecture the order $\omega\ln\omega$ soft factor given in the second and third lines of \eqref{S1_loop_gr} which is derivable from the analysis of one-loop scattering amplitude for the scattering generic spinning particles in a generic theory of quantum gravity extending the analysis of this article. The classical limit of this $\mathcal{O}(\omega\ln\omega)$ soft factor has already been derived in \cite{2106.10741} in the name of spin-dependent classical soft graviton theorem which provides evidence on the correctness of the conjecture.\footnote{For the scattering of non-spinning particles, the $G^2\omega\ln\omega$ waveform is consistent with the results from \cite{1812.08137,1901.10986,DiVecchia:2023frv}, as recently confirmed by Paolo Di Vecchia. When it comes to the scattering of spinning objects, it should also align with the findings of \cite{Jakobsen:2021lvp} when one takes soft limit, as will be reported in a revised version of \cite{Aoude:2023dui}, communicated by Carlo Heissenberg.} Note that the tree level subleading soft theorem result at order $\omega^0$ in \eqref{eq:S_tree_gr} is not universal as it receives correction at one-loop order, which is expected to be dependent on the theory as well as the value of detector resolution (IR regulator) \cite{1405.1015,1405.1410}. 

The two-loop contribution to the ``soft factor'' for single soft graviton emission in \eqref{Soft_theorem} reads
\be
(\mathbb{S}^{gr}_{\text{2-loop}})^{\ \alpha}_{\beta}&=&   \f{1}{2}\Big{\lbrace}K_{phase}^{reg}\Big{\rbrace}^{2}\ \f{\varepsilon_{\mu\nu}p_{i}^{\mu}p_{i}^{\nu}}{p_{i}\cdot k}\delta^\alpha_\beta+\ K_{phase}^{reg}\f{\varepsilon_{\mu\nu}p_{i}^{\mu}k_{\rho}}{p_{i}\cdot k}\Bigg(p_{i}^{\nu}\f{\p K_{gr}^{reg}}{\p p_{i\rho}}-p_{i}^{\rho}\f{\p K_{gr}^{reg}}{\p p_{i\nu}}\Bigg)\delta^\alpha_\beta \non\\
&&  + \f{1}{2}\sum_{i=1}^{N}\f{\varepsilon_{\mu\nu}k_{\rho}k_{\sigma}}{p_{i}\cdot k} \Bigg(p_{i}^{\mu}\f{\p K_{gr}^{reg}}{\p p_{i\rho}}-p_{i}^{\rho}\f{\p K_{gr}^{reg}}{\p p_{i\mu}}\Bigg)\Bigg(p_{i}^{\nu}\f{\p K_{gr}^{reg}}{\p p_{i\sigma}}-p_{i}^{\sigma}\f{\p K_{gr}^{reg}}{\p p_{i\nu}}\Bigg)\ \delta^\alpha_\beta \non\\
 &&\ +\ \mathcal{O}(\omega^n(\ln\omega)^2, n\geq 2)\ +\ \mathcal{O}(\omega^n\ln\omega, n\geq 1)\ +\mathcal{O}(\omega^n, n\geq 1)\ . \label{S2_loop_gr}
\ee
The above result was conjectured in \cite{2008.04376} as a two-loop exact result and can be obtained by analyzing two-loop amplitudes using the same methodology being developed in this paper. The classical limit of this order, denoted as $\mathcal{O}\big(\omega(\ln\omega)^2\big)$, was derived under the name of the sub-subleading classical soft graviton theorem in \cite{2008.04376}. This classical derivation offers substantial evidence supporting the validity of the above two-loop soft factor. Note that the $\mathcal{O}(\omega \ln \omega)$ soft factor in the second and third lines of \eqref{S1_loop_gr} at one-loop receives a correction at two-loop order. This correction is expected to depend on the theory of quantum gravity and the value of detector resolution (IR regulator), hence non-universal. From the analysis of the $n$-loop gravitational S-matrix, it is expected that the new leading non-analytic soft graviton factor, as the frequency $\omega$ approaches zero, behaves like $\omega^{n-1}(\ln\omega)^n$ relating it to tree level $N$-particle amplitude. The general structure of the order $\omega^{n-1}(\ln\omega)^n$ soft graviton theorem is provided in \cite{2008.04376}.

Note that the ``tree'', ``1-loop'' , ... subscripts in the soft factors in the expression \eqref{Soft_theorem} only specifies the first appearance of the soft factors in the perturbative analysis of the $(N+1)$-particle amplitudes at that order (tree or $n$-loop) and it multiplies to the corresponding tree-level $N$-particle amplitudes. But they also appears in the analysis of higher loop amplitudes as well. For example $\mathbb{S}^{gr}_{\text{tree}}$ also appears as a soft factor in the analysis of $(N+1)$-particle $n$-loop amplitude relating it to the $N$-particle $n$-loop amplitude for all $n\geq 1$. Similarly $\mathbb{S}^{gr}_{\text{1-loop}}$ also appears as a soft factor in the analysis of $(N+1)$-particle $n$-loop amplitude relating it to the $N$-particle $(n-1)$-loop amplitude for all $n\geq 2$, and $\mathbb{S}^{gr}_{\text{2-loop}}$ also appears as a soft factor in the analysis of $(N+1)$-particle $n$-loop amplitude relating it to the $N$-particle $(n-2)$-loop amplitude for all $n\geq 3$. These observations also apply to the soft photon theorem results provided below.

Now let us consider the same setup of scattering but turn off the gravitational interaction and turn on electromagnetic interaction between charged particles. We consider the finite energy scattered particles carry some electric charges $\lbrace e_i \rbrace$ and study scattering amplitude involving one soft photon emission with polarization and momentum $(\varepsilon , k )$. So in this case the soft expansion of $(N+1)$ particle amplitude takes the following form
\begin{align}
\mathcal{A}^{(N+1)}\big(\lbrace \epsilon_i,p_i ,e_i,\Sigma_i\rbrace ,\varepsilon ,k\big)=&\sum_{i=1}^{N}\epsilon_{i,\alpha} \left[(\mathbb{S}^{em}_{\text{tree}})^{\ \alpha}_{\beta}+(\mathbb{S}^{em}_{\text{1-loop}})^{\ \alpha}_{\beta}+(\mathbb{S}^{em}_{\text{2-loop}})^{\ \alpha}_{\beta}+\cdots\right]\nn\\
&\times\mathcal{A}_{(i)}^{(N)\beta}\left(p_i\right)\ , \label{soft_photon_theorem}
\end{align}
where $\mathcal{A}_{(i)}^{(N)\beta}\left(p_i\right)$ is defined through the relation \eqref{eq:stripped_amp_def}. In \eqref{soft_photon_theorem}, the expression of tree level ``soft factor'' for single soft photon emission reads \cite{Gell-Mann,Low1,low,saito,burnett,bell,duca,weinberg1,weinberg2,jackiw1,jackiw2,1611.07534,1809.01675}
\be
(\mathbb{S}^{em}_{\text{tree}})^{\ \alpha}_{\beta}=&&e_i\f{\varepsilon_{\mu}p_i^\mu }{p_i\cdot k}\delta^\alpha_\beta +e_i\f{\varepsilon_{\mu} k_\rho }{p_i\cdot k}\left(\left\lbrace p_i^\mu \f{\p }{\p p_{i\rho}}-p_i^\rho \f{\p}{\p p_{i\mu}}\right\rbrace \delta^\alpha_\beta +\big(\mathcal{N}_i^{\rho\mu}(-p_i)\big)_\beta^{\ \alpha}\right)\nn\\
&&\ +\mathcal{O}\left(\omega^n,n\geq 1\right),
\ee
where the generic expression for the non-universal term $ \mathcal{N}_{i}^{\rho\sigma}(-p_i)$ contributing to the subleading soft photon theorem has been derived in \cite{1809.01675} and its explicit form in provided in \eqref{N_non-universal}. $ \mathcal{N}_{i}^{\rho\mu}(-p_i)$ depends on the non-minimal coupling of two finite energy fields to a soft photon field through the field strength, and the general structure of the three-point 1PI vertex involving two hard particles and a soft photon. Extending the analysis of  \cite{1809.01675} it can be argued that a complete soft factorization of order $\omega^n$ for $n\geq 1$ is not possible in a generic theory of QED with all possible non-minimal couplings allowed by $U(1)$ gauge invariance, though a partial soft factorization is achievable by enforcing gauge invariance of the $(N+1)$-particle amplitude \cite{1802.03148,1801.05528}. The generalization of the tree-level soft factor $\mathbb{S}^{em}_{\text{tree}}$ for multiple soft photon emissions up to subleading order can be found in \cite{1809.01675}.

The one-loop contribution to the ``soft factor'' for single soft photon emission in \eqref{soft_photon_theorem} reads
\be
(\mathbb{S}^{em}_{\text{1-loop}})^{\ \alpha}_{\beta}&=& e_i\f{\varepsilon_{\mu} k_{\rho} }{p_i\cdot k} \left\lbrace p_i^\mu \f{\p K_{em}^{reg}}{\p p_{i\rho}}-p_i^\rho \f{\p K_{em}^{reg}}{\p p_{i\mu}}\right\rbrace \delta^\alpha_\beta\nn\\
&&\  +\mathcal{O}(\omega^n\ln\omega, n\geq 1)\ +\mathcal{O}(\omega^n, n\geq 0)\ , \label{eq:S_1loop_photon}
\ee
where
\be
&&K_{em}^{reg}= \f{i}{2}\sum_{\ell=1}^{N}\  \sum_{\substack{j=1\\j\neq \ell}}^{N}e_{\ell}e_{j}(p_{\ell}\cdot p_{j})\ \int_\omega^\Lambda \f{d^{4}\ell}{(2\pi)^{4}} \ \f{1}{\ell^{2}-i\epsilon}\ \f{1}{(p_{\ell}\cdot \ell +i\epsilon)\ (p_{j}\cdot\ell - i\epsilon)}\non\\
&\simeq & -\f{i}{2}\sum_{\ell =1}^{N}  \sum_{\substack{j=1\\j\neq \ell}}^{N}\f{e_{\ell}e_{j}}{4\pi} (\ln\omega)  \f{p_{\ell}\cdot p_{j}}{\sqrt{(p_{\ell}.p_{j})^{2}-p_{\ell}^{2}p_{j}^{2}}} \Bigg{\lbrace}\delta_{\eta_{\ell}\eta_{j},1}-\f{i}{2\pi}\ln\Bigg(\f{p_{\ell}.p_{j}+\sqrt{(p_{\ell}.p_{j})^{2}-p_{\ell}^{2}p_{j}^{2}}}{p_{\ell}.p_{j}-\sqrt{(p_{\ell}.p_{j})^{2}-p_{\ell}^{2}p_{j}^{2}}}\Bigg)\Bigg{\rbrace} .\non\\
\ee
The $\mathcal{O}(\ln\omega)$ soft factor for single soft photon emission in \eqref{eq:S_1loop_photon} has been derived in  \cite{1808.03288} as an one-loop exact result, analyzing one-loop S-matrices in the theory of minimally coupled scalar QED. There a correction to $\mathcal{O}(\ln\omega)$ soft photon factor due to gravitational interaction has also been derived. In this article we re-derive the $\mathcal{O}(\ln\omega)$ soft factor in a generic theory of quantum gravity for scattering of particles with arbitrary mass, charge and spin in presence of non-minimal coupling. This investigation will demonstrate the universal (independent of theory) nature of the $\mathcal{O}(\ln\omega)$ soft factor, while also extending the infrared divergence factorization prescription proposed in \cite{grammer,1808.03288} to encompass a broad range effective field theories for charged objects. The generalization of the one-loop soft factor $\mathbb{S}^{em}_{\text{1-loop}}$ for multiple soft photon emissions up to subleading order can be found in the section-(3.5) of \cite{Sahoo:2020csy}, and the final result has been provided in \eqref{eq:multi_soft_photon}.

The two-loop contribution to the ``soft factor'' for single soft photon emission in \eqref{soft_photon_theorem} reads
\be
(\mathbb{S}^{em}_{\text{2-loop}})^{\ \alpha}_{\beta}&=& \f{e_{i}}{2}\f{\varepsilon_{\mu}k_{\rho}}{p_{i}\cdot k}\ \Bigg(p_{i}^{\mu}\f{\p K_{em}^{reg}}{\p p_{i\rho}}\ -\ p_{i}^{\rho}\f{\p K_{em}^{reg}}{\p p_{i\mu}} \Bigg)\ \Bigg(  k_{\sigma}\f{\p K_{em}^{reg}}{\p p_{i\sigma}}\Bigg)\delta^\alpha_\beta \non\\
&&  +\ (\ln \omega)^{2}\ e_i\big( \varepsilon^{\mu} k^{\rho}-\varepsilon^\rho k^\mu \big) p_{i\mu}\mathcal{C}^{reg}_{\rho}\big(q_{i},p_{i};\lbrace e_{j}\rbrace ,\lbrace p_{j}\rbrace\big)\delta^\alpha_\beta\nn\\
&& +\ \mathcal{O}(\omega^n(\ln\omega)^2, n\geq 2)+\ \mathcal{O}(\omega^n\ln\omega, n\geq 1)\ +\mathcal{O}(\omega^n, n\geq 1)\ ,\label{eq:S_2loop_photon}
\ee
where
\begingroup
\allowdisplaybreaks
\be
&&\mathcal{C}^{reg}_{\rho}\big(q_{i},p_{i};\lbrace e_{j}\rbrace ,\lbrace p_{j}\rbrace\big)\non\\
 &=&-\sum_{\substack{j=1\\ j\neq i}}^{N}\sum_{\substack{\ell =1\\ \ell\neq i}}^{N}\f{e_{i}^{2}e_{j}e_{\ell}}{4}\ \lbrace p_{i}.p_{j}\delta_{\rho}^{\kappa}-p_{i}^{\kappa}p_{j\rho}\rbrace \f{\p}{\p p_{i\sigma}}\Big{\lbrace}\mathcal{I}(p_{i},p_{\ell})\times p_{i}.p_{\ell}\Big{\rbrace} \f{\p^2 \mathcal{I}(p_{i},p_{j})}{\p p_{i}^{\sigma}\p p_i^\kappa}\non\\
 &&\ +\sum_{\substack{j=1\\ j\neq i}}^{N}\sum_{\substack{\ell=1\\ \ell\neq j}}^{N}\f{e_{i}e_{j}^{2}e_{\ell}}{4}\ \lbrace p_{i}.p_{j}\delta_{\rho}^{\kappa}-p_{i}^{\kappa}p_{j\rho}\rbrace \f{\p}{\p p_{j\sigma}}\Big{\lbrace}\mathcal{I}(p_{j},p_{\ell})\times p_{j}.p_{\ell}\Big{\rbrace}\f{\p^2 \mathcal{I}(p_{i},p_{j})}{\p p_{i}^{\sigma}\p p_i^\kappa}\ ,
\ee
\endgroup
with
\be
\mathcal{I}(p_{i},p_{j})= -\f{1}{4\pi} \f{1}{\sqrt{(p_{i}.p_{j})^{2}-p_{i}^{2}p_{j}^{2}}}\left\lbrace\delta_{\eta_{i}\eta_{j},1}-\f{i}{2\pi}\ln\left(\f{p_{i}.p_{j}+\sqrt{(p_{i}.p_{j})^{2}-p_{i}^{2}p_{j}^{2}}}{p_{i}.p_{j}-\sqrt{(p_{i}.p_{j})^{2}-p_{i}^{2}p_{j}^{2}}}\right)\right\rbrace.
\ee
The $\mathcal{O}\left(\omega(\ln\omega)^2\right)$ soft factor for single soft photon emission in \eqref{eq:S_2loop_photon} has been derived in the section-4 of \cite{2008.04376} as a two-loop exact result, analyzing two-loop S-matrices in the theory of minimally coupled scalar QED. From the analysis of the $n$-loop QED S-matrix, it is expected that the new leading non-analytic soft factor for single photon emission, as the frequency $\omega$ approaches zero, behaves like $\omega^{n-1}(\ln\omega)^n$ and it relates to the tree level $N$-particle amplitude. The general structure of the order $\omega^{n-1}(\ln\omega)^n$ soft photon theorem is provided in \cite{2008.04376}.

The rest of the paper is organized as follows:
In section-\ref{S:strategy}, we establish our conventions and describe the general definition of IR-finite scattering amplitudes. We also discuss the EFT action involving massive spinning particles which transform in a generic reducible representation of the Lorentz group.
In section-\ref{S:soft_photon_thm}, we review the covariantization prescription and define one-loop IR-finite QED S-matrices involved in the derivation of the soft photon theorem. Starting from the IR-finite S-matrices, we derive the soft photon theorem up to subleading order.
In section-\ref{S:soft_graviton_theorem}, after reviewing Sen's covariantization prescription, we define the one-loop IR-finite quantum gravity S-matrices that are involved in the derivation of the soft graviton theorem. Starting from the IR-finite S-matrices, we derive the soft graviton theorem up to subleading order.
At the end of both section-\ref{S:soft_photon_thm} and \ref{S:soft_graviton_theorem}, we discuss the possible generalizations of our derivations to higher orders.
In section-\ref{S:summary_outlook}, we provide some open directions to explore in the future after reviewing what we have been achieved in this article.

\section{Setup and Strategy }\label{S:strategy}
\paragraph{Index convention:} We utilize the first few Latin alphabets $a, b, c, d, \ldots$ as Lorentz indices for the tangent space, ranging from 0 to 3. The Latin alphabets starting from $i, j, k, \ell, \ldots$ are employed as indices for identifying individual hard particles, ranging from 1 to N. The first few Greek alphabets $\alpha, \beta, \gamma, \delta, \ldots$ are used as polarization indices for spinning particles on the tangent space, while the Greek alphabets beginning with $\lambda, \mu, \nu, \rho, \sigma, \tau, \ldots$ serve as curved space indices, ranging from 0 to 3. In section-\ref{S:soft_photon_thm}, where we derive the soft photon theorem solely under electromagnetic interaction in a flat background, we employ both $a, b, c, d, \ldots$ and  $\lambda, \mu, \nu, \rho, \sigma, \tau, \ldots$ as flat space Lorentz indices.

\paragraph{Metric and unit conventions:} In our convention four dimensional Minkowski metric is $\eta_{ab}=\text{diag}(-1,+1,+1,+1)$. We work in the unit where speed of light $c=1$ and Planck constant $\hbar=1$ but keep the gravitational constant $G$ explicit. We define $\kappa\equiv \sqrt{8\pi G}$.

\paragraph{Setup of scattering event:}
Let us consider a scattering amplitude involving $N$ number of finite energy massive particles (hard particles) with charges, momenta, spins and polarizations  $\lbrace e_i, p_i , \Sigma_i,\epsilon_i \rbrace $ for $i=1,2,\cdots ,N$ and one low-energy (soft) outgoing photon/graviton with momenta and polarization $k,\varepsilon$, and denote this scattering amplitude by $\mathcal{A}^{(N+1)}$. In our convention we consider all the particles are incoming, so if some particles are outgoing we need to flip the sign of four momenta and electric charges for those particles. The energy of outgoing soft photon/graviton is denoted by $\omega$ so that $k^{\mu}=-\omega \mathbf{n}^{\mu}$ where $\mathbf{n}^\mu$ being the null vector whose spatial part denotes the direction of soft photon/graviton emission. Here we are only interested to evaluate $\mathcal{A}^{(N+1)}$ at one-loop order which involves Feynman diagrams involving one virtual photon/graviton running in the loop. Then we perform soft  expansion ($\omega<<|p_i|$) of $\mathcal{A}^{(N+1)}$ to relate it with the $N$ point amplitude which carries all the hard particles in the asymptotic state but no soft graviton, denoted by  $\mathcal{A}^{(N)}$.\footnote{Note that the soft limit can also be defined covariantly by demanding $\Big{|}\f{p_i.k}{p_i.p_j}\Big{|}<<1$ for all $i,j=1,\cdots ,N$.} Note that both the scattering amplitudes $\mathcal{A}^{(N)}$ and $\mathcal{A}^{(N+1)}$ are distributions in momenta as  $\mathcal{A}^{(N)}$ contains momentum conserving delta function $\delta^{(4)}\big(p_1+p_2+\cdots +p_N\big)$ and $\mathcal{A}^{(N+1)}$ contains momentum conserving delta function $\delta^{(4)}\big(p_1+p_2+\cdots +p_N+k\big)$. In four spacetime dimensions ($D=4$), both scattering amplitudes exhibit infrared (IR) divergences. Therefore, our first step is to separate out the IR divergent contributions from both the scattering amplitudes in an unambiguous manner. Then we can obtain the soft factor by examining the ratio of $\mathcal{A}^{(N+1)}$ and $\mathcal{A}^{(N)}$ after full/partial cancellation of the IR divergent contributions as we explain in later sections.
\paragraph{Feynman diagram conventions:} In all the Feynman diagrams describing scattering amplitudes, time flows from right to left and the particles involved in the scattering will always be treated ingoing. Solid lines in any diagrams corresponds to massive spinning particles and dashed lines represent photons/gravitons. If in a figure multiple Feynman diagrams appears, the counting of their numbers are always from left to right and from top to bottom. A Feynman diagram will be called an $n$-loop diagram only if the diagram contains $n$ number of loops where at-least one virtual photon/graviton is propagating in each loop. The loops involving only massive virtual particles will be taken care of inside the massive EFT 1PI vertices and renormalized propagators of the massive EFT. To determine the Feynman rules for vertices involving photons/gravitons and hard particles, we follow the covariantization technique developed in the photon/graviton background in the references \cite{1703.00024,1706.00759,1707.06803,1809.01675}. 

\paragraph{Handling IR divergences in the derivation of soft theorem:}
The traditional S-matrix in quantum electrodynamics and quantum gravity, in four spacetime dimensions, exhibits IR-divergence. This is due to the long-range nature of the interactions involved. Previous attempts to construct IR-finite S-matrices, beginning with the Kulish-Faddeev construction \cite{Faddeev-Kulish}, demonstrated explicit cancellation of IR divergences. However, a systematic method for extracting the unambiguous IR finite part remained absent, until Grammer and Yennie provided one in \cite{grammer}. A generalization of Grammer-Yennie prescription for perturbative QCD and quantum gravity can be found in \cite{senqcd} and  \cite{1808.03288} respectively. 

In the derivation of soft photon theorem, Grammer-Yennie prescription helps to factor out IR divergences from both the amplitudes $\mathcal{A}^{(N+1)}$ and $\mathcal{A}^{(N)}$ in the following way
\be
\mathcal{A}^{(N)}=\exp\lbrace K_{em}\rbrace\ \mathcal{A}^{(N)}_{\text{IR-finite}}\hspace{0.5cm} ,\hspace{0.5cm} \mathcal{A}^{(N+1)}=\exp\lbrace K_{em}\rbrace\ \mathcal{A}^{(N+1)}_{\text{IR-finite}}\ . \label{eq:QED_factorization}
\ee 
Above the exponential factor containing $K_{em}$ takes care of the full IR divergent contribution and the IR divergent contributions are exactly same for both the amplitudes. An explicit expression of  $K_{em}$ is provided in \eqref{K_em}. Basically Grammer-Yennie prescription provides a systematic procedure to compute IR-finite parts perturbatively for both the amplitudes. When the soft factor $\mathbb{S}^{em}$ is a multiplicative function instead of differential operator, we get
\be
\mathcal{A}^{(N+1)}=\mathbb{S}^{em} \ \mathcal{A}^{(N)}\ \Longrightarrow\ \mathcal{A}^{(N+1)}_{\text{IR-finite}}=\mathbb{S}^{em} \ \mathcal{A}^{(N)}_{\text{IR-finite}}\ .
\ee
Hence to derive the $\mathcal{O}(\ln\omega)$ and $\mathcal{O}\left(\omega(\ln\omega)^2\right)$ soft factors in \eqref{eq:S_1loop_photon} and \eqref{eq:S_2loop_photon} we can directly start from one and two-loop contributions of IR finite amplitude $\mathcal{A}^{(N+1)}_{\text{IR-finite}}$ and perform soft expansion.\footnote{In \cite{Himwich:2020rro} a similar analysis was conducted to deduce Weinberg's soft photon and graviton theorems from all loop order amplitudes. In this context, $\exp\lbrace K_{em}\rbrace$ is referred to as the soft S-matrix generated from a product of electromagnetic/gravitational Wilson line operators, while $\mathcal{A}^{(N)}_{\text{IR-finite}}$ is designated as the hard S-matrix. }

In the derivation of soft graviton theorem, $\mathcal{A}^{(N+1)}$ contains some extra divergent factors relative to $\mathcal{A}^{(N)}$ due to Feynman diagrams involving three graviton self-interaction vertices. An optimistic dream of the factorization IR divergence using the Grammer-Yennie decomposition proposed in \cite{1808.03288} would be (see also \cite{Naculich:2011ry, 1405.1410,1405.1015} for dimensional regularization)
\be
\mathcal{A}^{(N)}=\exp\lbrace K_{gr}\rbrace\ \mathcal{A}^{(N)}_{\text{IR-finite}}\hspace{0.5cm} ,\hspace{0.5cm} \mathcal{A}^{(N+1)}=\exp\lbrace K_{gr}+K_{phase}\rbrace\ \mathcal{A}^{(N+1)}_{\text{IR-finite}}\ , 
\ee 
where the IR-divergent expressions of $K_{gr}$ and $K_{phase}$ are given in \eqref{eq:K_gr} and \eqref{eq:K_phase_div}. The result mentioned above has only been verified rigorously up to one-loop order. Verifying it for all loop orders is a computationally challenging task that remains open for future investigation. Now when the soft factor $\mathbb{S}^{gr}$ is a multiplicative function instead of differential operator, we get
\be
\mathcal{A}^{(N+1)}=\mathbb{S}^{gr} \ \mathcal{A}^{(N)}\ \Longrightarrow\ \mathcal{A}^{(N+1)}_{\text{IR-finite}}=\exp
\lbrace -K_{phase}\rbrace\ \mathbb{S}^{gr} \ \mathcal{A}^{(N)}_{\text{IR-finite}}\ . \label{eq:S_K_commute}
\ee
Hence to derive an unambiguous soft factor by analyzing $\mathcal{A}^{(N+1)}_{\text{IR-finite}}$, we need to regulate the IR divergence of $K_{phase}$ using a cut off given by the detector resolution. This procedure can be followed to derive $\mathcal{O}(\ln\omega)$ and $\mathcal{O}\left(\omega(\ln\omega)^2\right)$ soft factors in \eqref{S1_loop_gr} and \eqref{S2_loop_gr} respectively by analyzing one and two loop IR finite amplitudes. But if we want to derive the $\mathcal{O}(\omega\ln\omega)$ soft factor in the second and third lines of \eqref{S1_loop_gr} we need to deal with the following additional subtleties:
\begin{enumerate}
\item Since the order $\omega\ln\omega$ soft factor in \eqref{S1_loop_gr} is a differential operator we can not really commute the soft factor and infrared divergent exponential to get the second equation of \eqref{eq:S_K_commute}. Hence to derive the $\mathcal{O}(\omega\ln\omega)$ soft factor we have to start with the full divergent scattering amplitude $\mathcal{A}^{(N+1)}$ instead of its IR finite part and at the end of the analysis we may be able to cancel the common IR divergent factor appears in both amplitudes in the soft theorem relation.
\item Note that the momentum conserving delta function associated with $\mathcal{A}^{(N+1)}$ is \\ $\delta^{(4)} \left(\sum\limits_{i=1}^N p_i+k\right)$. On the other hand the momentum conserving delta function associated with $\mathcal{A}^{(N)}$ is $\delta^{(4)} \left(\sum\limits_{i=1}^N p_i\right)$. Now Taylor series expansion of the first delta function around small $\omega$ produces a correction of order $\omega$. This correction, when multiplied with the $\mathcal{O}(\ln\omega)$ soft factor, yields an additional factor of order $\omega\ln\omega$ at one-loop order. Therefore, this additional contribution needs to be accounted for, if it contributes something non-vanishing at this order.
\end{enumerate}
In light of these additional intricacies, we have decided to postpone the derivation of the order  $\omega\ln\omega$ soft graviton factor in \eqref{S1_loop_gr} for future study and focus on deriving the order $\ln\omega$ soft graviton theorem here.

\paragraph{EFT involving massive particles with arbitrary spin:}
We begin with an effective field theory (EFT) that describes the dynamics of massive spinning particles. The one-particle irreducible (1PI) effective action for this EFT is obtained by integrating out all massive loops. The tree level amplitudes computed using this massive EFT action contain information about all the loop orders in the original un-integrated massive quantum field theory (QFT). However, if the un-integrated QFT includes massless fields, our initial approach using the 1PI effective action becomes invalid. Nevertheless, our prescription for covariantization and computation of loop amplitudes, as described below, remains valid. In such cases, the 1PI effective action should be regarded as the tree level action for the EFT. Let $\Phi_{\alpha}(x)$ denotes the set of all massive fields in real representation, present in the 1PI effective action\footnote{In $\Phi_{\alpha}(x)$ we also can include massless finite energy particles e.g. hard photon or graviton, in that case our 1PI effective action should be thought of as tree level action.} which transforms in a reducible representation of Lorentz group $SO(1,3)$ in the following way,
\be
SO(1,3):\ && x^a \rightarrow x^{\prime a}=\Lambda^a\ _b \ x^b = (\delta^a_b +\lambda^{a}\ _b)x^b +\mathcal{O}(\lambda^2)\non\\
&& \Phi_{\alpha}(x)\rightarrow \Phi'_{\alpha}(x)=\ \left[\exp\Big\lbrace -\f{i}{2}\lambda_{ab}\Sigma^{ab}\Big\rbrace\right]_\alpha^{\ \beta}\ \Phi_{\beta}(\Lambda^{-1}x)\ ,
\ee
where $\lambda_{ab}=-\lambda_{ba}$ is the infinitesimal Lorentz transformation parameter and $\Sigma^{ab}$ is the spin angular momentum generator of $SO(1,3)$ transformation in the real reducible representation. The subscript index $\alpha$ is used as a combined notation for denoting different fields in the theory as well as the spin/polarization indices of each of the fields. Under global $U(1)_{EM}$ the field $\Phi(x)$ transforms in the following way, 
\be
U(1)_{EM}:\ \Phi_{\alpha}(x)\rightarrow \Phi'_{\alpha}(x)=\ \Big[\exp\big\lbrace i \mathcal{Q} \theta\big\rbrace \Big]_\alpha^{\ \beta}\ \Phi_{\beta}(x)\ ,
\ee
where $\theta$ is the parameter of global $U(1)_{EM}$ transformation and $\mathcal{Q}$ is the generator of global $U(1)_{EM}$ transformation in the real representation of $\Phi(x)$. Usually we associate $U(1)_{EM}$ global charge to complex fields but since we want to covariantize the theory simultaneously in the background of gravity and gauge theory together following \cite{1809.01675}, working in terms of real field components is convenient. For example instead of a complex scalar field we work with two real scalar fields considering them in a two component vector which rotates under $SO(2)$ and $\mathcal{Q}$ is the generator of $SO(2)$ transformation. In the set of fields denoted by $\Phi_{\alpha}(x)$, there may be some elementary fields in the irreducible representation of Lorentz group which does not transform under global $U(1)_{EM}$, for those fields the elements of the charge matrix $\mathcal{Q}$ will be zero.

Let us start with the general form of the quadratic part of the massive particle 1PI effective action\footnote{If the original theory contains some massless fields, then this action should be thought of as quadratic part of the tree level gauge fixed action. Because in presence of massless fields, the 1PI effective action of the theory may be non-local and the kinetic operator $\mathcal{K}(q)$ may not be polynomially expandable around $q^\mu =0$, which is the key assumption for the validity of the covariantization prescription discussed below. }
\be
S^{(2)}&=&\f{1}{2}\int \f{d^4 q_1}{(2\pi)^4}\f{d^4 q_2}{(2\pi)^4}\ (2\pi)^4\delta^{(4)}(q_1+q_2)\ \Phi_{\alpha}(q_1)\mathcal{K}^{\alpha\beta}(q_2)\Phi_{\beta}(q_2)\non\\
&=&\f{1}{2}\int \f{d^4 q_1}{(2\pi)^4}\f{d^4 q_2}{(2\pi)^4}\ (2\pi)^4\delta^{(4)}(q_1+q_2)\ \Phi^{T}(q_1)\mathcal{K}(q_2)\Phi(q_2)\label{S2}\ ,
\ee
where $\mathcal{K}(q)$ is the renormalized momentum space kinetic operator which satisfy the following condition:
\be
&&\mathcal{K}^{\alpha\beta}(q)=\pm \ \mathcal{K}^{\beta\alpha}(-q)\ ,\non\\
&& \mathcal{K}(q)=\pm \ \Big[\mathcal{K}(-q)\Big]^{T}\ .\label{K_property}
\ee
In the second lines of \eqref{S2} and \eqref{K_property} we introduced the index free notation, which we follow through out the article. In the RHS of the above equation, $+$ sign is for bosonic field and $-$ sign is for fermionic field. For simplicity we work with the $+$ sign considering $\Phi(x)$ being Grassmannian even, but the final result of soft theorem computation will be same for both bosonic and fermionic fields. The Feynman propagator for the $i$-th particle with renormalized mass $m_i$ from \eqref{S2} becomes
\be
\Delta^{i}_{\alpha\beta}(q)\ =\ i\big[\mathcal{K}_i^{-1}(q)\big]_{\alpha\beta}\equiv (q^2+m_i^2-i\epsilon)^{-1}\ \Xi_{i\alpha\beta}(q)\ ,\label{propagator}
\ee
where $\mathcal{K}_i(q)$ is the kinetic term for the set of fields representing the $i$-th particle after proper diagonalization of the quadratic part of the action $S^{(2)}$. The above equation also defines $\Xi_i(q)$ as the residue of the pole of the propagator for $i$-th particle. The relation between $\mathcal{K}_i$ and $\Xi_i$ and their momentum derivatives satisfy the following relations in the index free notation, which will be useful for later computation \cite{1703.00024}
\begingroup
\allowdisplaybreaks
\be
&&\mathcal{K}_i(q)\Xi_i(q)=i(q^2+m_i^2 -i\epsilon)\label{K1}\ ,\\
&& \f{\p \mathcal{K}_i(q)}{\p q^{a}}\Xi_i(q)=-\mathcal{K}_i(q)\f{\p \Xi_i(q)}{\p q^{a}}+2iq_a\label{K2}\ ,\\
&&  \f{\p^2 \mathcal{K}_i(q)}{\p q^{a} \p q^{b}}\Xi_i(q)=-\f{\p \mathcal{K}_i(q)}{\p q^a}\f{\p \Xi_i(q)}{\p q^{b}}-\f{\p \mathcal{K}_i(q)}{\p q^b}\f{\p \Xi_i(q)}{\p q^{a}}-\mathcal{K}_i(q)\f{\p^2 \Xi_i(q)}{\p q^{a}\p q^b}+2i\eta_{ab}\label{K3}\ ,\\
&&\Xi_i(q)\mathcal{K}_i(q)=i(q^2+m_i^2-i\epsilon)\label{K4}\ ,\\
&& \f{\p \Xi_i(q)}{\p q^{a}}\mathcal{K}_i(q)=-\Xi_i(q)\f{\p \mathcal{K}_i(q)}{\p q^{a}}+2iq_a\label{K5}\ ,\\
&&  \f{\p^2 \Xi_i(q)}{\p q^{a} \p q^{b}}\mathcal{K}_i(q)=-\f{\p \Xi_i(q)}{\p q^a}\f{\p \mathcal{K}_i(q)}{\p q^{b}}-\f{\p \Xi_i(q)}{\p q^b}\f{\p \mathcal{K}_i(q)}{\p q^{a}}-\Xi_i(q)\f{\p^2 \mathcal{K}_i(q)}{\p q^{a}\p q^b}+2i\eta_{ab}\ .\label{K6}
\ee
\endgroup
The Lorentz covariance of $\mathcal{K}_i$ and $\Xi_i$ implies the following two relations
\be
&&\big(\Sigma_i^{ab}\big)^{T}\mathcal{K}_i(q)=-\mathcal{K}_i(q)\Sigma_i^{ab}+q^a \f{\p \mathcal{K}_i(q)}{\p q_b}-q^b \f{\p \mathcal{K}_i(q)}{\p q_a}\ ,\label{SigmaK}\\
&& \Sigma_i^{ab}\ \Xi_i(q)=-\Xi_i(q)\ \big(\Sigma_i^{ab})^{T}-q^a \f{\p \Xi_i(q)}{\p q_b}+q^b \f{\p \Xi_i(q)}{\p q_a}\ ,\label{SigmaXi}
\ee
where $\Sigma_i$ is the spin angular momentum generator for $i$-th component field inside $\Phi(x)$. Taking derivatives with respect to momenta the above expressions become
\be 
&&\big(\Sigma_i^{ab}\big)^{T}\f{\p \mathcal{K}_i(q)}{\p q_c}=-\f{\p \mathcal{K}_i(q)}{\p q_c}\Sigma_i^{ab}+q^a \f{\p^2 \mathcal{K}_i(q)}{\p q_b \p q_c}-q^b \f{\p^2 \mathcal{K}_i(q)}{\p q_a \p q_c}+\eta^{ac}\f{\p \mathcal{K}_i(q)}{\p q_b}-\eta^{bc}\f{\p \mathcal{K}_i(q)}{\p q_a},\non\\
&& \Sigma_i^{ab}\ \f{\p \Xi_i(q)}{\p q_c}=-\f{\p \Xi_i(q)}{\p q_c}\ \big(\Sigma_i^{ab})^{T}-q^a \f{\p^2 \Xi_i(q)}{\p q_b \p q_c}+q^b \f{\p^2 \Xi_i(q)}{\p q_a\p q_c}-\eta^{ac}\f{\p \Xi_i(q)}{\p q_b}+\eta^{bc}\f{\p \Xi_i(q)}{\p q_a} .\non\\
\ee
Invariance of \eqref{S2} under global $U(1)_{EM}$ transformation implies
\be
\mathcal{Q}_\gamma\ ^\alpha \mathcal{K}^{\gamma\beta} +\mathcal{K}^{\alpha\gamma}\mathcal{Q}_\gamma\ ^\beta\ =0\ \Rightarrow\ \mathcal{Q}^{T}\mathcal{K}+\mathcal{K}\mathcal{Q}=0\ . \label{KQ_relation}
\ee
This also imposes constraint on the numerator of the propagator which reads
\be
\mathcal{Q}\Xi +\Xi \mathcal{Q}^{T}=0\ .\label{XiQ_relation}
\ee
The above two equations are also valid for component fields in real representation. We can take momentum derivatives on the above two relations to find useful expressions. When the $i$-th spinning particle is on-shell with momentum $q_i$ and polarization tensor $\epsilon_i(q_i)$ it satisfies
\be
&&\mathcal{K}_i^{\alpha\beta}(q_i)\epsilon_{i\beta}(q_i)=0 \ \Rightarrow \ \mathcal{K}_i(q_i)\epsilon_i(q_i)=0\hspace{0.5cm} \text{and}\hspace{0.5cm} \epsilon_i^{T}(q_i)\mathcal{K}_i^{T}(q_i)=0\label{on-shell_condition}\ .
\ee

\section{Soft photon theorem at one-loop}\label{S:soft_photon_thm}
In this section we derive subleading soft photon theorem analyzing one-loop amplitudes for a quantum mechanical scattering process involving $N$ number of massive charged particles with arbitrary spin. In \cite{1808.03288}, the order $\ln\omega$ soft factor has been derived analyzing one-loop amplitude in a theory of minimally coupled scalar-QED in presence of scalar contact interaction and the soft photon factor is determined in terms of the charges and asymptotic momenta of scattered particles, and the direction cosine of soft photon emission. Here in this section we show that even for arbitrary spinning particle scattering in a generic theory of QED with non-minimal interaction, the order $\ln\omega$ soft factor derived in  \cite{1808.03288} is universal (theory independent). This section should be thought of as a warm up of the next section where we are going to derive one-loop soft graviton theorem for spinning particle scattering in a generic theory of quantum gravity. 
\subsection{Covariantization and Feynman rules} \label{S:Feynman_rules}
In \cite{1809.01675}, the quadratic action $S^{(2)}$ in \eqref{S2} has been covariantized simultaneously in photon and graviton background to determine 1PI vertices involving two hard spinning particles and one or two photons/gravitons up to subleading order in the expansion of the momenta of photons/gravitons. Without going into too much details, here we summarise the outcomes of the covariantization prescription in photon background and write down the Feynman rules for vertices involving one and two photons. We derive the vertices for off-shell photon with Feynman gauge fixing term, such that the Feynman propagator for virtual photon reads
\be
\Delta^{F}_{\mu\nu}(\ell)=\f{-i}{\ell^2-i\epsilon}\ \eta_{\mu\nu}\ .
\ee
In position space the kinetic operator in \eqref{S2} contains derivatives over the field $\Phi_{\beta}(x)$, which have to replace by covariant derivatives under covariantization in presence of photon field $A_{\mu}(x)$. For example in the case of one and two derivatives the covariantization rule in position space becomes
\begingroup
\allowdisplaybreaks
\be
\p_\mu \Phi_{\beta}\rightarrow\ D_\mu\Phi_\beta &&=\ (\p_\mu \delta_\beta^\gamma -i\mathcal{Q}_{\beta}\ ^\gamma A_\mu)\Phi_\gamma\ , \label{single_derivative}\\
\p_\mu\p_\nu \Phi_{\beta}\rightarrow\ D_{(\mu} D_{\nu)}\Phi_\beta &&=\ \big[\p_{(\mu} \delta_\beta^\gamma -i\mathcal{Q}_{\beta}\ ^\gamma A_{(\mu}\big]\big[\p_{\nu)} \delta_\gamma^\delta -i\mathcal{Q}_{\gamma}\ ^\delta A_{\nu)}\big]\Phi_\delta\nn\\
&&=\p_\mu\p_\nu\Phi_\beta -i\mathcal{Q}_\beta\ ^{\gamma}(A_\mu\p_\nu +A_\nu\p_\mu)\Phi_\gamma\nn\\
&&\ -\f{i}{2}\mathcal{Q}_\beta\ ^{\gamma}(\p_\mu A_\nu +\p_\nu A_\mu)\Phi_\gamma -A_\mu A_\nu \mathcal{Q}_\beta\ ^{\gamma}\ \mathcal{Q}_\gamma\ ^{\delta}\Phi_\delta\ .\label{double_derivative}
\ee
\endgroup
Above we use the following symmetrization convention $E_{(\mu}F_{\nu)}=\f{1}{2}\big(E_\mu F_\nu +E_\nu F_\mu\big)$ for two vectors $E$ and $F$. For determining minimal interaction vertices $\Gamma^{(3)}$ involving two massive spinning particles and one photon up to one derivative on photon field, and $\Gamma^{(4)}$ involving two massive spinning particles and two photons with no derivative on any of the photon fields, the information of covariantization for single and two derivatives as done above would be enough. In momentum space these covariantization rules generate the following minimally interacting actions starting from \eqref{S2}:
\be
S^{(3)}&=&\f{1}{2}\int \f{d^4 q_1}{(2\pi)^4}\f{d^4 q_2}{(2\pi)^4}\f{d^4\ell}{(2\pi)^4}\ (2\pi)^4\delta^{(4)}(q_1+q_2+\ell)\Phi_{\alpha}(q_1)\Big[-A^\mu(\ell)\f{\p\mathcal{K}^{\alpha\gamma}(q_2)}{\p q_2^\mu}\mathcal{Q}_{\gamma}\ ^{\beta}\non\\
&&\ -\f{1}{4}\big(\ell^\mu A^\nu(\ell)+\ell^\nu A^\mu(\ell)\big)\f{\p^2 \mathcal{K}^{\alpha\gamma}(q_2)}{\p q_2^\mu \p q_2^\nu}\mathcal{Q}_{\gamma}\ ^{\beta}\ +\ \mathcal{O}(\ell^2)\Big]\Phi_{\beta}(q_2)\ ,\label{S3_photon}
\ee
and
\be
S^{(4)}&=&\f{1}{2}\int \f{d^4 q_1}{(2\pi)^4}\f{d^4 q_2}{(2\pi)^4}\f{d^4\ell_1}{(2\pi)^4}\f{d^4\ell_2}{(2\pi)^4}\ (2\pi)^4\delta^{(4)}(q_1+q_2+\ell_1+\ell_2)\non\\
&&\Phi_{\alpha}(q_1)\Big[ \f{1}{2} A^\mu(\ell_1) A^\nu(\ell_2)\f{\p^2 \mathcal{K}^{\alpha\delta}(q_2)}{\p q_2^\mu \p q_2^\nu}\mathcal{Q}_{\delta}\ ^{\gamma}\mathcal{Q}_{\gamma}\ ^{\beta}\ +\ \mathcal{O}(\ell_1,\ell_2)\Big]\Phi_{\beta}(q_2)\ .\label{S4_photon}
\ee
Above $A_\mu(\ell)$ is the Fourier transform of gauge field defined through the following relation: $A_\mu(x)\equiv \int \f{d^4\ell}{(2\pi)^4}\ e^{i\ell\cdot x}A_\mu(\ell)$. At the order of one derivative on the gauge field, we can have non-minimal coupling of photon with matter fields interacting via field strength. In momentum space, the general form of non-minimal interaction takes the following form
\be
\bar{S}^{(3)}= \f{1}{2} &&\int\f{d^{4}q_{1}}{(2\pi)^{4}}\f{d^{4}q_{2}}{(2\pi)^{4}}\f{d^4\ell}{(2\pi)^4}\ (2\pi)^{4}\delta^{(4)}(q_{1}+q_{2}+\ell)\non\\
 &&\ \Phi_{\alpha}(q_{1})\ \Big[F_{\mu\nu}(\ell)\  \mathcal{B}^{\alpha\beta,\mu\nu}(q_{2})\ +\mathcal{O}(\ell^2)\Big] \Phi_{\beta}(q_{2})\ . \label{non-minimal}
\ee
where
\be
F_{\mu\nu}(\ell)=i\big[\ell_{\mu}A_{\nu}(\ell)-\ell_{\nu}A_{\mu}(\ell)\big]\ ,
\ee
and $\mathcal{B}(q_2)$ satisfies the following relations
\be
\mathcal{Q}_{\gamma}\ ^\alpha \mathcal{B}^{\gamma\beta,\mu\nu}(q_{2})+\mathcal{B}^{\alpha\gamma,\mu\nu}(q_{2})\mathcal{Q}_\gamma\ ^{\beta}=0\ &&\Rightarrow\ \mathcal{Q}^{T}\mathcal{B}^{\mu\nu}(q_2)+\mathcal{B}^{\mu\nu}(q_2)\mathcal{Q}=0\ ,\label{B_1}\\
\mathcal{B}^{\alpha\beta,\mu\nu}(q_{2})\ =\ -\mathcal{B}^{\alpha\beta,\nu\mu}(q_{2})\ &&\Rightarrow \ \mathcal{B}^{\mu\nu}(q_2)=-\mathcal{B}^{\nu\mu}(q_2)\ ,\label{B_2}\\
\mathcal{B}^{\alpha\beta,\mu\nu}(q_{2})\ =\ \pm\ \mathcal{B}^{\beta\alpha,\mu\nu}(-q_{1}-\ell)\ &&\Rightarrow \mathcal{B}^{\mu\nu}(q_{2})\ =\ \pm\ \big(\mathcal{B}^{\mu\nu}(-q_{1}-\ell)\big)^T \label{B_3}\ .
\ee
In the last equation above $+$ sign is for Grassmannian even field and $-$ sign is for Grassmannian odd field. Again during the derivation we consider $\Phi$ field components being Grassmannian even, but the final result will be valid for both Grassmannian even and odd fields. We introduced index free notations for all the equations above.

Starting from \eqref{S3_photon} and \eqref{non-minimal}, the interaction vertex describing two ingoing spinning particles with momenta $q$ and $-(q+\ell)$, polarization index $\alpha$ and $\beta$, and one ingoing photon with momentum $\ell$ becomes
\be
&&\Gamma^{(3)\alpha\beta}_\mu \big(q,-q-\ell,\ell\big)\non\\
&=& \f{i}{2}\Bigg[ \f{\p\mathcal{K}^{\alpha\gamma}(-q-\ell)}{\p q^\mu}\mathcal{Q}_{\gamma}\ ^{\beta}-\f{1}{2}\ell^\nu \f{\p^2\mathcal{K}^{\alpha\gamma}(-q-\ell)}{\p q^\mu \p q^\nu}\mathcal{Q}_{\gamma}\ ^{\beta} -2i\ell^\nu \mathcal{B}^{\alpha\beta ,}\ _{\mu\nu}(-q-\ell)\non\\
&&\ -\f{\p\mathcal{K}^{\beta\gamma}(q)}{\p q^\mu}\mathcal{Q}_{\gamma}\ ^{\alpha}-\f{1}{2}\ell^\nu \f{\p^2\mathcal{K}^{\beta\gamma}(q)}{\p q^\mu \p q^\nu}\mathcal{Q}_{\gamma}\ ^{\alpha} -2i\ell^\nu \mathcal{B}^{\beta\alpha ,}\ _{\mu\nu}(q) +\mathcal{O}(\ell^2)\Bigg]\ ,\label{Gamma3_phootn_polarization}
\ee
Expanding in small $\ell$ limit and using \eqref{K_property},\eqref{KQ_relation}, \eqref{B_3} the above vertex reduce to the following polarization index suppressed form
\be
\Gamma^{(3)}_\mu \big(q,-q-\ell,\ell\big)
&=& i\Bigg[ \f{\p \mathcal{K}(-q)}{\p q^\mu}\mathcal{Q}+\f{1}{2}\ell^\nu \f{\p^2 \mathcal{K}(-q) }{\p q^\mu \p q^\nu}\mathcal{Q}-2i\ell^\nu \mathcal{B}_{\mu\nu}(-q)\ +\ \mathcal{O}(\ell^2)\Bigg].\label{Gamma3_photon}
\ee
Similarly starting from \eqref{S4_photon} the polarization index suppressed four point interaction vertex involving two incoming spinning particles with momenta $q$ and $-(q+\ell_1+\ell_2)$ and two incoming photons with momenta $\ell_1$ and $\ell_2$ becomes
\be
\Gamma^{(4)}_{\mu\nu}\big(q,-q-\ell_1-\ell_2,\ell_1, \ell_2\big)&=&\ i\f{\p^2 \mathcal{K}(-q)}{\p q^\mu \p q^\nu}\mathcal{Q}\mathcal{Q}+\ \mathcal{O}(\ell_1,\ell_2)\ .\label{Gamma4_photon}
\ee
We denote the scattering amplitude describing $N$-number of spinning hard particle scattering in massive EFT by $\Gamma^{(N)}$, which can be expressed as a polarization tensor contracted form in the following way
\be
\Gamma^{(N)}& =& \Big\lbrace \prod_{\substack{i=1}}^{N}\epsilon_{\alpha_i}(p_i) \Big\rbrace \ \Gamma^{(N)\alpha_1\alpha_2\cdots \alpha_N}\nn\\
&= & \epsilon_{\alpha_i}(p_i)\epsilon_{\alpha_j}(p_j)\ \Gamma_{(ij)}^{(N)\alpha_i\alpha_j}(p_i,p_j)\nn\\
& = &\ \epsilon_{\alpha_i}(p_i)\Gamma_{(i)}^{(N)\alpha_i}(p_i)\ ,
\ee
where
\be
 \Gamma_{(ij)}^{(N)\alpha_i\alpha_j}(p_i,p_j)&\equiv &\Big\lbrace\prod_{\substack{k=1\\ k\neq i,j}}^{N}\epsilon_{\alpha_k}(p_k) \Big\rbrace \ \Gamma^{(N)\alpha_1\alpha_2\cdots \alpha_N}\ ,\\
 \Gamma_{(i)}^{(N)\alpha_i}(p_i) &\equiv & \Big\lbrace\prod_{\substack{j=1\\ j\neq i}}^{N}\epsilon_{\alpha_j}(p_j) \Big\rbrace \ \Gamma^{(N)\alpha_1\alpha_2\cdots \alpha_N}.
\ee
Above $\epsilon_{\alpha_i}(p_i)$ denotes the polarization tensor for $i$-th spinning particle with momentum $p_i$. Here we also should remember that $\Gamma^{(N)}$ is a distribution as it contains a momentum conserving delta function $\delta^{(4)}\big(p_1+p_2+\cdots +p_N\big)$. We denote a part of amputated Green's function involving $N$ number of spinning hard particles and one  photon with momentum $\ell$ by $\widetilde{\Gamma}_\mu^{(N+1)}(\ell)$, which describes sum of the contributions of the Feynman diagrams where the photon is not connected to any external leg. Since $\widetilde{\Gamma}_\mu^{(N+1)}(\ell)$ does not contain any loop involving massless particles, we can write down the following relation between $\widetilde{\Gamma}_\mu^{(N+1)}(\ell)$ and $\Gamma^{(N)}$ using the same covariantization prescription described above
\be
\widetilde{\Gamma}_\mu^{(N+1)\alpha_1\cdots \alpha_N}(\ell)=-\sum_{i=1}^{N} \mathcal{Q}_{\beta_i}\ ^{\alpha_i}\ \f{\p}{\p p_i^\mu}\ \Gamma^{(N)\alpha_1\cdots \alpha_{i-1}\beta_i\alpha_{i+1}\cdots \alpha_N}\ +\ \mathcal{O}(\ell) \ .\label{Gamma_tilde_photon}
\ee
In the above expression $\Gamma^{(N)}$ contains momentum conserving delta function $\delta^{(4)}\big(p_1+p_2+\cdots +p_N\big)$ and $\widetilde{\Gamma}^{(N+1)}$ contains momentum conserving delta function $\delta^{(4)}\big(p_1+p_2+\cdots +p_N+\ell\big)$.

\subsection{Grammer-Yennie decomposition and IR-finite amplitudes}
In this section we discuss the Grammer-Yennie decomposition introduces in \cite{grammer,1808.03288} and show how it helps to factorize the IR-divergent exponential from the IR-finite part of the amplitudes as proposed in \eqref{eq:QED_factorization}.
 In Feynman gauge we decompose  the photon propagator with momentum $\ell$ flowing from the leg $i$ to the leg $j$ for $i\neq j$
\be
\Delta_{F}^{\mu\nu}(\ell)=-i \f{\eta^{\mu\nu}}{ \ell^2 - i\epsilon} &=& -\f{i}{ \ell^2-i\epsilon} 
\Big\lbrace K_{(ij)}^{\mu\nu} + G_{(ij)}^{\mu\nu}\Big\rbrace\ , \label{KG_photon_decomposition}
\ee
where
\be \label{KG_photon}
K_{(ij)}^{\mu\nu} = \ell^\mu \ell^\nu {(2p_i-\ell)\cdot(2p_j+\ell) \over (2p_i\cdot\ell -\ell^2+i\epsilon) 
(2p_j\cdot \ell+\ell^2-i\epsilon)} \ ,
\quad G_{(ij)}^{\mu\nu} = \eta^{\mu\nu} - K_{(ij)}^{\mu\nu}\, .
\ee
Note that $p_i$ and $p_j$ refer to the external momenta flowing into the legs $i$ and $j$, and not necessarily the momenta of the lines to which the photon propagator attaches (which may have additional contribution from external soft photon momentum or internal virtual photon momentum). For $i=j$ we do not carry out any decomposition i.e. for the virtual photon involved in the self energy loop we do not carry out the KG-decomposition of \eqref{KG_photon_decomposition}. Also if one or both ends of the virtual photon propagator are attached to any internal massive particle propagator carrying sum of two or more external massive particle's momenta, or vertices of the massive EFT involving more than two massive particles in a loop diagram, then we do not need to perform KG-decomposition as those loop diagrams are IR-finite. In \eqref{KG_photon_decomposition} the propagator part containing $K_{(ij)}$ will be denoted by K-photon propagator and the propagator part containing $G_{(ij)}$ will be denoted by G-photon propagator throughout this section.

\begin{center}
\begin{figure}
	\includegraphics[scale=0.44]{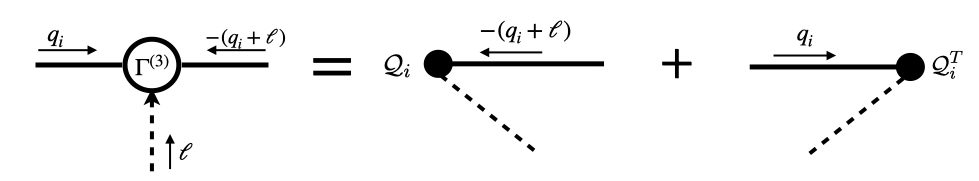}
	\caption{This figure is a Feynman diagrammatic representation of the expression in \eqref{Ward_identity_photon}. Solid lines represent the massive spinning particles, dashed lines represent the ingoing virtual photon with momentum $\ell$ and the arrow in the photon line represents that it is a K-photon 
	 (pure gauge part contracted). The solid blobs in the RHS represent a new kind of vertices and the Feynman rules for those vertices are just $\mathcal{Q}_i$ and $\mathcal{Q}_i^T$ as written next to the vertices.}\label{f:KG_photon_figure}
\end{figure}
\end{center}

\paragraph{Ward identities involving K-photon:}
 Since K-photon propagator is proportional to $\ell^\mu\ell^\nu$ i.e. pure gauge, we can study the Ward identity for an off-shell un-amputated three particle Green's function with one leg being the K-photon as drawn in Fig.\ref{f:KG_photon_figure}. The LHS of Fig.\ref{f:KG_photon_figure} after contracting with $\ell^\mu$ representing K-photon can be expressed as
\begingroup
\allowdisplaybreaks
\be
&&\f{1}{q_i^2+m_i^2-i\epsilon}\ \Xi_i(-q_i)\ \ell^\mu \Gamma^{(3)}_{\mu}(q_i,-q_i-\ell ,\ell)\ \Xi_i(-q_i-\ell)\ \f{1}{(q_i+\ell)^2+m_i^2 -i\epsilon}\non\\
&=& i\f{1}{q_i^2+m_i^2-i\epsilon}\f{1}{(q_i+\ell)^2+m_i^2 -i\epsilon}\Bigg[-i(2q_i.\ell +\ell^2)\Xi_i(-q_i)\mathcal{Q}_i^T\non\\
&&\ +\Xi_i(-q_i)\mathcal{K}_i(-q_i)\Bigg\lbrace \ell^\mu \f{\p \Xi_i(-q_i)}{\p q_i^\mu}+\f{1}{2}\ell^\mu \ell^\nu \f{\p^2 \Xi_i(-q_i)}{\p q_i^\mu \p q_i^\nu}\ +\ \mathcal{O}(\ell^3)\Bigg\rbrace \mathcal{Q}^{T}_i\Bigg]\ .
\ee
\endgroup
To write down the Feynman rule in the first line above for the diagram in Fig.\ref{f:KG_photon_figure}, we follow the convention that our time arrow runs from right to left. We shall follow the same time arrow convention for all the Feynman diagrams in this article. To evaluate the first line, with the expression in \eqref{Gamma3_photon} we use the properties \eqref{K2},\eqref{K3},\eqref{KQ_relation} and \eqref{XiQ_relation}. Now replacing  $(2q_i.\ell +\ell^2)=\lbrace (q_i+\ell)^2+m_i^2\rbrace -\lbrace q_i^2+m_i^2\rbrace$ within the square bracket above and using the property \eqref{K4}, the RHS of the above expression translates to
\be
&&\f{\Xi_i(-q_i)\mathcal{Q}_i^{T}}{q_i^2+m_i^2-i\epsilon}-\f{\Xi_i(-q_i)\mathcal{Q}_i^{T}}{(q_i+\ell)^2+m_i^2-i\epsilon}\non\\
&&-\f{1}{(q_i+\ell)^2+m_i^2 -i\epsilon}
\Bigg\lbrace \ell^\mu \f{\p \Xi_i(-q_i)}{\p q_i^\mu}+\f{1}{2}\ell^\mu \ell^\nu \f{\p^2 \Xi_i(-q_i)}{\p q_i^\mu \p q_i^\nu}\ +\ \mathcal{O}(\ell^3)\Bigg\rbrace \mathcal{Q}^{T}_i\ .
\ee
If we un-do the small $\ell$ expansion in the second line of the above expression\footnote{Instead of \eqref{Gamma3_photon}, if we use the unexpanded expression \eqref{Gamma3_phootn_polarization} for $\Gamma^{(3)}$ vertex, we do not need to un-do the  small $\ell$ expansion to derive the result below. Actually the relation in \eqref{Ward_identity_photon} is an exact relation valid up to all order in $\ell$ expansion with any arbitrary non-minimal coupling contributing to $\Gamma^{(3)}$, as it is a direct consequence of the Ward Identity.}  and use the relation \eqref{XiQ_relation} the  Ward identity turns out to be the following expression which has been diagrammatically represented in Fig.\ref{f:KG_photon_figure}.
\be
&&\f{1}{q_i^2+m_i^2-i\epsilon}\ \Xi_i(-q_i)\ \ell^\mu \Gamma^{(3)}_{\mu}(q_i,-q_i-\ell ,\ell)\ \Xi_i(-q_i-\ell)\ \f{1}{(q_i+\ell)^2+m_i^2 -i\epsilon}\non\\
&=&\mathcal{Q}_i\f{\Xi_i(-q_i-\ell)}{(q_i+\ell)^2+m_i^2-i\epsilon}\ +\ \f{\Xi_i(-q_i)}{q_i^2+m_i^2-i\epsilon}\mathcal{Q}_i^{T}\ .\label{Ward_identity_photon}
\ee
Important to note that the solid blob vertices in Fig.\ref{f:KG_photon_figure} carry only the information of charge of the particle with which the K-photon is interacting. The Feynman rules for the blob vertices are independent of the momenta or any other information of the theory.

\begin{center}
\begin{figure}[h!]
	\includegraphics[scale=0.43]{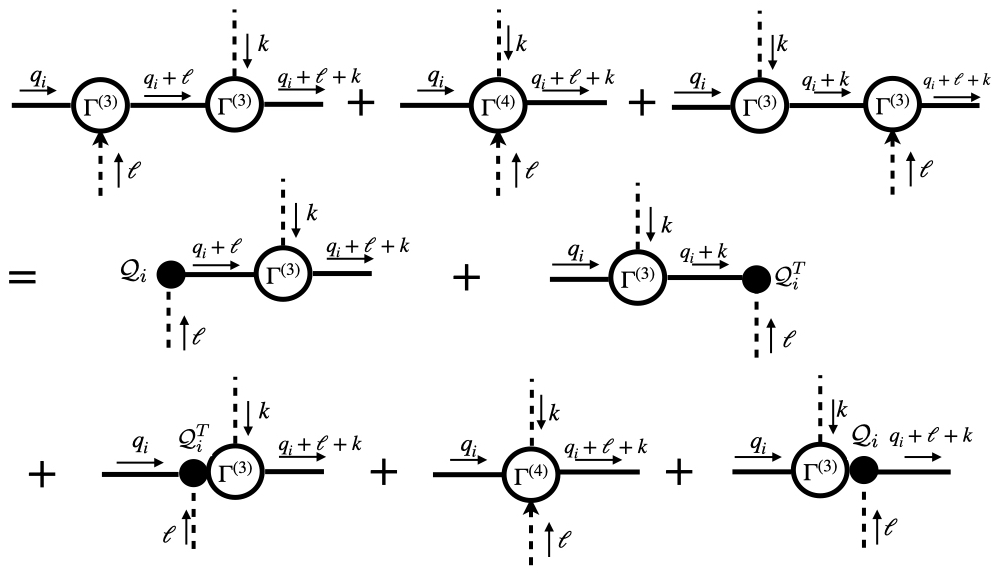}
	\caption{Diagrams in the first line represents the contribution to the four point un-amputated Green's function with one photon and one K-photon. By using the Ward identity in Fig.\ref{f:KG_photon_figure} we get the diagrams in second and third lines. }\label{f:KG_decomposition_2_photons}
\end{figure}
\end{center}

\begin{center}
\begin{figure}
	\includegraphics[scale=0.43]{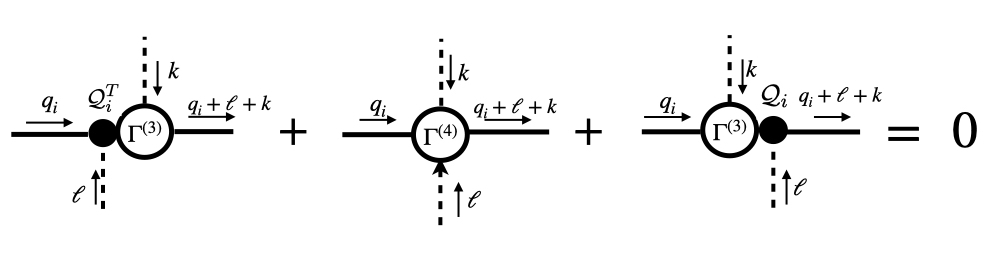}
	\caption{Identity involving last three Feynman diagrams in Fig.\ref{f:KG_decomposition_2_photons} whose mathematical expression has been written in \eqref{last_three_diagrams}. }\label{f:KG_decomposition_photon_subdiagram}
\end{figure}
\end{center}

We also need to study the consequence of Ward identity due to insertion of a K-photon in presence of an external off-shell photon with momentum $k$ and Lorentz index $\nu$. The set of Feynman diagrams describing the four point un-amputated Green's function with one photon and one K-photon has been drawn in the first line of Fig.\ref{f:KG_decomposition_2_photons}. Using the Ward identity described in Fig.\ref{f:KG_photon_figure} for first and third diagrams in the first line of Fig.\ref{f:KG_decomposition_2_photons}, we find the diagrams drawn after the equality in Fig.\ref{f:KG_decomposition_2_photons}.  Now if we can show that the sum of the contribution of the three diagrams in the last line of Fig.\ref{f:KG_decomposition_2_photons} vanishes, then the Ward identity of Fig.\ref{f:KG_photon_figure} is also valid in presence of an external photon line. The three diagrams in the last line of Fig.\ref{f:KG_decomposition_2_photons} has been drawn again in Fig.\ref{f:KG_decomposition_photon_subdiagram} and the contribution becomes
\be
&&\f{1}{q_i^2+m_i^2-i\epsilon}\ \Xi_i(-q_i)\Big[ \mathcal{Q}_i^{T}\ \Gamma^{(3)}_\nu (q_i+\ell ,-q_i-\ell-k, k) +\ \ell^\mu\ \Gamma^{(4)}_{\mu\nu}(q_i,-q_i-\ell-k,\ell, k)\non\\
&& +\ \Gamma^{(3)}_\nu (q_i, -q_i-k,k)\ \mathcal{Q}_i \Big]\ \Xi_i(-q_i-\ell-k)\f{1}{(q_i+\ell+k)^2 +m_i^2 -i\epsilon}\ .\label{last_three_diagrams}
\ee
Substituting the expressions for the vertices from \eqref{Gamma3_photon} and \eqref{Gamma4_photon}, the expression inside the square bracket of \eqref{last_three_diagrams}  turns out to be
\begingroup
\allowdisplaybreaks
\be
&&i\mathcal{Q}_i^{T}\Bigg[ \f{\p \mathcal{K}_i(-q_i-\ell)}{\p q_i^\nu }\mathcal{Q}_i +\f{1}{2}k^\mu \f{\p^2 \mathcal{K}_i(-q_i-\ell)}{\p q_i^\mu \p q_i^\nu}\mathcal{Q}_i -2ik^\mu \mathcal{B}^i_{\nu\mu}(-q_i-\ell)\Bigg]\non\\
&& \ +\ i\ell^\mu \ \f{\p^2 \mathcal{K}_i(-q_i)}{\p q_i^\mu \p q_i^\nu}\ \mathcal{Q}_i\mathcal{Q}_i\non\\
&&\ +i\Bigg[ \f{\p \mathcal{K}_i(-q_i)}{\p q_i^\nu }\mathcal{Q}_i +\f{1}{2}k^\mu \f{\p^2 \mathcal{K}_i(-q_i)}{\p q_i^\mu \p q_i^\nu}\mathcal{Q}_i -2ik^\mu \mathcal{B}^i_{\nu\mu}(-q_i)\Bigg]\mathcal{Q}_i\ .
\ee
\endgroup
 \begin{center}
\begin{figure}
	\includegraphics[scale=0.43]{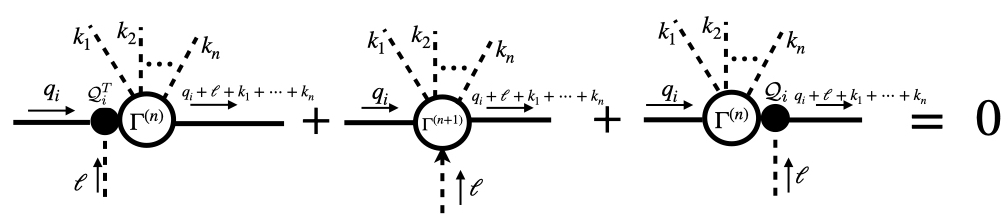}
	\caption{Generalization of the identity in Fig.\ref{f:KG_decomposition_photon_subdiagram} in presence of $n$-number of photons. }\label{f:KG_decomposition_photon_generalized}
\end{figure}
\end{center}
By Taylor expanding the first line above for small $\ell$ and only keeping terms up to linear order in $\ell$ or $k$, we can use the identities in \eqref{KQ_relation} and \eqref{B_1} to show that the sum of the contribution in the three lines above vanishes up to linear order in $\ell$ or $k$. This proves the diagrammatic identity in Fig.\ref{f:KG_decomposition_photon_subdiagram}. By utilizing the expressions of vertices from \eqref{Gamma3_photon} and \eqref{Gamma4_photon}, which are given up to linear and zeroth order in $\ell$ or $k$ respectively, it may initially appear that the validity of the result in Fig.\ref{f:KG_decomposition_photon_subdiagram} is limited to linear order in $\ell$ or $k$. However, it is important to note that the results depicted in Fig.\ref{f:KG_decomposition_photon_subdiagram} and Fig.\ref{f:KG_photon_figure} hold true for all orders in the expansion of small $\ell$ and $k$. These results play a crucial role in establishing the gauge invariance of any amplitude involving external photons in quantum electrodynamics, as they are connected to the Ward-Takahashi identity of QED. For further details on the spinor-QED case, please look at  \cite{grammer} and section-(7.4) of \cite{Peskin:1995ev}. For an un-amputated Green's function with two massive spinning particles and  arbitrary number of external photon legs, one insertion of K-photon in all possible way finally reduces to sum over sets of diagrams where the K-photon is connected in the end of the spinning particle legs with solid blob vertices as discussed above. This strong statement can be proved using the identity in Fig.\ref{f:KG_photon_figure} and the generalized identity in Fig.\ref{f:KG_decomposition_photon_generalized}. The identity in Fig.\ref{f:KG_decomposition_photon_generalized} is a straight forward generalization of the example discussed in Fig.\ref{f:KG_decomposition_photon_subdiagram}, which has been tested with the covariantized vertices up to linear order in photon momenta expansion for $\Gamma^{(n)}$ and $\Gamma^{(n+1)}$ vertices with $n=3$.

\paragraph{IR-finite amplitudes:}
As we defined earlier, $\mathcal{A}^{(N)}$ represents the all loop scattering amplitude with $N$ number of external massive spinning particles, and $\mathcal{A}^{(N+1)}$ represents the all loop scattering amplitude with $N$ number of external massive spinning particles and one external photon. If the massive spinning particles carry definite charges $\lbrace e_i\rbrace$ then the following identity holds
\be
\mathcal{Q}_{\alpha_i}\ ^{\beta_i}\ \epsilon_{i\beta_i}(p_i)=\ e_i\ \epsilon_{i\alpha_i}(p_i)\ \Rightarrow\ \mathcal{Q} \epsilon_i(p_i)=e_i\epsilon_i(p_i)\ ,\  \epsilon^T_i(p_i)\mathcal{Q}^T=e_i\epsilon^T_i(p_i)\ . \label{charge_eigen_value}  
\ee
The K-photon insertion Ward identities of Fig.\ref{f:KG_photon_figure} and Fig.\ref{f:KG_decomposition_photon_generalized} imply the exponentiation of the one-loop K-photon contribution $K_{em}$ in \eqref{eq:QED_factorization}, as proven in \cite{grammer}. However, the proof of this exponentiation in \cite{grammer} is valid only when the tree-level amplitude of the massive EFT $\left(\Gamma^{(N)}\right)$ is independent of the momenta of scattering particles i.e. it is described by a momenta-independent contact interaction between $N$ number of massive fields, as considered in \cite{1808.03288}. The validity of the exponentiation of $K_{em}$ in a generic theory of QED, incorporating all possible interactions without considering to any kind of approximation (like assuming that virtual photon momenta are significantly smaller than external massive particle momenta), is too good to be true.\footnote{We are thankful to P.V. Athira for extensive discussion on this topic.} But in the limit of small virtual photon momenta the IR divergent piece of $K_{em}$ exponentiate, which is known as the leading Eikonal exponentiation.

\begin{center}
\begin{figure}
	\includegraphics[scale=0.40]{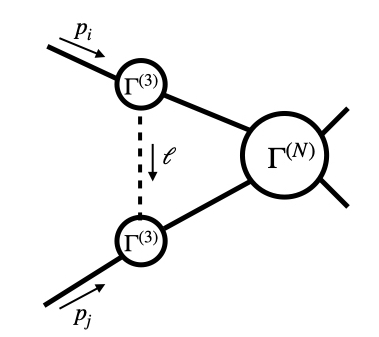}
	\caption{Diagram contributes to $\mathcal{A}^{(N)}_{G,1}$, where the virtual photon propagator is the G-photon propagator connected between two external hard particle lines. The diagram also contributes to $\mathcal{A}^{(N)}_{K-\text{finite},1}$ when evaluated with K-photon propagator and subtracted $K_{em}\Gamma^{(N)}$ from it's contribution. }\label{f:A^N_G}
\end{figure}
\end{center}

The final outcome of the Grammer--Yennie decomposition of virtual photon propagator in \eqref{KG_photon_decomposition} is
\be
&&\mathcal{A}^{(N)}\equiv \exp\lbrace K_{em}\rbrace\ \mathcal{A}^{(N)}_{\text{IR-finite}} \label{A_N}\ ,\\
&&\mathcal{A}^{(N+1)}\equiv \exp\lbrace K_{em}\rbrace\ \mathcal{A}^{(N+1)}_{\text{IR-finite}}\ ,\label{A_N+1}
\ee 
where
\be
K_{em}&= & \ \f{i}{2}\sum_{i=1}^{N}\  \sum_{\substack{j=1\\j\neq i}}^{N}e_{i}e_{j}\ \int \f{d^{4}\ell}{(2\pi)^{4}} \ \f{1}{\ell^{2}-i\epsilon}\ {(2p_i-\ell)\cdot(2p_j+\ell) \over (2p_i\cdot\ell -\ell^2+i\epsilon) 
(2p_j\cdot \ell+\ell^2-i\epsilon)}\label{K_em}\ .
\ee
In equations \eqref{A_N} and \eqref{A_N+1}, $\mathcal{A}^{(N)}_{\text{IR-finite}}$ and $\mathcal{A}^{(N+1)}_{\text{IR-finite}}$ represent the infrared finite components of the $N$-particle and $N$-particle-1-photon amplitudes, respectively. These components are obtained by removing the exponentiated IR-divergent parts from the original divergent amplitudes defined through the relations \eqref{A_N} and \eqref{A_N+1}. Both $\mathcal{A}^{(N)}_{\text{IR-finite}}$ and $\mathcal{A}^{(N+1)}_{\text{IR-finite}}$ comprise contributions from the corresponding tree-level amplitudes and loop amplitudes up to all orders in perturbation theory. However, there is a condition: if both ends of a virtual photon propagator are connected to external massive spinning particle lines (which may already contain additional real or virtual photon lines), then this photon propagator should be replaced by a G-photon propagator when we evaluate them for the IR-finite parts. Additionally the same set of diagrams need to be evaluated with K-photon propagator as well and then have to subtract by a factor of $K_{em}$ times the IR finite amplitude at one less loop level. In our convention the tree level amplitudes are given by
\be
\mathcal{A}^{(N)}_{\text{IR-finite,0}}= \Gamma^{(N)}\ ,\ \mathcal{A}^{(N+1)}_{\text{IR-finite,0}}=\Gamma^{(N+1)}\ ,\label{eq:AN_tree}
\ee
where in the subscript `$0$' corresponds to $0$-loop i.e. tree level. At one-loop order, $\mathcal{A}^{(N)}_{\text{IR-finite}}$ and $\mathcal{A}^{(N+1)}_{\text{IR-finite}}$ are given by:
\be
\mathcal{A}^{(N)}_{\text{IR-finite,1}}&\equiv &\Big[\mathcal{A}^{(N)}_{G,1}+\mathcal{A}^{(N)}_{K-\text{finite},1}+\mathcal{A}^{(N)}_{\text{self,1}}+\mathcal{A}^{(N)}_{\text{non-div,1}}\Big]\ ,\label{eq:AN_finite}\\
\mathcal{A}^{(N+1)}_{\text{IR-finite,1}}&\equiv &\Big[\mathcal{A}^{(N+1)}_{G,1}+\mathcal{A}^{(N+1)}_{K-\text{finite},1}+\mathcal{A}^{(N+1)}_{\text{self,1}}+\mathcal{A}^{(N+1)}_{\text{non-div,1}}\Big]\ \label{eq:AN+1_finite},
\ee
where in the subscript `$1$' corresponds to $1$-loop. Above 
\begingroup
\allowdisplaybreaks
\begin{enumerate}
	\item $\mathcal{A}^{(N)}_{G,1}$ corresponds to the diagram in Fig.\ref{f:A^N_G} which we need to evaluate with G-photon propagator.
	\item  $\mathcal{A}^{(N)}_{K-\text{finite},1}$ corresponds to the contribution from the diagram in Fig.\ref{f:A^N_G}, evaluated with K-photon propagator and then subtracted the contribution of $K_{em}\times \Gamma^{(N)}$ at the integrand level.
	\item $\mathcal{A}^{(N)}_{\text{self,1}}$ represents the set of Feynman diagrams in Fig.\ref{f:A^N_self} evaluated with full photon propagator.
	\item $\mathcal{A}^{(N)}_{\text{non-div,1}}$ represents the set of diagrams in Fig.\ref{f:A^N_non-div}, evaluated with full photon propagator.
	\item $\mathcal{A}^{(N+1)}_{G,1}$ corresponds to the set of diagrams in Fig.\ref{f:A^N+1_G} when we evaluate them with G-photon propagator.
	\item $\mathcal{A}^{(N+1)}_{K-\text{finite},1}$ corresponds to the contribution from the diagrams in Fig.\ref{f:A^N+1_G}, evaluated with K-photon propagator and then subtracted the contribution $K_{em}\times \Gamma^{(N+1)}$ at the integrand level.
	\item $\mathcal{A}^{(N+1)}_{\text{self,1}}$ represents the set of Feynman diagrams in Fig.\ref{f:A^N+1_self}, evaluated with full photon propagator.
	\item $\mathcal{A}^{(N+1)}_{\text{non-div,1}}$ represents the set of diagrams in Fig.\ref{f:A^N+1_non-div}, evaluated with full photon propagator.
\end{enumerate}
\endgroup
\subsection{Derivation of soft photon theorem}\label{s:photon_one_loop_computation}
The goal here will be to derive the order $\omega^{-1}$ and $\ln\omega$ soft factors from the ratio of $\mathcal{A}^{(N+1)}_{\text{IR-finite}}$ and $\mathcal{A}^{(N)}_{\text{IR-finite}}$ when the external photon energy is small i.e. $\omega<<|p_i^\mu |$.

\subsubsection{IR-finite one loop $N$-particle amplitude} 
First we want to analyze all the Feynman diagrams contributing to \eqref{eq:AN_finite} and show that the result is IR-divergence free. We also provide an explicit integral expression of IR-finite 1-loop amplitude for  $\mathcal{A}^{(N)}_{\text{IR-finite,1}}$.  Using the Feynman rules derived in subsection-\ref{S:Feynman_rules}, the diagram in Fig.\ref{f:A^N_G} with G-photon propagator contributes to the following
\begingroup
\allowdisplaybreaks
\be
\mathcal{A}^{(N)}_{G,1}&=&\sum_{\substack{i,j=1\\i> j}}^{N} \int \f{d^4\ell}{(2\pi)^4}\ \f{-i}{\ell^2-i\epsilon} \f{1}{(p_i-\ell)^2+m_i^2-i\epsilon} \f{1}{(p_j+\ell)^2+m_j^2 -i\epsilon} \non\\
&&\Big[ \epsilon_i^{T}(-p_i)\Gamma^{(3)}_\mu (p_i,-p_i+\ell,-\ell)\Xi_i(-p_i+\ell)\Big] \ G_{(ij)}^{\mu\nu}(\ell)\ \non\\
&&\times\Big[\epsilon_j^{T}(-p_j)\Gamma^{(3)}_\nu (p_j,-p_j-\ell,\ell)\Xi_j(-p_j-\ell)\Big]\ \Gamma^{(N)}_{(ij)}(p_i-\ell,p_j+\ell)\non\\
&=& \sum_{\substack{i,j=1\\i> j}}^{N}8e_i e_j\epsilon_i^{T}\epsilon_j^{T} \int \f{d^4\ell}{(2\pi)^4}\ \f{-i}{\ell^2-i\epsilon} \f{1}{(p_i-\ell)^2+m_i^2-i\epsilon} \f{1}{(p_j+\ell)^2+m_j^2 -i\epsilon} \non\\
&&\times \Big\lbrace\Big[ p_i^{\mu}\ell^{\nu}\mathcal{N}^{j}_{\mu\nu}(-p_j)\ -\ p_j^{\mu}\ell^{\nu}\mathcal{N}^{i}_{\mu\nu}(-p_i)\Big]\  \Gamma^{(N)}_{(ij)}(p_i,p_j)+\mathcal{O}(\ell\ell)\Big\rbrace\ , \label{A_G^N_result}
\ee
\endgroup
where to get the last two lines we have used the identity in \eqref{zeta_4} for both $i$-th and $j$-th particles and Taylor expanded the numerator in the limit $|\ell^\mu|<<|p_i^\mu| ,|p_j^\mu|$. The expression for $\mathcal{N}^i(-p_i)$ is given in \eqref{N_non-universal}. On the other hand the contribution of $\mathcal{A}^{(N)}_{K-\text{finite},1}$ from the diagram in Fig.\ref{f:A^N_G} becomes
\be
\mathcal{A}^{(N)}_{K-\text{finite},1}&=& \sum_{\substack{i,j=1\\i> j}}^{N} \int \f{d^4\ell}{(2\pi)^4}\ \f{-i}{\ell^2-i\epsilon} \f{1}{(p_i-\ell)^2+m_i^2-i\epsilon} \f{1}{(p_j+\ell)^2+m_j^2 -i\epsilon} \non\\
&&\Big[ \epsilon_i^{T}(-p_i)\Gamma^{(3)}_\mu (p_i,-p_i+\ell,-\ell)\Xi_i(-p_i+\ell)\Big] \ K_{(ij)}^{\mu\nu}(\ell)\ \non\\
&&\times\Big[\epsilon_j^{T}(-p_j)\Gamma^{(3)}_\nu (p_j,-p_j-\ell,\ell)\Xi_j(-p_j-\ell)\Big]\ \Gamma^{(N)}_{(ij)}(p_i-\ell,p_j+\ell)\non\\
&& -\ K_{em}\times \Gamma^{(N)}\ .
\ee
Now evaluating this expression using the identity in \eqref{zeta_4} and Taylor expanding $\Gamma^{(N)}_{(ij)}(p_i-\ell,p_j+\ell)$ in the limit $|\ell^\mu|<<|p_i^\mu| ,|p_j^\mu|$ we get
\be
\mathcal{A}^{(N)}_{K-\text{finite},1}&=&\ i\ 
\sum_{\substack{i,j=1\\i> j}}^{N}e_{i}e_{j}\ \epsilon_i^{T}\epsilon_j^{T}\int \f{d^{4}\ell}{(2\pi)^{4}} \ \f{1}{\ell^{2}-i\epsilon}\ {(2p_i-\ell)\cdot(2p_j+\ell) \over (2p_i\cdot\ell -\ell^2+i\epsilon) 
(2p_j\cdot \ell+\ell^2-i\epsilon)}\non\\
&&\times \Big[ -\ell^{\rho}\f{\p }{\p p_i^\rho}\Gamma^{(N)}_{(ij)}(p_i,p_j) +\ell^\rho\f{\p }{\p p_j^\rho}\Gamma^{(N)}_{(ij)}(p_i,p_j) +\mathcal{O}(\ell\ell)\Big]\ .\label{A_K^N_result}
\ee
\begin{center}
\begin{figure}
	\includegraphics[scale=0.43]{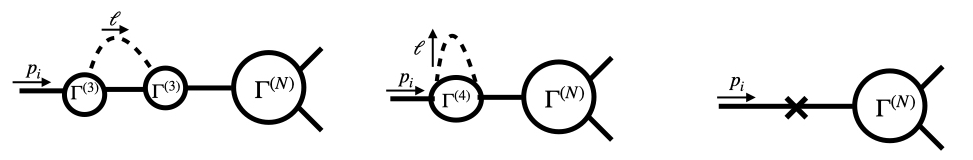}
	\caption{Diagram contributes to $\mathcal{A}^{(N)}_{\text{self,1}}$, where the virtual photon propagator is the full photon propagator. In the last diagram the cross corresponds to counter term, which cancels the UV divergences.}\label{f:A^N_self}
\end{figure}
\end{center}

Diagrams in Fig.\ref{f:A^N_self} renormalizes the massive spinning particle propagators in presence of electromagnetic interaction, and all the loops are IR-finite. Say the three diagrams in Fig.\ref{f:A^N_self} contributes to 
\be
\epsilon_i^{T}F_1 \f{\Xi_i(-p_i)}{p_i^2+m_i^2}\Gamma^{(N)}_{(i)}(p_i)\  , \ \epsilon_i^{T}F_2 \f{\Xi_i(-p_i)}{p_i^2+m_i^2}\Gamma^{(N)}_{(i)}(p_i)\  , \ \epsilon_i^{T}C\f{\Xi_i(-p_i)}{p_i^2+m_i^2}\Gamma^{(N)}_{(i)}(p_i) \label{eq:F_i_definitions}
\ee
respectively. Where $F_1,F_2, C$ are unknown constant matrices, which are related using on-shell renormalization condition
\be
C=-F_1 -F_2\ . \label{renormalization_condition}
\ee
Hence the on-shell renormalization condition implies
\be
\mathcal{A}^{(N)}_{\text{self,1}}=0\ .\label{A_self^N_result}
\ee
\begin{center}
\begin{figure}
	\includegraphics[scale=0.43]{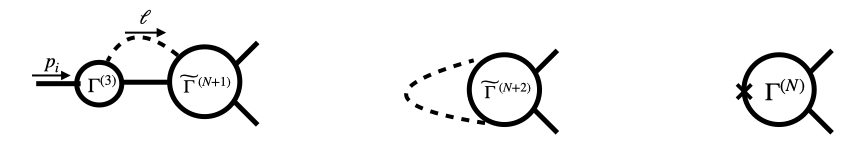}
	\caption{Diagram contributes to $\mathcal{A}^{(N)}_{\text{non-div,1}}$, where the virtual photon propagator is the full photon propagator whose one or both end connected to $\Gamma^{(N)}$. In the last diagram the cross corresponds to counter term, which cancels the UV divergences.  }\label{f:A^N_non-div}
\end{figure}
\end{center}
In Fig.\ref{f:A^N_non-div} we draw the sets of diagrams where one or both ends of the photon loop are attached to some internal massive virtual line or massive EFT vertex inside $\Gamma^{(N)}$. These diagrams are also IR-finite. For example the first diagram in Fig.\ref{f:A^N_non-div} with full photon propagator becomes
\be
&&\int \f{d^4\ell}{(2\pi)^4}\ \f{-i}{\ell^2 -i\epsilon}\ \f{1}{(p_i-\ell)^2+m_i^2 -i\epsilon}\ \eta^{\mu\nu}\non\\
&&\ \times \epsilon_i^{T}(-p_i)\Big[ \Gamma^{(3)}_\mu (p_i,-p_i+\ell ,-\ell)\Xi_i(-p_i+\ell)\widetilde{\Gamma}^{(N+1)}_{(i)\nu}(p_i-\ell ;\ell)\Big]\ .
\ee
Now using the identity in \eqref{zeta_4} for $i$-th particle and the Feynman rule of \eqref{Gamma_tilde_photon} it is evident that in the limit when the loop momentum $\ell^\mu\rightarrow 0$ the integration behaves as $\int \f{d^4\ell}{|\ell|^3}$, hence is IR-finite. This is the reason we call the set of diagrams in Fig.\ref{f:A^N_non-div} as $\mathcal{A}^{(N)}_{\text{non-div,1}}$ as those are IR-divergence free. We do not need to evaluate the contribution $\mathcal{A}^{(N)}_{\text{non-div,1}}$ explicitly for deriving soft photon theorem. Now summing over the contributions of \eqref{A_G^N_result}, \eqref{A_K^N_result}, \eqref{A_self^N_result} and $\mathcal{A}^{(N)}_{\text{non-div,1}}$, we get
\be
\mathcal{A}^{(N)}_{\text{IR-finite,1}}&=& \sum_{\substack{i,j=1\\i> j}}^{N}8e_i e_j\epsilon_i^{T}\epsilon_j^{T} \int \f{d^4\ell}{(2\pi)^4}\ \f{-i}{\ell^2-i\epsilon} \f{1}{(p_i-\ell)^2+m_i^2-i\epsilon} \f{1}{(p_j+\ell)^2+m_j^2 -i\epsilon} \non\\
&&\times \Big\lbrace\Big[ p_i^{\mu}\ell^{\nu}\mathcal{N}^{j}_{\mu\nu}(-p_j)\ -\ p_j^{\mu}\ell^{\nu}\mathcal{N}^{i}_{\mu\nu}(-p_i)\Big]\  \Gamma^{(N)}_{(ij)}(p_i,p_j)+\mathcal{O}(\ell\ell)\Big\rbrace\non\\
&&+i\ \sum_{\substack{i,j=1\\i> j}}^{N}e_{i}e_{j}\ \epsilon_i^{T}\epsilon_j^{T}\int \f{d^{4}\ell}{(2\pi)^{4}} \ \f{1}{\ell^{2}-i\epsilon}\ {(2p_i-\ell)\cdot(2p_j+\ell) \over (2p_i\cdot\ell -\ell^2+i\epsilon) 
(2p_j\cdot \ell+\ell^2-i\epsilon)}\non\\
&&\times \Big[ -\ell^{\rho}\f{\p }{\p p_i^\rho}\Gamma^{(N)}_{(ij)}(p_i,p_j) +\ell^\rho\f{\p }{\p p_j^\rho}\Gamma^{(N)}_{(ij)}(p_i,p_j) +\mathcal{O}(\ell\ell)\Big]\non\\
&&+\ \mathcal{A}^{(N)}_{\text{non-div,1}}\ . \label{A_IR-finite^N_result}
\ee
From the above expression, it is clear that the loop integrals become infrared finite in the limit as $\ell$ approaches zero. Therefore, the Grammer-Yennie prescription offers a clear definition of an IR-finite S-matrix. An explicit expression of IR-finite S-matrix at one-loop order is provided in the above expression in a general theory of quantum electrodynamics (QED).\begin{center}
\begin{figure}
	\includegraphics[scale=0.43]{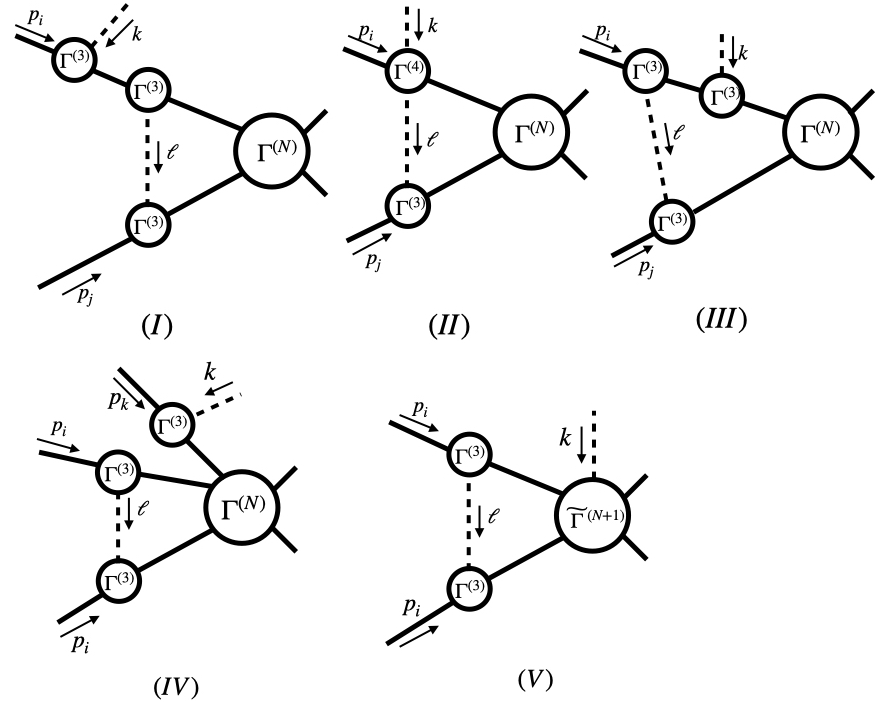}
	\caption{ Set of 1-loop diagrams contributing to $\mathcal{A}^{(N+1)}_{G,1}$, where the virtual photon propagator is the G-photon propagator connected between two external particle lines. Here we omitted the diagrams involving counter terms to remove UV divergences. We need to sum over all possible external legs while evaluating the contributions from these diagrams. When we evaluate these diagrams  with K-photon propagator and subtracted $K_{em}\Gamma^{(N+1)}$ from it's contribution, it also contributes to $\mathcal{A}^{(N+1)}_{K-\text{finite},1}$ . }\label{f:A^N+1_G}
\end{figure}
\end{center}
\subsubsection{IR-finite one loop $(N+1)$-particle amplitude in the soft limit} 
Here we analyze all the Feynman diagrams contributing to \eqref{eq:AN+1_finite} in the soft limit i.e. $\omega\rightarrow 0$. 
Let us start analyzing the first diagram in Fig.\ref{f:A^N+1_G} with G-photon propagator, which has the following expression after using Feynman rules
\begingroup
\allowdisplaybreaks
\be
A_{I}&\equiv &\ \sum_{i=1}^{N}\sum_{\substack{j=1\\j\neq i}}^{N}\int \f{d^4\ell}{(2\pi)^4}\ \f{-i}{\ell^2-i\epsilon} \f{1}{(p_i+k)^2+m_i^2-i\epsilon}\f{1}{(p_i+k-\ell)^2+m_i^2-i\epsilon}\f{1}{(p_j+\ell)^2+m_j^2 -i\epsilon} \non\\
&&\Big[ \epsilon_i^{T}(-p_i)\varepsilon^{\rho}(k)\Gamma^{(3)}_\rho (p_i,-p_i-k,k)\Xi_i(-p_i-k)\Gamma^{(3)}_\mu(p_i+k,-p_i-k+\ell ,-\ell)\Xi_i(-p_i-k+\ell)\Big]\non\\
&&\times \ G_{(ij)}^{\mu\nu}(\ell)\ \Big[\epsilon_j^{T}(-p_j)\Gamma^{(3)}_\nu (p_j,-p_j-\ell,\ell)\Xi_j(-p_j-\ell)\Big]\ \Gamma^{(N)}_{(ij)}(p_i+k-\ell,p_j+\ell)\ .\label{A_I}
\ee
\endgroup
Substituting the results from \eqref{zeta_1} and \eqref{zeta_4} in the above expression and after some manipulation we get 
\begingroup
\allowdisplaybreaks
\be
A_{I}
&=& \sum_{i=1}^{N}\sum_{\substack{j=1\\j\neq i}}^{N}e_i^2 e_j \epsilon_i^{T}\epsilon_j^{T}\f{1}{(p_i+k)^2+m_i^2-i\epsilon}\int \f{d^4\ell}{(2\pi)^4}\ \f{-i}{\ell^2-i\epsilon} \f{1}{(p_i+k-\ell)^2+m_i^2-i\epsilon}\f{1}{(p_j+\ell)^2+m_j^2 -i\epsilon} \non\\
&&\Bigg[ 8\varepsilon.p_i p_j.k +16\varepsilon.p_i p_i^{\rho}\ell^\sigma \mathcal{N}^j_{\rho\sigma}(-p_j)-16\varepsilon.p_i p_j^\rho \ell^\sigma\mathcal{N}^i_{\rho\sigma}(-p_i)+8p_i.p_j (\varepsilon^\rho k^\sigma -\varepsilon^\sigma k^\rho )\mathcal{N}^i_{\rho\sigma}(-p_i)\non\\
&&+4ip_i.k \varepsilon^\rho p_j^\sigma \f{\p \mathcal{K}_i(-p_i)}{\p p_i^\rho}\f{\p \Xi_i(-p_i)}{\p p_i^\sigma}-\f{4p_i.p_j}{2p_i.\ell -\ell^2+i\epsilon}\Big\lbrace 4\varepsilon.p_i \ell.k +4p_i.\ell (\varepsilon^\rho k^\sigma -\varepsilon^\sigma k^\rho )\mathcal{N}^i_{\rho\sigma}(-p_i)\non\\
&&+2ip_i.k\varepsilon^\rho \ell^\sigma\f{\p \mathcal{K}_i(-p_i)}{\p p_i^\rho}\f{\p \Xi_i(-p_i)}{\p p_i^\sigma}\Big\rbrace \ +\mathcal{O}(\ell\ell, k\ell,kk)\Bigg] \Big\lbrace\Gamma^{(N)}_{(ij)}(p_i,p_j)+\mathcal{O}(\ell,k)\Big\rbrace\ .
\ee
\endgroup
Inside the square bracket of the numerator in the above expression, we only keep the terms up to linear order in $\ell$ or $k$, as our vertices are derived only up to that order. Note that the above expression is IR-finite in the limit $\ell^\mu\rightarrow 0$ and in the region of integration $|\ell^\mu|<<\omega<<|p_i^\mu |,|p_j^\mu|$ it contributes at order $\mathcal{O}(\omega^0)$. Now to extract $\ln\omega$ contribution we approximate the integrand in the integration range $\omega<<|\ell^\mu|<<|p_i^\mu |,|p_j^\mu|$, by doing so we can approximate
 \be
 \f{1}{ (p_i+k-\ell)^2+m_i^2-i\epsilon}\simeq \f{1}{ (p_i-\ell)^2+m_i^2-i\epsilon}\Big[1+\f{p_i.k}{p_i.\ell +i\epsilon}\Big]\ .
 \ee
 With this approximation the order $\omega^{-1}$ and $\ln\omega$ contribution turns out to be,
 \begingroup
\allowdisplaybreaks
 \be
 A_{I}
&=& \sum_{i=1}^{N}\sum_{\substack{j=1\\j\neq i}}^{N}e_i\f{\varepsilon.p_i}{p_i.k}\ 8e_i e_j \epsilon_i^{T}\epsilon_j^{T}\int_{reg}  \f{d^4\ell}{(2\pi)^4}\ \f{-i}{\ell^2-i\epsilon} \f{1}{(p_i-\ell)^2+m_i^2-i\epsilon} \non\\
&&\f{1}{(p_j+\ell)^2+m_j^2 -i\epsilon} \Big[p_i^{\rho}\ell^\sigma \mathcal{N}^j_{\rho\sigma}(-p_j)- p_j^\rho \ell^\sigma\mathcal{N}^i_{\rho\sigma}(-p_i)\Big]\Gamma^{(N)}_{(ij)}(p_i,p_j)\non\\
&&+i\sum_{i=1}^{N}\sum_{\substack{j=1\\j\neq i}}^{N}e_i^2 e_j \f{1}{p_i.k}\epsilon_i^{T}\epsilon_j^{T}\int_{reg} \f{d^4\ell}{(2\pi)^4}\ \f{1}{\ell^2-i\epsilon} \f{1}{p_i.\ell+i\epsilon}\f{1}{p_j.\ell -i\epsilon} \non\\
&&\Bigg[ \varepsilon.p_i p_j.k +\f{p_i.k}{p_i.\ell+i\epsilon}\Big\lbrace2\varepsilon.p_i p_i^{\rho}\ell^\sigma \mathcal{N}^j_{\rho\sigma}(-p_j)-2\varepsilon.p_i p_j^\rho \ell^\sigma\mathcal{N}^i_{\rho\sigma}(-p_i)\Big\rbrace \non\\
&&+\f{i}{2}p_i.k \varepsilon^\rho p_j^\sigma \f{\p \mathcal{K}_i(-p_i)}{\p p_i^\rho}\f{\p \Xi_i(-p_i)}{\p p_i^\sigma}-\f{p_i.p_j}{p_i.\ell +i\epsilon}\Big\lbrace \varepsilon.p_i \ell.k \non\\
&&+\f{i}{2}p_i.k\varepsilon^\rho \ell^\sigma\f{\p \mathcal{K}_i(-p_i)}{\p p_i^\rho}\f{\p \Xi_i(-p_i)}{\p p_i^\sigma}\Big\rbrace \ \Bigg]\ \Gamma^{(N)}_{(ij)}(p_i,p_j)\ +\ \mathcal{O}(\omega^0).\label{A_I_result}
 \ee
 \endgroup
Above the subscript ``$reg$'' in the loop integral corresponds to the restricted loop-momentum range $\omega<<|\ell^\mu|<<|p_i^\mu |,|p_j^\mu|$. 

Using Feynman rules the second diagram in Fig.\ref{f:A^N+1_G} with G-photon propagator becomes
\be
A_{II}&\equiv &\ \sum_{i=1}^{N}\sum_{\substack{j=1\\j\neq i}}^{N}\int \f{d^4\ell}{(2\pi)^4}\ \f{-i}{\ell^2-i\epsilon} \f{1}{(p_i+k-\ell)^2+m_i^2-i\epsilon}\f{1}{(p_j+\ell)^2+m_j^2 -i\epsilon} \non\\
&&\Big[ \epsilon_i^{T}(-p_i)\varepsilon^{\rho}(k)\Gamma^{(4)}_{\rho\mu} (p_i,-p_i-k+\ell,k,-\ell)\Xi_i(-p_i-k+\ell)\Big]\ G_{(ij)}^{\mu\nu}(\ell) \non\\
&&\times \  \Big[\epsilon_j^{T}(-p_j)\Gamma^{(3)}_\nu (p_j,-p_j-\ell,\ell)\Xi_j(-p_j-\ell)\Big]\ \Gamma^{(N)}_{(ij)}(p_i+k-\ell,p_j+\ell).\label{A_II}
\ee
After substituting the results from \eqref{zeta_2} and \eqref{zeta_4} and keeping terms which can contribute up to order $\ln\omega$ in the integration range $reg\equiv (\omega<<|\ell^\mu|<<|p_i^\mu|, |p_j^\mu|)$ we get,
\begingroup
\allowdisplaybreaks
\be
A_{II}&=&-i\sum_{i=1}^{N}\sum_{\substack{j=1\\j\neq i}}^{N}e_i^2e_j\ \epsilon_i^{T}\epsilon_j^{T}\int_{reg} \f{d^4\ell}{(2\pi)^4} \f{1}{\ell^2-i\epsilon} \f{1}{p_i.\ell+i\epsilon}\f{1}{p_j.\ell -i\epsilon} \non\\
&&\Bigg[\varepsilon.p_j+\f{i}{2}\varepsilon^\rho p_j^\sigma\Big\lbrace \f{\p \mathcal{K}_i(-p_i)}{\p p_i^\rho}\f{\p \Xi_i(-p_i)}{\p p_i^\sigma}+\f{\p \mathcal{K}_i(-p_i)}{\p p_i^\sigma}\f{\p \Xi_i(-p_i)}{\p p_i^\rho}\Big\rbrace -\f{p_i.p_j}{p_i.\ell+i\epsilon} \non\\
&&\Bigg( \varepsilon.\ell+\f{i}{2}\varepsilon^\rho \ell^\sigma\Big\lbrace \f{\p \mathcal{K}_i(-p_i)}{\p p_i^\rho}\f{\p \Xi_i(-p_i)}{\p p_i^\sigma}+\f{\p \mathcal{K}_i(-p_i)}{\p p_i^\sigma}\f{\p \Xi_i(-p_i)}{\p p_i^\rho}\Big\rbrace\Bigg)\Bigg]\ \Gamma^{(N)}_{(ij)}(p_i,p_j)\nn\\
&&\ +\ \mathcal{O}(\omega^0)\ .\label{A_II_result}
\ee
\endgroup
\begin{center}
\begin{figure}
	\includegraphics[scale=0.43]{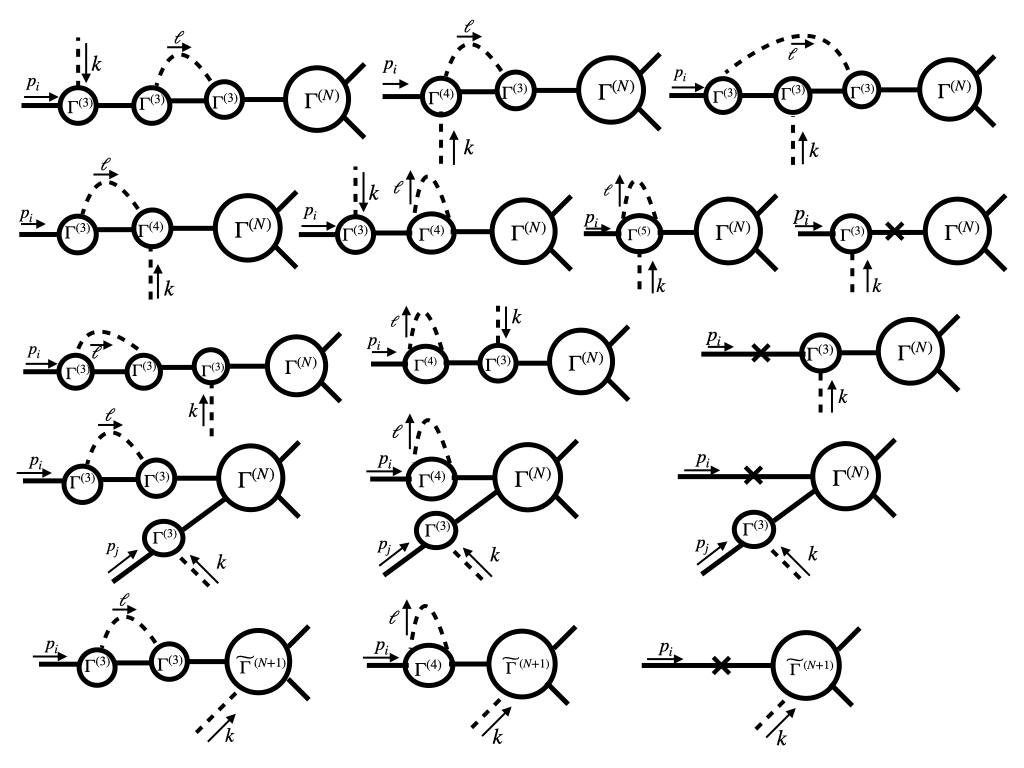}
	\caption{ Set of 1-loop diagrams contributing to $\mathcal{A}^{(N+1)}_{\text{self,1}}$, where the virtual photon propagator is the full photon propagator connecting two different points on the same massive spinning particle leg.  The cross appears in some diagrams above corresponds to counter term, which cancels the UV divergences in the renormalization prescription.}\label{f:A^N+1_self}
\end{figure}
\end{center}

Using Feynman rules the third diagram in Fig.\ref{f:A^N+1_G} with G-photon propagator becomes
\begingroup
\allowdisplaybreaks
\be
A_{III}&\equiv &\ \sum_{i=1}^{N}\sum_{\substack{j=1\\j\neq i}}^{N}\int \f{d^4\ell}{(2\pi)^4}\ \f{-i}{\ell^2-i\epsilon} \f{1}{(p_i-\ell)^2+m_i^2-i\epsilon}\f{1}{(p_i+k-\ell)^2+m_i^2-i\epsilon}\f{1}{(p_j+\ell)^2+m_j^2 -i\epsilon} \non\\
&&\Big[ \epsilon_i^{T}(-p_i)\varepsilon^{\rho}(k)\Gamma^{(3)}_\mu (p_i,-p_i+\ell,-\ell)\Xi_i(-p_i+\ell)\Gamma^{(3)}_\rho(p_i-\ell,-p_i-k+\ell ,k)\Xi_i(-p_i-k+\ell)\Big]\non\\
&&\times \ G_{(ij)}^{\mu\nu}(\ell)\ \Big[\epsilon_j^{T}(-p_j)\Gamma^{(3)}_\nu (p_j,-p_j-\ell,\ell)\Xi_j(-p_j-\ell)\Big]\ \Gamma^{(N)}_{(ij)}(p_i+k-\ell,p_j+\ell)\ .\label{A_III}
\ee
After substituting the results from \eqref{zeta_3} and \eqref{zeta_4} and keeping terms which can contribute up to order $\ln\omega$ in the integration range $\omega<<|\ell^\mu|<<|p_i^\mu|,|p_j^\mu|$ we get,
\be
A_{III}&=& -i\sum_{i=1}^{N}\sum_{\substack{j=1\\j\neq i}}^{N}e_i^2e_j\ \epsilon_i^{T}\epsilon_j^{T}\int_{reg} \f{d^4\ell}{(2\pi)^4} \f{1}{\ell^2-i\epsilon} \f{1}{(p_i.\ell+i\epsilon)^{2}}\f{1}{p_j.\ell -i\epsilon} \non\\
&&\Bigg[ 2\varepsilon.p_i p_i^\rho \ell^\sigma \mathcal{N}^{j}_{\rho\sigma}(-p_j)-2\varepsilon.p_i p_j^\rho \ell^\sigma \mathcal{N}^i_{\rho\sigma}(-p_i)-\f{i}{2}p_i.\ell p_j^\rho \varepsilon^\sigma \f{\p \mathcal{K}_i(-p_i)}{\p p_i^\rho}\f{\p \Xi_i(-p_i)}{\p p_i^\sigma}\non\\
&&\ +\f{i}{2}p_i.p_j \ell^\rho \varepsilon^\sigma \f{\p \mathcal{K}_i(-p_i)}{\p p_i^\rho}\f{\p \Xi_i(-p_i)}{\p p_i^\sigma}\Bigg] \Gamma^{(N)}_{(ij)}(p_i,p_j) +\ \mathcal{O}(\omega^0)\ .\label{A_III_result}
\ee
\endgroup
The fourth diagram in Fig.\ref{f:A^N+1_G} with G-graviton contributes at order $\omega^{-1}$ but won't contribute at order $\ln\omega$. The order $\omega^{-1}$ contribution turns out to be,
\be
 A_{IV}
&=& \sum_{k=1}^{N}e_k\f{\varepsilon.p_k}{p_k.k}\ \sum_{\substack{i=1\\ i\neq k}}^{N}\sum_{\substack{j=1\\j\neq k\\j>i}}^{N}8e_i e_j \epsilon_i^{T}\epsilon_j^{T}\int_{reg}  \f{d^4\ell}{(2\pi)^4}\ \f{-i}{\ell^2-i\epsilon} \f{1}{(p_i-\ell)^2+m_i^2-i\epsilon} \non\\
&&\f{1}{(p_j+\ell)^2+m_j^2 -i\epsilon} \Big[p_i^{\rho}\ell^\sigma \mathcal{N}^j_{\rho\sigma}(-p_j)- p_j^\rho \ell^\sigma\mathcal{N}^i_{\rho\sigma}(-p_i)\Big]\Gamma^{(N)}_{(ij)}(p_i,p_j)\ +\ \mathcal{O}(\omega^0)\ .
\ee
The fifth diagram in Fig.\ref{f:A^N+1_G} start contributing from order $\omega^0$ in the soft expansion when we evaluate it with G-photon propagator i.e. $A_{V}=0+\mathcal{O}(\omega^{0})$. Here we are not writing down the non-vanishing contribution of $A_{V}$ at order $\omega^{0}$ explicitly, as it is not essential for deriving the order $\omega^{-1}$ and $\ln\omega$ soft factors. Now summing over the external particle legs, total contribution of $\mathcal{A}^{(N+1)}_{G,1}$ at orders $\omega^{-1}$ and $\ln\omega$ turns out to be
\begingroup
\allowdisplaybreaks
\be
&&\mathcal{A}^{(N+1)}_{G,1}\nn\\
&=&\ A_{I}+A_{II}+A_{III}+A_{IV}+A_{V}\non\\
&=& \sum_{k=1}^{N}e_k\f{\varepsilon.p_k}{p_k.k}\sum_{i=1}^{N}\  \sum_{\substack{j=1\\j>  i}}^{N} 8e_i e_j \epsilon_i^{T}\epsilon_j^{T}\int_{reg} \f{d^4\ell}{(2\pi)^4}\ \f{-i}{\ell^2-i\epsilon} \f{1}{(p_i-\ell)^2+m_i^2-i\epsilon} \non\\
&&\f{1}{(p_j+\ell)^2+m_j^2 -i\epsilon} \Big[p_i^{\rho}\ell^\sigma \mathcal{N}^j_{\rho\sigma}(-p_j)- p_j^\rho \ell^\sigma\mathcal{N}^i_{\rho\sigma}(-p_i)\Big]\Gamma^{(N)}_{(ij)}(p_i,p_j)\non\\
&&+\ i\ \sum_{i=1}^{N}\  \sum_{\substack{j=1\\j\neq i}}^{N}e_i^2e_j\ \epsilon_i^{T}\epsilon_j^{T}\int_{reg} \f{d^4\ell}{(2\pi)^4} \f{1}{\ell^2-i\epsilon}\ \f{1}{p_i.\ell+i\epsilon }\ \f{1}{p_j.\ell -i\epsilon} \non\\
&&\ \times\Bigg[ \f{\varepsilon.p_i}{p_i.k}p_j.k -\f{\varepsilon.p_i}{p_i.k}\ p_i.p_j\f{ \ell.k}{p_i.\ell+i\epsilon}-\varepsilon.p_j +\varepsilon.\ell \f{p_i.p_j}{p_i.\ell+i\epsilon}\Bigg]\Gamma^{(N)}_{(ij)}(p_i,p_j)\ +\mathcal{O}(\omega^0)\ .
\ee
\endgroup
Interestingly in the above expression all the theory dependent pieces involving $\mathcal{K}_i\ , \Xi_i\ , \mathcal{B}^i$ at order $\ln\omega$ cancels out at the integrand level, when we some over the contributions \eqref{A_I_result},\eqref{A_II_result},\eqref{A_III_result}. This confirms the fact that the subleading soft photon theorem at order $\ln\omega$ is universal. The above result can also be rewritten in the following compact form
\be
\mathcal{A}^{(N+1)}_{G,1} &=&\sum_{k=1}^{N}e_k\f{\varepsilon.p_k}{p_k.k} \ \mathcal{A}^{(N)}_{G,1}\nn\\
&+&\sum_{i=1}^{N}e_i\f{\varepsilon_\mu k_\nu}{p_i\cdot k}\Big\lbrace\Big(p_i^{\mu}\f{\p }{\p p_{i\nu}}-p_i^{\nu}\f{\p }{\p p_{i\mu}}\Big)K_{em}^{reg}\Big\rbrace\ \Gamma^{(N)}\ +\mathcal{O}(\omega^0)\ ,\label{N+1_G_contribution}
\ee
where the expression of $\mathcal{A}^{(N)}_{G,1}$ is given in \eqref{A_G^N_result}, and $K_{em}^{reg}$ is the approximated form of the integral $K_{em}$ in \eqref{K_em} in the integration range $\omega<<|\ell^\mu|<<|p_i^\mu|,|p_j^\mu|$. The integration has been explicitly evaluated in \cite{1808.03288} and the result reads
\begingroup
\allowdisplaybreaks
\be
&&K_{em}^{reg}\equiv \f{i}{2}\sum_{\ell=1}^{N}\  \sum_{\substack{j=1\\j\neq \ell}}^{N}e_{\ell}e_{j}\ \int_{reg} \f{d^{4}\ell}{(2\pi)^{4}} \ \f{1}{\ell^{2}-i\epsilon}\ \f{(p_{\ell}\cdot p_{j})}{(p_{\ell}\cdot \ell +i\epsilon)\ (p_{j}\cdot\ell - i\epsilon)}\non\\
&\simeq & -\f{i}{2}\sum_{\ell =1}^{N}\  \sum_{\substack{j=1\\j\neq \ell}}^{N}\f{e_{\ell}e_{j}}{4\pi}\ (\ln\omega) \ \f{p_{\ell}\cdot p_{j}}{\sqrt{(p_{\ell}.p_{j})^{2}-p_{\ell}^{2}p_{j}^{2}}} \Bigg{\lbrace}\delta_{\eta_{\ell}\eta_{j},1}-\f{i}{2\pi}\ln\Bigg(\f{p_{\ell}.p_{j}+\sqrt{(p_{\ell}.p_{j})^{2}-p_{\ell}^{2}p_{j}^{2}}}{p_{\ell}.p_{j}-\sqrt{(p_{\ell}.p_{j})^{2}-p_{\ell}^{2}p_{j}^{2}}}\Bigg)\Bigg{\rbrace} .\non\\ 
\ee
\endgroup
Above $\eta_j=+1$ if $j$-th particle is ingoing and $\eta_j=-1$ if $j$-th particle is outgoing and under $\simeq$ sign we only have written the order $\ln\omega$ contribution while evaluating the integral.
\begin{center}
\begin{figure}
	\includegraphics[scale=0.55]{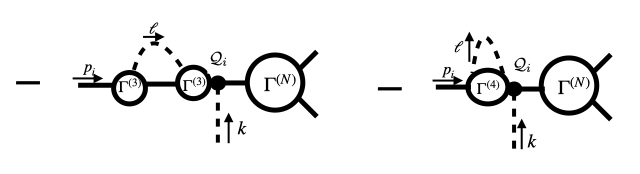}
	\caption{ The first diagram above corresponds to the sum over the contributions of the first four diagrams in Fig.\ref{f:A^N+1_self} when we replace the photon polarization $\varepsilon^\mu$ by $k^\mu$. The second diagram above corresponds to the sum over the contributions of the fifth and sixth diagrams in Fig.\ref{f:A^N+1_self} when we replace the photon polarization $\varepsilon^\mu$ by $k^\mu$.}\label{f:self_energy_sum}
\end{figure}
\end{center}

To evaluate the contribution of $\mathcal{A}^{(N+1)}_{K-\text{finite},1}$, first we compute the diagrams in Fig.\ref{f:A^N+1_G} with the virtual photon propagator replaced by the K-photon propagator. Afterward, we subtract the contribution of $K_{em}\Gamma^{(N+1)}$ from this computed result. The final contribution of $\mathcal{A}^{(N+1)}_{K-\text{finite},1}$ within the regulated range of integration $\omega<<|\ell^\mu|<<|p_i^\mu|,|p_j^\mu|$ becomes
\be
\mathcal{A}^{(N+1)}_{K-\text{finite},1}&=&  \sum_{k=1}^{N}e_k\f{\varepsilon.p_k}{p_k.k} \mathcal{A}^{(N)}_{K-\text{finite},1} +\mathcal{O}(\omega^0)\ ,\label{A_K^N+1_result}
\ee
where the expression of $\mathcal{A}^{(N)}_{K-\text{finite},1}$ is given in \eqref{A_K^N_result}. Note that the above result contributes at order $\omega^{-1}$ and does not contribute to order $\ln\omega$ soft theorem.

The self-energy kind of diagrams contributing to $\mathcal{A}^{(N+1)}_{\text{self,1}}$ in Fig.\ref{f:A^N+1_self} are not necessary to compute explicitly as these diagrams sum up to zero using on-shell renormalization condition \eqref{renormalization_condition} as we are going to discuss below. The contribution from the sum of the first four diagrams in Fig.\ref{f:A^N+1_self} can be described by the following general structure $\epsilon_i^{T}\varepsilon.p_i f_{1}(p_i.k)\Gamma_{(i)}^{(N)}(p_i+k)$. Similarly the contribution from the sum of fifth and sixth diagrams in Fig.\ref{f:A^N+1_self} can be described by the following structure $\epsilon_i^{T}\varepsilon.p_i f_{2}(p_i.k)\Gamma_{(i)}^{(N)}(p_i+k)$. Where $f_1(p_i.k)$ and $f_2(p_i.k)$ are two unknown functions with specific polarization/spin indices, which we determine below by replacing $\varepsilon^\mu \rightarrow k^\mu$ and using Ward identity. Using the diagrammatic identities of Fig.\ref{f:KG_photon_figure} and Fig.\ref{f:KG_decomposition_photon_subdiagram},  the first four diagrams in Fig.\ref{f:A^N+1_self} after replacing $\varepsilon^\mu \rightarrow k^\mu$ reduces to the first diagram in Fig.\ref{f:self_energy_sum}. Similarly using the diagrammatic identities of Fig.\ref{f:KG_photon_figure} and Fig.\ref{f:KG_decomposition_photon_generalized},  the sum of fifth and sixth diagrams in Fig.\ref{f:A^N+1_self} after replacing $\varepsilon^\mu \rightarrow k^\mu$ reduces to the second diagram in Fig.\ref{f:self_energy_sum}. Now using the constant matrices $F_1,F_2$ introduced in \eqref{eq:F_i_definitions} for the diagrams in Fig.\ref{f:A^N_self} and comparing the general structure above we get
\be
f_1(p_i.k)=-\f{1}{2(p_i.k)^2} F_1\mathcal{Q}_i\Xi_i(-p_i-k)\ ,\ f_1(p_i.k)=-\f{1}{2(p_i.k)^2} F_2\mathcal{Q}_i\Xi_i(-p_i-k)\ .
\ee
 Now we substituting the above result in the general structures for the sum of diagrams mentioned above. Finally summing over the contributions of the first seven diagrams in Fig.\ref{f:A^N+1_self} we get
 \be
 \epsilon_i^{T}\f{\varepsilon.p_i}{2(p_i.k)^2}\Big[-F_1\mathcal{Q}_i -F_2\mathcal{Q}_i +\mathcal{Q}_i^{T}C\Big] \Xi_i(-p_i-k)\Gamma^{(N)}_{(i)}(p_i+k)\ .
 \ee
 Note that $F_1$, $F_2$ satisfy the same property under the operation of charge matrix $\mathcal{Q}$  as $\Xi_i$ satisfies in \eqref{XiQ_relation}. Hence using this relation the terms inside the square bracket in the above expression reduces to $\mathcal{Q}_i^{T}(F_1+F_2+C)$ which vanishes using the on-shell renormalization condition \eqref{renormalization_condition}. Note that in the above expression we neglected the possible order $\omega^0$ contribution in the soft expansion of counter term diagram. Using the same on-shell renormalization condition the sum of the rest of the diagrams in Fig.\ref{f:A^N+1_self} also vanishes up to possible order $\omega^0$ contribution . Hence in the soft limit, the sum over all the Feynman diagrams in Fig.\ref{f:A^N+1_self} contribute as
 \be
 \mathcal{A}^{(N+1)}_{\text{self,1}} \ =\ 0\ +\mathcal{O}(\omega^0)\ .\label{N+1_self_contribution}
 \ee
\begin{center}
\begin{figure}
	\includegraphics[scale=0.43]{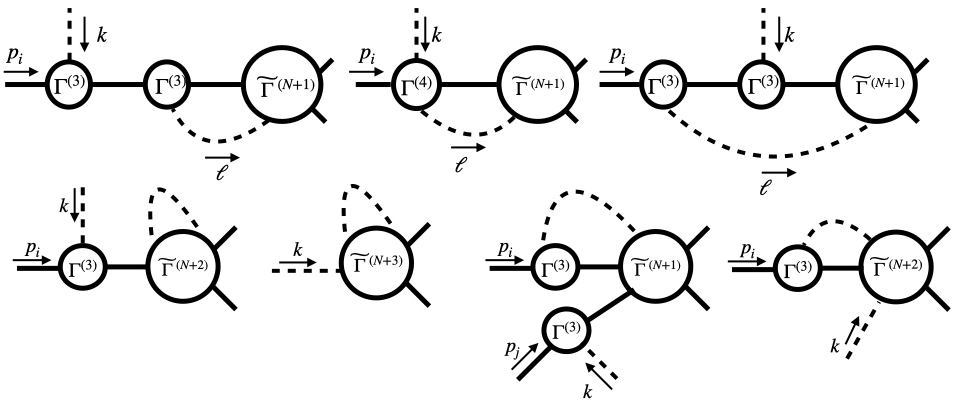}
	\caption{ Set of 1-loop diagrams contributing to $\mathcal{A}^{(N+1)}_{\text{non-div,1}}$ consists of diagrams where the virtual photon propagator is the full photon propagator, with at least one leg connected to an internal massive particle propagator or massive EFT vertices. Diagrams involving counter terms to remove UV divergences have been omitted.}\label{f:A^N+1_non-div}
\end{figure}
\end{center}

The diagrams in Fig.\ref{f:A^N+1_non-div} contributing to $\mathcal{A}^{(N+1)}_{\text{non-div,1}}$ are IR-divergence free, as in $\ell^\mu\rightarrow 0$ limit and finite $k^\mu$ the third, sixth and seventh diagrams behaves like $\int \f{d^4\ell}{|\ell|^3}$, and first, second, fourth and fifth diagrams behaves like $\int \f{d^4\ell}{|\ell|^2}$. On the other hand the sum of first, fourth and sixth diagrams in Fig.\ref{f:A^N+1_non-div} contribute to leading soft factor at order $\omega^{-1}$. Now in the integration region $\omega<<|\ell^\mu|<<|p_i^\mu |$, individually the first and third diagrams in Fig.\ref{f:A^N+1_non-div} behave like $\int_{reg} \f{d^4\ell}{\ell^2-i\epsilon}\f{1}{(p_i.\ell+i\epsilon )^2} $ after the expansion of the propagators, hence those have the possibility of contributing at order $\ln\omega$. But when we sum over the contributions of the first, second and third diagrams, the order $\ln\omega$ contributing coefficient of $\int_{reg} \f{d^4\ell}{\ell^2-i\epsilon}\f{1}{(p_i.\ell+i\epsilon )^2} $ cancels each other, and  left out part starts contributing from order $\omega^0$. Hence leaving first, fourth and sixth diagrams, all the other diagrams start contributing at order $\omega^0$ in the soft expansion. After summing over all the contributions, we get
\be
\mathcal{A}^{(N+1)}_{\text{non-div,1}}=\sum_{i=1}^{N} e_i\f{\varepsilon.p_i}{p_i.k}\ \mathcal{A}^{(N)}_{\text{non-div,1}}\ +\mathcal{O}(\omega^0)\ .\label{N+1_non-div_contribution}
\ee
\paragraph{Soft photon theorem result at one-loop:}
Summing over the contributions of \\
 \eqref{N+1_G_contribution},\eqref{A_K^N+1_result},\eqref{N+1_self_contribution},\eqref{N+1_non-div_contribution} in the soft limit, we get the following soft theorem expression 
\be
\mathcal{A}^{(N+1)}_{\text{IR-finite,1}}&=&\Big[\mathcal{A}^{(N+1)}_{G,1}+ \mathcal{A}^{(N+1)}_{K-\text{finite},1}+\mathcal{A}^{(N+1)}_{\text{self,1}}+\mathcal{A}^{(N+1)}_{\text{non-div,1}}\Big]\non\\
&\overset{\omega\rightarrow 0}{ =} &\ \sum_{i=1}^{N}e_i\f{\varepsilon.p_i}{p_i.k} \ \mathcal{A}^{(N)}_{\text{IR-finite},1}\nn\\
&+&\sum_{i=1}^{N}e_i\f{\varepsilon_\mu k_\nu}{p_i\cdot k}\Big\lbrace\Big(p_i^{\mu}\f{\p }{\p p_{i\nu}}-p_i^{\nu}\f{\p }{\p p_{i\mu}}\Big)K_{em}^{reg}\Big\rbrace\ \mathcal{A}^{(N)}_{\text{IR-finite,0}} +\mathcal{O}(\omega^0)\ . \label{final_result_QED}
\ee
This result agrees with the loop corrected subleading soft photon theorem, originally derived in \cite{1808.03288} for minimally coupled scalar QED. The derivation of this result further confirms the universality of the $\ln\omega$ soft factor in scattering events involving particles with arbitrary spins in a generic theory of quantum electrodynamics, which allows for arbitrary non-minimal couplings. Moreover, the obtained result also verifies the well-known fact that Weinberg's leading soft photon theorem remains unaltered by loop corrections. Therefore, even from the analysis presented above, we observe that Weinberg's soft theorem still holds, relating two one-loop IR-finite amplitudes.

\subsection{Discussion on generalization}\label{S:DiscussionI}
In order to obtain the order $\omega\ln\omega$ soft factor from the aforementioned one-loop amplitude, we require the vertices: $\Gamma^{(3)}$ as given in equation \eqref{Gamma3_photon} up to order $\mathcal{O}(\ell^2)$, $\Gamma^{(4)}$ as given in equation \eqref{Gamma4_photon} up to order $\mathcal{O}(\ell_1, \ell_2)$, and $\widetilde{\Gamma}^{(N+1)}$ up to order $\mathcal{O}(\ell)$. However, when dealing with non-minimal couplings as expressed in equation \eqref{non-minimal}, it is currently unknown how to derive $\widetilde{\Gamma}^{(N+1)}$ in terms of $\Gamma^{(N)}$ at order $\mathcal{O}(\ell)$. Additionally, at order $\mathcal{O}(\ell^2)$, new sets of non-minimal couplings will contribute to $\Gamma^{(3)}$. Due to these reasons, we are unable to derive the order $\omega\ln\omega$ soft photon theorem in this article, and it is not clear whether such a soft factorization at order $\omega\ln\omega$ is possible or not.

In the theory of scalar-QED minimally coupled to gravity the order $\ln\omega$ correction to \eqref{final_result_QED} due to gravitational interaction has also been derived in \cite{1808.03288} by analyzing one-loop amplitudes. In the theory of scalar-QED soft photon theorem at order $\omega (\ln\omega)^2$ has also been derived in \cite{2008.04376} analyzing two-loop amplitudes and the soft factor is provided in \eqref{eq:S_2loop_photon}. Multiple soft photon theorem up to subleading order in soft expansion has also been derived in section-(3.5) of \cite{Sahoo:2020csy}, by analyzing one-loop amplitudes and the result reads:
\be
&&\mathcal{A}^{(N+M)}_{\text{IR-finite,1}}\non\\
&&\overset{\omega\rightarrow 0}{ =} \ \prod_{\ell=1}^{M}\Bigg\lbrace \sum_{i=1}^{N}e_i\f{\varepsilon_\ell\cdot p_i}{p_i\cdot k_\ell}\Bigg\rbrace \ \mathcal{A}^{(N)}_{\text{IR-finite},1}\nn\\
&&\ +\sum_{\ell=1}^{M}\Bigg[\prod_{\substack{m=1\\ m\neq \ell}}^{M}\Bigg\lbrace \sum_{j=1}^{N}e_j\f{\varepsilon_m\cdot p_j}{p_j\cdot k_m}\Bigg\rbrace \Bigg] \sum_{i=1}^{N}e_i\f{\varepsilon_{\ell\mu} k_{\ell\nu}}{p_i\cdot k_\ell}\Big\lbrace\Big(p_i^{\mu}\f{\p }{\p p_{i\nu}}-p_i^{\nu}\f{\p }{\p p_{i\mu}}\Big)K_{em}^{reg}\Big\rbrace\ \mathcal{A}^{(N)}_{\text{IR-finite,0}} \non\\
&&\ +\ \mathcal{O}\big(\omega^{-M+1}\big)\ . \label{eq:multi_soft_photon}
\ee
Above $\mathcal{A}^{(N+M)}_{\text{IR-finite,1}}$ represents the IR-finite part of 1-loop scattering amplitude involving $N$-number of hard particles and $M$-number of soft photons with polarizations and momenta $\lbrace \varepsilon_\ell ,k_\ell=-\omega \mathbf{n}_\ell\rbrace$ for $\ell=1,\cdots, M$ as external states.

\section{Soft graviton theorem at one-loop}
\label{S:soft_graviton_theorem}
In this section we derive subleading soft graviton theorem analyzing one-loop amplitudes for a quantum mechanical scattering process involving $N$ number of massive particles with arbitrary spin and one graviton, extending the analysis of \cite{1808.03288} for a generic theory of quantum gravity. This derivation will establish the universal (theory independent) nature of  $\ln\omega$ soft graviton factor.

\subsection{Sen's Covariantization prescription and Feynman rules}\label{S:Covariantization_gravity}
The covariantization of the quadratic part of the massive EFT action \eqref{S2} in the soft gravitational background has been carried out in \cite{1703.00024, 1706.00759, 1707.06803}. This development has been utilized to derive vertices involving two massive spinning particles and one or two on-shell soft gravitons. However, when performing loop computations, we require the same vertices involving off-shell gravitons. Due to the off-shell nature of the gravitons, it is not possible to independently impose the traceless and transverse conditions on the gravitational fluctuation consistently with diffeomorphism. To address this issue, we make slight modifications to the covariantization prescription proposed in \cite{1703.00024, 1706.00759, 1707.06803}, as described below. In this work, we use a different parametrization of gravitational fluctuation, which is closely related to the one presented in \cite{9411092}. This alternative parametrization allows us to derive the vertices within the covariantization prescription while ensuring compatibility with the {\it{de Donder}} gauge choice. Let us define the deviation of background metric from flat Minkowski metric as
\be
h_{\mu\nu}(x)\  \equiv \  \f{1}{2\kappa}\big(g_{\mu\nu}(x)-\eta_{\mu\nu}\big)\ ,\label{hmunu_definition}
\ee
where $\kappa=\sqrt{8\pi G}\ $ with $G$ being the four dimensional Newton's constant. In {\it{de Donder}} gauge $\p^\mu h_{\mu\nu}=\f{1}{2}\p_\nu h^{\rho}_\rho$ the graviton propagator becomes
\be
\Delta_F^{\mu\nu,\rho\sigma}(\ell)=\f{-i}{\ell^2 -i\epsilon}\ \f{1}{2}\big(\eta^{\mu\rho}\eta^{\nu\sigma}+\eta^{\mu\sigma}\eta^{\nu\rho}-\eta^{\mu\nu}\eta^{\rho\sigma}\big)\ , \label{eq:full_gr_propagator}
\ee
where $(\mu\nu)$ and $(\rho\sigma)$ are Lorentz indices of the two ends of the graviton propagator. The  polarization tensor for on-shell graviton with momentum $k$ will be denoted by $\varepsilon_{\mu\nu}(k)$ which satisfies the traceless and transverse conditions
\be
\eta^{\mu\nu}\varepsilon_{\mu\nu}(k)=0\ \hspace{1cm},\hspace{1cm} k^\mu \varepsilon_{\mu\nu}(k)=0\ .
\ee
With the definition of gravitational fluctuation \eqref{hmunu_definition}, we express the following quantities as a power series expansion of $\kappa$: 
\begingroup
\allowdisplaybreaks
\be
\sqrt{-\text{det}\ g}&=&1+\kappa h+\f{\kappa^2}{2}(h^2 -2h^{\mu\nu}h_{\mu\nu})+\f{\kappa^3}{6}(h^3+8h^{\mu\rho} h_{\rho\nu} h^\nu_{\ \mu}-6hh^{\mu\nu}h_{\mu\nu})+\mathcal{O}(\kappa^4)\ ,\non\\
g^{\mu\nu}&=&\eta^{\mu\nu}-2\kappa h^{\mu\nu}+4\kappa^2 h^{\mu\rho}h_\rho^{\ \nu} -8\kappa^3 h^{\mu\tau}h_{\tau\rho}h^{\rho\nu}+\mathcal{O}(\kappa^4)\ ,\non\\
e_\mu^{\ a} &=&\delta_\mu^a +\kappa h_\mu^{\ a}-\f{\kappa^2}{2}h_\rho^{\ a}h_\mu^{\ \rho}+\f{\kappa^3}{2}h_\rho^{\ a}h_\nu^{\ \rho}h_\mu^{\ \nu}+\mathcal{O}(\kappa^4)\ ,\non\\
E_a^{\ \mu} &=&\delta_a^\mu -\kappa h_a^{\ \mu}+\f{3}{2}\kappa^2 h_a^{\ \rho}h_\rho^{\ \mu}-\f{5}{2}\kappa^3 h_a^{\ \rho}h_{\rho\nu}h^{\nu\mu}+\mathcal{O}(\kappa^4)\ ,\non\\
\Gamma^{\lambda}_{\mu\nu}&=&\kappa (\eta^{\lambda\sigma}-2\kappa h^{\lambda\sigma}+4\kappa^2 h^{\lambda\rho}h_\rho^{\ \sigma})\big( \p_\mu h_{\nu\sigma}+\p_\nu h_{\sigma\mu}-\p_\sigma h_{\mu\nu}\big)+\mathcal{O}(\kappa^4)\ ,\non\\
\omega_\mu^{ab}&\equiv & \eta^{bc}\ e_\nu^{\ a}D_\mu E_c^{\ \nu}=\eta^{bc}\ e_\nu^{\ a}\p_\mu E_c^{\ \nu}+\eta^{bc}\ e_\nu^{\ a}\Gamma_{\mu\rho}^\nu E_c^{\ \rho}\non\\
&=& \kappa (\p^b h_\mu^{\ a}-\p^a h_\mu^{\ b})+\kappa^2\Big( \f{1}{2}h^{b\nu}\p_\mu h_\nu^{\ a}-\f{1}{2}h^{a\nu}\p_\mu h_\nu^{\ b}+h^{b\nu}\p^a h_{\mu\nu}-h^{a\nu}\p^b h_{\mu\nu}\non\\
&&\ +h^{a\nu}\p_\nu h_\mu^{\ b}-h^{b\nu}\p_\nu h_\mu^{\ a}\Big)+ \mathcal{O}(\kappa^3)\ ,\non\\
R_{\mu\nu\rho\sigma}&=& \kappa(\p_\rho \p_\nu h_{\mu\sigma}+\p_\mu \p_\sigma h_{\nu\rho}-\p_\mu\p_\rho h_{\nu\sigma}-\p_\sigma \p_\nu h_{\mu\rho})+\mathcal{O}(\kappa^2)\ .\label{metric_perturbation}
\ee
\endgroup
In the RHS of all the above expressions both curved space indices ($\mu, \nu, \rho, \sigma, \ldots$) and tangent space indices ($a, b, c, \ldots$) are raised or lowered by using the Minkowski metric $\eta$. The trace of the metric fluctuation is defined as $h\equiv h_{\mu\nu}\eta^{\mu \nu}$. Additionally, the symbols $e_\mu^{\ a}$ represent the vierbein, $E_a^{\ \mu}$ represents the inverse vierbein, $\omega_\mu^{ab}$ represents the spin-connection, $\Gamma^{\lambda}_{\mu\nu}$ represents the Christoffel connection, and $R_{\mu\nu\rho\sigma}$ represents the Riemann tensor.

\paragraph{Covariantization:}
In the covariantization prescription, we derive the action that describes the interaction between two spinning particles and one graviton up to second derivative order on gravitational fluctuations. Additionally, we obtain the interaction between two spinning particles and two gravitons up to first derivative order on gravitational fluctuations in the derivative expansion. At the second derivative order, the interacting action of two spinning particles and one graviton also includes a generic non-minimal coupling through the curvature tensor, which is inherently general coordinate invariant. Under covariantization prescription tangent space derivatives on $\Phi_\beta$ should be replaced by covariant derivative in the curved space after multiplication of inverse vierbeins in the following way
\be
\p_{a_1}\p_{a_2}\cdots \p_{a_n}\Phi_\beta \ \rightarrow\ E_{a_1}^{\ \mu_1}E_{a_2}^{\ \mu_2}\cdots E_{a_n}^{\ \mu_n}\ D_{(\mu_1}D_{\mu_2}\cdots D_{\mu_n)}\Phi_\beta\ , \label{covariantize_gravity}
\ee
where the expression of one, two and three covariant derivatives on $\Phi_\beta$ are given by
\begingroup
\allowdisplaybreaks
\be
D_\mu \Phi_{\beta}&=&\p_\mu \Phi_\beta +\left[\f{1}{2}\omega_\mu^{ab}(\Sigma_{ab})_\beta^{\ \gamma}\Phi_\gamma \right]\ ,\\
D_{(\mu} D_{\nu)}\Phi_{\beta}&=&\p_\mu \p_\nu \Phi_\beta +\f{1}{2}\Big(\omega_\mu^{ab}(\Sigma_{ab})_\beta^{\ \gamma}\p_\nu\Phi_\gamma +\omega_\nu^{ab}(\Sigma_{ab})_\beta^{\ \gamma}\p_\mu\Phi_\gamma\Big)+\left[\f{1}{2}\p_{(\mu}\omega_{\nu)}^{ab}(\Sigma_{ab})_\beta^{\ \gamma}\Phi_\gamma\right]\non\\
&&+\f{1}{4}\omega_{(\mu}^{ab}\omega_{\nu)}^{cd}(\Sigma_{ab})_\beta^{\ \gamma}(\Sigma_{cd})_\gamma^{\ \delta}\Phi_\delta -\left[\Gamma^\rho_{\mu\nu}\p_\rho \Phi_\beta \right] -\f{1}{2}\Gamma^{\rho}_{\mu\nu}\omega_\rho^{ab}(\Sigma_{ab})_\beta^{\ \gamma}\Phi_\gamma \ ,\\
D_\mu D_\nu D_\rho \Phi_\beta &=& \p_\mu \p_\nu \p_\rho \Phi_\beta +\f{1}{2}\p_\mu \omega_\nu^{ab}(\Sigma_{ab})_\beta^{\ \gamma}D_\rho\Phi_\gamma +\f{1}{2} \omega_\nu^{ab}(\Sigma_{ab})_\beta^{\ \gamma}\p_\mu D_\rho\Phi_\gamma \non\\
&&-\left[ \p_\mu \Gamma^\sigma_{\nu\rho}\p_\sigma \Phi_\beta\right]-\f{1}{2} \p_\mu \Gamma^\sigma_{\nu\rho}\ \omega_\sigma^{ab}(\Sigma_{ab})_\beta^{\ \gamma}\Phi_\gamma  - \Gamma^\sigma_{\nu\rho}\p_\mu D_\sigma\Phi_\beta \non\\
&&+\f{1}{2}\omega_\mu^{ab}(\Sigma_{ab})_\beta^{\ \gamma}D_\nu D_\rho \Phi_\gamma -\Gamma^\sigma_{\mu\nu}D_\sigma D_\rho \Phi_\beta -\Gamma^\sigma_{\mu\rho}D_\nu D_\sigma \Phi_\beta\ .
\ee
\endgroup
The last equation above needs to symmetrize in $\mu,\nu,\rho$ indices before substituting in \eqref{covariantize_gravity}. In the above expressions the terms within square brackets are new at each derivative order in the covariantization prescription, which are important in deriving the interaction between two spinning particles and one graviton up to two derivative order on gravitational fluctuation, and the interaction between two spinning particles and two gravitons up to one derivative order on gravitational fluctuation. To the derivative order we are interested to find the interacting parts of the action we do not need to know the new terms coming from covariantization of more than three derivatives on $\Phi$. Also while covariantizing the action \eqref{S2}, we need to include $\sqrt{-\text{det }g}$ as a covariant measure of volume. 

The Fourier transform of gravitational fluctuation $h_{\mu\nu}(x)$ will be denoted by $h_{\mu\nu}(\ell)$ and the relation between them is given by
\be
h_{\mu\nu}(x)\equiv \int \f{d^4\ell}{(2\pi)^4} \ e^{i\ell.x}\ h_{\mu\nu}(\ell)\ .
\ee
\paragraph{Interaction part of the action:}
Under the above prescribed covariantization procedure the interactive part of the action describing interaction between two massive spinning particles and one graviton up to quadratic order in graviton momentum becomes
\begingroup
\allowdisplaybreaks
\be
S^{(3)}&=&\ \f{\kappa}{2}\int \f{d^4 q_1}{(2\pi)^4}\f{d^4 q_2}{(2\pi)^4}\f{d^4\ell}{(2\pi)^4}\ (2\pi)^4\delta^{(4)}(q_1+q_2+\ell)\Phi_{\alpha}(q_1)\Bigg[ h(\ell)\mathcal{K}^{\alpha\beta}(q_2)-h_a^\mu(\ell) q_{2\mu}\f{\p \mathcal{K}^{\alpha\beta}(q_2)}{\p q_{2a}}\non\\
&&+\f{1}{2} \big\lbrace \ell^b h^a_\mu(\ell)-\ell^a h_\mu^b(\ell)\big\rbrace \f{\p \mathcal{K}^{\alpha\gamma}(q_2)}{\p q_{2\mu}}(\Sigma_{ab})_\gamma^{\ \beta}+\f{1}{4} \ell_\mu\big\lbrace \ell^b h^a_\nu(\ell)-\ell^a h_\nu^b(\ell)\big\rbrace\ \f{\p^2 \mathcal{K}^{\alpha\gamma}(q_2)}{\p q_{2\mu} \p q_{2\nu}}(\Sigma_{ab})_\gamma^{\ \beta}\non\\
&& -\f{1}{2}\big\lbrace \ell_\mu h^\rho_\nu(\ell) +\ell_\nu h^\rho_\mu(\ell) -\ell^\rho h_{\mu\nu}(\ell)\big\rbrace  q_{2\rho}\f{\p^2 \mathcal{K}^{\alpha\beta}(q_2)}{\p q_{2\mu} \p q_{2\nu}}\non\\
&&-\f{1}{6}\ell_\mu \big\lbrace \ell_\nu h_\rho^\sigma(\ell) +\ell_\rho h_\nu^\sigma(\ell) -\ell^\sigma h_{\nu\rho}(\ell)\big\rbrace q_{2\sigma}\f{\p^3 \mathcal{K}^{\alpha\beta}(q_2)}{\p q_{2\mu} \p q_{2\nu} \p q_{2\rho}} +\mathcal{O}(\ell^3)\Bigg]\Phi_{\beta}(q_2)\ .\label{S3_graviton}
\ee
\endgroup
By covariantization procedure we only get the interacting action representing minimal coupling of graviton with matter field. 
On top of it at the quadratic order in graviton momentum we also need to add generic non-minimal interaction term which describes the interaction between two spinning particles and one graviton through lineariszed Riemann tensor. A generic form of the action describing such kind of non-minimal interaction is given by
\be
\overline{S}^{(3)}= \f{1}{2} &&\int\f{d^{4}q_{1}}{(2\pi)^{4}}\f{d^{4}q_{2}}{(2\pi)^{4}}\f{d^4\ell}{(2\pi)^4}\ (2\pi)^{4}\delta^{(4)}(q_{1}+q_{2}+\ell)\non\\
 &&\ \Phi_{\alpha}(q_{1})\ \Big[R_{\mu\nu\rho\sigma}(\ell)\  \mathcal{G}^{\alpha\beta,\mu\nu\rho\sigma}(q_{2})\ +\mathcal{O}(\ell^3)\Big] \Phi_{\beta}(q_{2})\ , \label{non-minimal-gravity}
\ee
where $R_{\mu\nu\rho\sigma}(\ell)$ is the Fourier transform of the linearized Riemann tensor in \eqref{metric_perturbation} which reads
\be
R_{\mu\nu\rho\sigma}(\ell)=-\kappa(\ell_\rho \ell_\nu h_{\mu\sigma}+\ell_\mu \ell_\sigma h_{\nu\rho}-\ell_\mu\ell_\rho h_{\nu\sigma}-\ell_\sigma \ell_\nu h_{\mu\rho})\ .
\ee
In the non-minimal action \eqref{non-minimal-gravity}, $\mathcal{G}$ satisfies the following property
\be
\mathcal{G}^{\alpha\beta,\mu\nu\rho\sigma}(q_2)&=&\pm \ \mathcal{G}^{\beta\alpha,\mu\nu\rho\sigma}(-q_1-\ell)\ .
\ee
In the above equation $+$ sign is for Grassmannian even field and $-$ sign is for Grassmannian odd field. Again during the derivation we consider $\Phi$ field components being Grassmannian even, but the final result will be valid for both Grassmannian even and odd fields.

Following the covariantization procedure, the part of the action describing interaction between two massive spinning particles and two gravitons up to linear order in graviton momenta becomes
\begingroup
\allowdisplaybreaks
\be
S^{(4)}&=&\f{\kappa^2}{2}\int \f{d^4 q_1}{(2\pi)^4}\f{d^4 q_2}{(2\pi)^4}\f{d^4\ell_1}{(2\pi)^4}\f{d^4\ell_2}{(2\pi)^4}\ (2\pi)^4\delta^{(4)}(q_1+q_2+\ell_1+\ell_2)\non\\
&&\Phi_{\alpha}(q_1)\Bigg[ \f{1}{2}\big\lbrace h(\ell_1)h(\ell_2)-2h^{\mu\nu}(\ell_1)h_{\mu\nu}(\ell_2)\big\rbrace \mathcal{K}^{\alpha\beta}(q_2)+h(\ell_1)\Bigg(-h_a^\mu(\ell_2) q_{2\mu}\f{\p \mathcal{K}^{\alpha\beta}(q_2)}{\p q_{2a}}\non\\
&&+\f{1}{2} \big\lbrace \ell_2^b h^a_\mu(\ell_2)-\ell_2^a h_\mu^b(\ell_2)\big\rbrace \f{\p \mathcal{K}^{\alpha\gamma}(q_2)}{\p q_{2\mu}}(\Sigma_{ab})_\gamma^{\ \beta}-\f{1}{2}\big\lbrace \ell_{2\mu} h^\rho_\nu(\ell_2) +\ell_{2\nu} h^\rho_\mu(\ell_2) -\ell_2^\rho h_{\mu\nu}(\ell_2)\big\rbrace \non\\
&& \times q_{2\rho} \f{\p^2 \mathcal{K}^{\alpha\beta}(q_2)}{\p q_{2\mu} \p q_{2\nu}}\Bigg)+\f{3}{2}h_a^\rho(\ell_1)h_\rho^\mu (\ell_2)q_{2\mu}\f{\p \mathcal{K}^{\alpha\beta}(q_2)}{\p q_{2a}} +\f{1}{2}h_a^\mu(\ell_1)h_b^\nu(\ell_2)q_{2\mu}q_{2\nu}\f{\p^2 \mathcal{K}^{\alpha\beta}(q_2)}{\p q_{2a} \p q_{2b}}\non\\
&&-\f{1}{2}h_c^\mu(\ell_1)\big\lbrace \ell_2^b h_\mu^a(\ell_2)-\ell_2^a h_\mu^b(\ell_2)\big\rbrace \f{\p \mathcal{K}^{\alpha\gamma}(q_2)}{\p q_{2c}}(\Sigma_{ab})_\gamma^{\ \beta}-\f{1}{4}h_c^\mu(\ell_1)q_{2\mu}\big\lbrace \ell_2^b h_\nu^a(\ell_2)-\ell_2^a h_\nu^b(\ell_2)\big\rbrace\non\\
&&\times  \f{\p^2 \mathcal{K}^{\alpha\gamma}(q_2)}{\p q_{2c} \p q_{2\nu}}(\Sigma_{ab})_\gamma^{\ \beta} +\f{1}{2}h_a^\mu(\ell_1)\big\lbrace \ell_{2\mu}h_\nu^\rho (\ell_2)+\ell_{2\nu}h_\mu^\rho(\ell_2)-\ell_2^\rho h_{\mu\nu}(\ell_2)\big\rbrace q_{2\rho}\f{\p^2 \mathcal{K}^{\alpha\beta}(q_2)}{\p q_{2a} \p q_{2\nu}}\non\\
&&+\f{1}{6}h_a^\mu(\ell_1)q_{2\mu}\big\lbrace \ell_{2\nu}h_\sigma^\rho(\ell_2)+\ell_{2\sigma}h_\nu^\rho(\ell_2)-\ell_2^\rho h_{\nu\sigma}(\ell_2)\big\rbrace q_{2\rho} \f{\p^3 \mathcal{K}^{\alpha\beta}(q_2)}{\p q_{2a}\p q_{2\nu} \p q_{2\sigma}}\non\\
&& +\f{1}{2}\Big\lbrace \f{1}{2}h^{b\sigma}(\ell_1)\ell_{2\mu}h_\sigma^{a}(\ell_2)-\f{1}{2}h^{a\sigma}(\ell_1)\ell_{2\mu}h_\sigma^{b}(\ell_2)+h^{b\sigma}(\ell_1)\ell_2^a h_{\mu\sigma}(\ell_2)-h^{a\sigma}(\ell_1)\ell_2^b h_{\mu\sigma}(\ell_2)\non\\
&&+h^{a\sigma}(\ell_1)\ell_{2\sigma}h_\mu^b(\ell_2)-h^{b\sigma}(\ell_1)\ell_{2\sigma}h_\mu^a(\ell_2)\Big\rbrace \f{\p \mathcal{K}^{\alpha\gamma}(q_2)}{\p q_{2\mu}}(\Sigma_{ab})_\gamma^{\ \beta}\ +h^{\rho\sigma}(\ell_1)\big\lbrace \ell_{2\mu}h_{\nu\sigma}(\ell_2)\non\\
&&+\ell_{2\nu}h_{\mu\sigma}(\ell_2)-\ell_{2\sigma}h_{\mu\nu}(\ell_2)\big\rbrace q_{2\rho}\f{\p^2 \mathcal{K}^{\alpha\beta}(q_2)}{\p q_{2\mu}\p q_{2\nu}} +\mathcal{O}(\ell_1^2, \ell_1\ell_2,\ell_2^2)\Bigg]\Phi_{\beta}(q_2)\ .\label{S4_graviton}
\ee
\endgroup

We also need to provide a purely gravitational effective action, constructed out off curvature tensors in derivative expansion. This action describes the self-interaction of the graviton field and provide dynamics to graviton. For the analysis we are conducting here, it suffices to consider the leading term of the EFT action, which corresponds to the Einstein-Hilbert action and is expressed as follows:
\be
S_{\text{EFT}}&=& \f{1}{2\kappa^2}\int d^4x\sqrt{-\text{det}(g_{\mu\nu})}\ \left(R +\mathcal{O}(R^2)\right)\ .\label{eq:S_EFT}
\ee

\paragraph{Feynman rules for the vertices:}
Starting from the interacting parts of the action given in \eqref{S3_graviton} and \eqref{non-minimal-gravity}, Feynman rule for the vertex describing interaction between two spinning particles with momenta $q$ and $-(q+\ell)$, and one graviton with Lorentz indices $\mu\nu$ and momentum $\ell$ turns out to be
\begingroup
\allowdisplaybreaks
\be
&&\Gamma^{(3)}_{\mu\nu}(q,-q-\ell,\ell)\non\\
&=&\ i\kappa \Bigg[ \eta_{\mu\nu}\mathcal{K}(-q)+\f{1}{2}\eta_{\mu\nu}\ell^\rho \f{\p \mathcal{K}(-q)}{\p q^\rho}+\f{1}{4}\eta_{\mu\nu}\ell^\rho \ell^\sigma \f{\p^2 \mathcal{K}(-q)}{\p q^\rho \p q^\sigma}- q_{(\mu} \f{\p \mathcal{K}(-q)}{\p q^{\nu )}}-\f{1}{2}\ell_{(\mu} \f{\p \mathcal{K}(-q)}{\p q^{\nu)}}\non\\
&&-\f{1}{2}q_{(\mu}\ell^\rho \f{\p^2 \mathcal{K}(-q)}{\p q^{\nu)} \p q^\rho} -\f{1}{4}q_{(\mu}\ell^\rho \ell^\sigma \f{\p^3 \mathcal{K}(-q)}{\p q^{\nu)} \p q^\rho \p q^\sigma}-\f{1}{2}\ell_{(\mu} \ell^\rho \f{\p^2 \mathcal{K}(-q)}{\p q^{\nu)} \p q^\rho}-\f{1}{2}\ell^b\f{\p \mathcal{K}(-q)}{\p q^{(\mu}}\Sigma_{\nu )b} \non\\
&&+\f{1}{2}\ell^b \ \Sigma^T_{(\nu b} \f{\p \mathcal{K}(-q)}{\p q^{\mu)}} -\f{1}{4}\ell^b \ell^\rho\ \f{\p^2 \mathcal{K}(-q)}{\p q^{(\mu} \p q^\rho}\Sigma_{\nu) b} +\f{1}{4}\ell^b \ell^\rho \Sigma^T_{(\nu b}\f{\p^2 \mathcal{K}(-q)}{\p q^{\mu)} \p q^\rho}\non\\
&& +\f{1}{4}\Bigg\lbrace \ell_{\mu} \ell^\rho \f{\p^2 \mathcal{K}(-q)}{\p q^\rho \p q^{\nu}}+\ell_{\nu} \ell^\rho \f{\p^2 \mathcal{K}(-q)}{\p q^\rho \p q^{\mu}}-\ell^2   \f{\p^2 \mathcal{K}(-q)}{\p q^\mu \p q^\nu}\Bigg\rbrace \non\\
&&+\f{1}{12}\ell^\sigma\Bigg\lbrace q_{\mu} \ell^\rho \f{\p^3 \mathcal{K}(-q)}{\p q^\rho \p q^{\nu} \p q^\sigma}+q_{\nu} \ell^\rho \f{\p^3 \mathcal{K}(-q)}{\p q^\rho \p q^{\mu} \p q^\sigma}-q.\ell   \f{\p^3 \mathcal{K}(-q)}{\p q^\mu \p q^\nu \p q^\sigma}\Bigg\rbrace \non\\
&& -\ell^\rho \ell^\sigma \Big\lbrace \mathcal{G}_{(\mu\rho\sigma \nu)}(-q)+\mathcal{G}_{\sigma(\nu\mu)\rho}(-q)-\mathcal{G}_{(\mu\rho\nu)\sigma }(-q) -\mathcal{G}_{\sigma(\nu\rho\mu)}(-q)\Big\rbrace +\mathcal{O}(\ell^3)\Bigg]\ ,\label{Gamma3_graviton}
\ee
\endgroup
where we suppressed the massive particle spin/polarization indices. The above expression is symmetrized under $\mu \leftrightarrow\nu$ exchange and in our convention momenta of the particles are always flowing towards the interaction vertex i.e. ingoing. Analogously starting from the interacting part of the action given in \eqref{S4_graviton}, Feynman rule for the vertex describing interaction between two spinning particles with momenta $q$ and $-(q+\ell_1+\ell_2)$, and two gravitons with Lorentz indices $(\mu\nu), (\rho\sigma)$ and momenta $\ell_1, \ell_2$ respectively, turns out to be
\begingroup
\allowdisplaybreaks
\be
&&\Gamma^{(4)}_{\mu\nu ,\rho\sigma}(q,-q-\ell_1-\ell_2,\ell_1,\ell_2)\non\\
&=&i\kappa^2 \Bigg[ (\eta_{\mu\nu}\eta_{\rho\sigma}-2\eta_{\mu\rho}\eta_{\nu\sigma})\Big\lbrace\mathcal{K}(-q)+\f{1}{2}(\ell_1+\ell_2)^\kappa \f{\p \mathcal{K}(-q)}{\p q^\kappa}\Big\rbrace -\eta_{\mu\nu}\Bigg\lbrace q_\rho \f{\p \mathcal{K}(-q)}{\p q^\sigma}\non\\
&&+\f{1}{2}(\ell_1+\ell_2)_\rho \f{\p \mathcal{K}(-q)}{\p q^\sigma}+\f{1}{2}q_\rho (\ell_1+\ell_2)^\kappa \f{\p^2 \mathcal{K}(-q)}{\p q^\sigma \p q^\kappa}+\f{1}{2}\ell_2^b \f{\p \mathcal{K}(-q)}{\p q^\rho}\Sigma_{\sigma b} -\f{1}{2}\ell_2^b \Sigma^T_{\sigma b}\f{\p \mathcal{K}(-q)}{\p q^\rho}\Bigg\rbrace\non\\
&&-\eta_{\rho\sigma}\Bigg\lbrace q_\mu \f{\p \mathcal{K}(-q)}{\p q^\nu}+\f{1}{2}(\ell_1+\ell_2)_\mu \f{\p \mathcal{K}(-q)}{\p q^\nu}+\f{1}{2}q_\mu (\ell_1+\ell_2)^\kappa \f{\p^2 \mathcal{K}(-q)}{\p q^\nu \p q^\kappa}+\f{1}{2}\ell_1^b \f{\p \mathcal{K}(-q)}{\p q^\mu}\Sigma_{\nu b}\non\\
&& -\f{1}{2}\ell_1^b \Sigma^T_{\nu b}\f{\p \mathcal{K}(-q)}{\p q^\mu}\Bigg\rbrace +\f{3}{2}\eta_{\mu\rho}\Bigg\lbrace q_\sigma \f{\p \mathcal{K}(-q)}{\p q^\nu}+\f{1}{2}(\ell_1+\ell_2)_\sigma \f{\p \mathcal{K}(-q)}{\p q^\nu}+\f{1}{2}q_\sigma (\ell_1+\ell_2)^\kappa \f{\p^2 \mathcal{K}(-q)}{\p q^\kappa \p q^\nu}\non\\
&&+q_\nu \f{\p \mathcal{K}(-q)}{\p q^\sigma} +\f{1}{2}(\ell_1+\ell_2)_\nu \f{\p \mathcal{K}(-q)}{\p q^\sigma}+\f{1}{2}q_\nu (\ell_1+\ell_2)^\kappa \f{\p^2 \mathcal{K}(-q)}{\p q^\kappa \p q^\sigma}\Bigg\rbrace +q_\mu q_\rho \f{\p^2 \mathcal{K}(-q)}{\p q^\nu \p q^\sigma}\non\\
&&+\f{1}{2}\Big\lbrace q_\mu (\ell_1 +\ell_2)_\rho +q_\rho (\ell_1+\ell_2)_\mu \Big\rbrace \f{\p^2 \mathcal{K}(-q)}{\p q^\nu \p q^\sigma}+\f{1}{2}q_\mu q_\rho (\ell_1 +\ell_2)^\kappa\f{\p^3 \mathcal{K}(-q)}{\p q^\nu \p q^\sigma \p q^\kappa}\non\\
&&+\f{1}{2}\eta_{\mu\rho}\Bigg\lbrace \ell_2^b \f{\p \mathcal{K}(-q)}{\p q^\nu}\Sigma_{\sigma b}-\ell_2^b \Sigma^T_{\sigma b}\f{\p \mathcal{K}(-q)}{\p q^\nu} +\ell_1^b \f{\p \mathcal{K}(-q)}{\p q^\sigma}\Sigma_{\nu b} -\ell_1^b \Sigma^T_{\nu b}\f{\p \mathcal{K}(-q)}{\p q^\sigma}\Bigg\rbrace\non\\
&&+\f{1}{4}q_\mu \ell_2^b \Bigg\lbrace \f{\p^2 \mathcal{K}(-q)}{\p q^\nu \p q^\rho}\Sigma_{\sigma b} -\Sigma_{\sigma b}^T \f{\p^2 \mathcal{K}(-q)}{\p q^\nu \p q^\rho}\Bigg\rbrace +\f{1}{4}q_\rho \ell_1^b \Bigg\lbrace \f{\p^2 \mathcal{K}(-q)}{\p q^\sigma \p q^\mu}\Sigma_{\nu b}-\Sigma_{\nu b}^T \f{\p^2 \mathcal{K}(-q)}{\p q^\sigma \p q^\mu}\Bigg\rbrace \non\\
&&-\f{1}{4}\eta_{\nu\sigma}(\ell_2-\ell_1)^\kappa \Bigg\lbrace \f{\p \mathcal{K}(-q)}{\p q^\kappa}\Sigma_{\rho\mu}-\Sigma_{\rho\mu}^{T}\f{\p \mathcal{K}(-q)}{\p q^\kappa}\Bigg\rbrace -\f{1}{2}\eta_{\nu\sigma}\ell_2^a \Bigg\lbrace \f{\p \mathcal{K}(-q)}{\p q^\rho}\Sigma_{a\mu} -\Sigma_{a\mu}^T \f{\p \mathcal{K}(-q)}{\p q^\rho}\Bigg\rbrace \non\\
&&-\f{1}{2}\eta_{\nu\sigma}\ell_1^a \Bigg\lbrace \f{\p \mathcal{K}(-q)}{\p q^\mu}\Sigma_{a\rho} -\Sigma_{a\rho}^T \f{\p \mathcal{K}(-q)}{\p q^\mu}\Bigg\rbrace -\f{1}{2}\ell_{2\nu}\Bigg\lbrace \f{\p \mathcal{K}(-q)}{\p q^\sigma}\Sigma_{\mu\rho}-\Sigma_{\mu\rho}^T \f{\p \mathcal{K}(-q)}{\p q^\sigma}\Bigg\rbrace \non\\
&&-\f{1}{2}\ell_{1\sigma}\Bigg\lbrace \f{\p \mathcal{K}(-q)}{\p q^\nu}\Sigma_{\rho\mu}-\Sigma_{\rho\mu}^T \f{\p \mathcal{K}(-q)}{\p q^\nu}\Bigg\rbrace +\mathcal{O}(\ell_1^2, \ell_1\ell_2,\ell_2^2)\Bigg]\ ,\label{Gamma4_graviton}
\ee
\endgroup
where we suppressed the massive particle spin/polarization indices. We could have symmetrize the above expression under $\mu\leftrightarrow\nu$ and $\rho\leftrightarrow\sigma$ exchanges. However, it is unnecessary because any contraction involved with this vertex in the computation of loop diagrams will already exhibit symmetry under these exchanges.

Now we want to evaluate the amputated Green's function involving $N$ number of massive spinning particles and one graviton with momentum $\ell$ and Lorentz indices $\mu\nu$, where the graviton is not connected to any external particle leg. Using covariantization prescription it has been evaluated in appendix-\ref{appendix-Gamma-tilde}, where additional complication has been taken care off due to the presence of different momentum conserving delta functions between $\widetilde{\Gamma}^{(N+1)}$ and $\Gamma^{(N)}$. The final result reads from \eqref{tilde-gamma-final-appendix} turns out to be
 \be
\widetilde{\Gamma}^{(N+1)\alpha_1\cdots \alpha_N}_{\mu\nu}(\ell)&=& \kappa \sum_{i=1}^{N}\Bigg[-\delta_{\beta_i}^{\alpha_i}\ p_{i(\mu}\f{\p }{\p p_i^{\nu)}} +(\Sigma_{i(\nu b})_{\beta_i}^{\ \alpha_i}\ell^b \ \f{\p }{\p p_{i}^{\mu)}}\ -\f{1}{2}\delta_{\beta_i}^{\alpha_i}\Bigg\lbrace \ell^\rho p_{i\mu}\f{\p^2}{\p p_i^\rho \p p_i^\nu}\non\\
&&+\ell^\rho p_{i\nu}\f{\p^2}{\p p_i^\rho \p p_i^\mu}-p_i.\ell \f{\p^2}{\p p_i^\mu \p p_i^\nu }\Bigg\rbrace \Bigg]\Gamma^{(N)\alpha_1\cdots \alpha_{i-1}\beta_i\alpha_{i+1}\cdots \alpha_N}+\mathcal{O}(\ell^2). \label{Gamma_tilde_graviton}
\ee

The three graviton vertex follows from \eqref{eq:S_EFT} with the ingoing graviton momenta $k$, $\ell $ and $-(\ell+k)$, and the Lorentz indices carried by them $(\mu\nu)$, $(\rho\sigma)$ and $(\tau\kappa)$ respectively takes the following form: 
\be
&&V^{(3)}_{\mu\nu, \rho\sigma,\tau\kappa}(k,\ell, -\ell-k)\non\\
&=& i\, \kappa\ \Big[  -(k.\ell+\ell^2+k^2)\left( \eta_{\mu\nu}\eta_{\rho\tau}\eta_{\sigma\kappa}+
\eta_{\rho\sigma}\eta_{\mu\tau}\eta_{\nu\kappa}+\eta_{\tau\kappa}\eta_{\mu\rho}\eta_{\nu\sigma}\right)\non\\
&&+\ 4\big(k.\ell+\ell^2+k^2\big)\eta_{\rho\nu}\eta_{\sigma\tau}\eta_{\kappa\mu}+\f{1}{2}\big(k.\ell+\ell^2 +k^2\big)\eta_{\mu\nu}\eta_{\rho\sigma}\eta_{\tau\kappa}\non\\
&&-\ 2\, \big(k_{\tau}\ell_{\kappa}\eta_{\mu\rho}\eta_{\nu\sigma}-\ell_{\mu}(k+\ell)_{\nu}\eta_{\rho\tau}\eta_{\sigma\kappa}-(k+\ell)_{\rho} k_{\sigma}\eta_{\mu\tau}\eta_{\nu\kappa}\big)\non\\
&&-2\big((k_\mu\ell_\tau +\ell_\mu\ell_\tau-k_{\tau}\ell_{\mu})\eta_{\nu\rho}\eta_{\sigma\kappa}+(2\ell_{\mu}k_{\rho}+\ell_\mu\ell_\rho +k_\mu k_\rho)\eta_{\tau\nu}\eta_{\kappa\sigma}\non\\
&&+(k_{\rho}
k_{\tau} +\ell_\rho k_\tau -\ell_\tau k_\rho)\eta_{\mu\sigma}\eta_{\nu\kappa}\big)\Big]\ +\mathcal{O}(k^4, k^3\ell, k^2\ell^2, k\ell^3,\ell^4)\ . \label{eq:V3_exp}
\ee

\subsection{KG-decomposition and IR-finite amplitudes}
In Grammer-Yennie prescription we decompose the internal graviton propagator \eqref{eq:full_gr_propagator} with momentum $\ell$ flowing from massive spinning particle $i$ to $j (\neq i)$ in the following way \cite{1808.03288} (see also \cite{Gervais:2017zky})
\be \label{eq:KG_decom_gr}
\Delta_F^{\mu\nu,\rho\sigma}(\ell)\ =\ \f{-i}{\ell^2 -i\epsilon}\ \f{1}{2}\left[K_{(ij)}^{\mu\nu,\rho\sigma}(\ell,p_i,p_j)+G_{(ij)}^{\mu\nu,\rho\sigma}(\ell,p_i,p_j)\right],
 \ee
 where
\be
K^{\mu\nu,\rho\sigma}_{(ij)}(\ell ,p_{i},p_{j}) &=& \mathcal{C}(\ell ,p_{i},p_{j})\ \big[(p_{i}-\ell )^{\mu}\ell^{\nu}
+(p_{i}-\ell )^{\nu}\ell^{\mu}\big]\ \big[(p_{j}+\ell )^{\rho}\ell^{\sigma}+(p_{j}+\ell )^{\sigma}\ell ^{\rho}\big] ,\non\\
G^{\mu\nu,\rho\sigma}_{(ij)}(\ell ,p_{i},p_{j}) &=&\big(\eta^{\mu\rho}\eta^{\nu\sigma}+\eta^{\mu\sigma}\eta^{\nu\rho}-\eta^{\mu\nu}\eta^{\rho\sigma}\big)-K^{\mu\nu,\rho\sigma}_{(ij)}(\ell ,p_{i},p_{j})\ ,\label{eq:KG_gr_def}
\ee
with
\be
\mathcal{C}(\ell ,p_{i},p_{j})\  &&=\ \f{(-1)}{\{p_{i}.(p_{i}-\ell )-i\epsilon\}
\ \{p_{j}.(p_{j}+\ell )-i\epsilon\} \{\ \ell .(\ell -2p_{i})-i\epsilon\} \ \{\ell .(\ell +2p_{j})-i\epsilon\}}
\non\\
&&\times \Big[2(p_{i}.p_{j})^{2}-p_{i}^{2}p_{j}^{2}-\ell ^{2}(p_{i}.p_{j})+2(p_{i}.p_{j})(p_{i}.\ell )-2(p_{i}.p_{j})(p_{j}.\ell )
\Big]\, .
\ee
Note that $p_i$ and $p_j$ above refer to the external momenta flowing into the legs $i$ and $j$, and not necessarily the momenta of the lines to which the graviton propagator attaches (which may have additional contribution from external soft graviton momentum or internal virtual graviton momentum). For virtual gravitons whose one or both ends are attached to a 3-graviton vertex instead of a massive particle, or to some internal massive particle line or vertex inside $\widetilde{\Gamma}^{(N+1)}$, we do not carry out any Grammer-Yennie decomposition as those won't contribute to IR divergences. In \eqref{eq:KG_decom_gr} the propagator part containing $K_{(ij)}$ will be denoted by K-graviton propagator and the propagator part containing $G_{(ij)}$ will be denoted by G-graviton propagator throughout this section.
\begin{center}
\begin{figure}
	\includegraphics[scale=0.5]{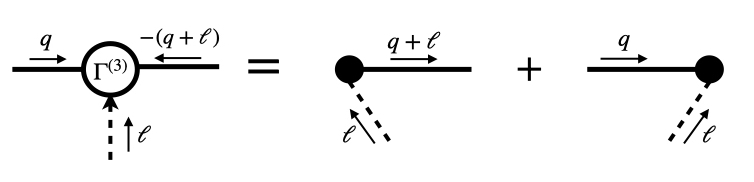}
	\caption{This figure is a Feynman diagrammatic representation of the expression in \eqref{KG_decomposition_graviton}. Solid lines represent the massive spinning particles, dashed lines represent the ingoing virtual graviton with momentum $\ell$ and the arrow in the graviton line represents that it is a K-graviton proportional to $\zeta_\mu \ell_\nu +\zeta_\nu \ell_\mu$. The solid blobs in the RHS represent a new kind of vertices and the Feynman rules for the left blob vertex is $-\kappa\lbrace 2 \zeta.(q+\ell) +\zeta^\nu \ell^b \Sigma_{\nu b}\rbrace$ and for the right blob vertex is $\kappa\lbrace 2\zeta.q -\zeta^\nu \ell^b \Sigma_{\nu b}^T \rbrace$.}\label{f:KG_graviton_figure}
\end{figure}
\end{center}
\paragraph{Ward identities involving K-graviton:} From the definition of K-graviton propagator in \eqref{eq:KG_gr_def}, it is clear that K-graviton is proportional to a pure gauge of structure $\zeta^\mu\ell^\nu +\zeta^\nu \ell^\mu$, with $\zeta=p_i-\ell$ when it flows from $i$-th leg. Let us study the Ward identity for an off-shell un-amputated three particle Green’s function involving two massive fields and one K-graviton. It is diagrammatically represented in Fig.\ref{f:KG_graviton_figure}.

The LHS of the Fig.\ref{f:KG_graviton_figure} takes the following form:
\be 
\f{1}{q^2 +m^2-i\epsilon}\f{1}{(q+\ell)^2+m^2-i\epsilon}(\zeta^\mu \ell^\nu +\zeta^\nu \ell^\mu)\Big[ \Xi(-q)\Gamma^{(3)}_{\mu\nu}(q,-q-\ell,\ell)\Xi(-q-\ell)\Big]\ .
\ee 
To compute the above expression we used the result of \eqref{Gamma3Xi_gravity} and then simplified using the identities in \eqref{K1},\eqref{SigmaXi} and derivatives of \eqref{SigmaXi}. After all the simplification we get
\begingroup
\allowdisplaybreaks
\be 
&&\f{i\kappa}{(q+\ell)^2+m^2-i\epsilon}\Bigg\lbrace 2i\zeta.\ell \Xi(-q)+2i\zeta.q\ell^\rho \f{\p \Xi(-q)}{\p q^\rho}+i\zeta.\ell \ell^\rho \ell^\sigma \f{\p^2 \Xi(-q)}{\p q^\rho \p q^\sigma}+2i\zeta.\ell \ell^\rho \f{\p \Xi(-q)}{\p q^\rho}\non\\
&& +\f{i}{2}\ell^b \ell^\rho \zeta^\nu \ell^\mu \Sigma_{\nu b}\f{\p^2 \Xi(-q)}{\p q^\mu \p q^\rho}+\f{i}{3}\zeta.q \ell^\rho \ell^\sigma\ell^\nu \f{\p^3 \Xi(-q)}{\p q^\nu \p q^\rho \p q^\sigma}+2i\zeta.q \Xi(-q)+i\zeta^\nu \ell^b\Sigma_{\nu b}\Xi(-q)\non\\
&&+i\zeta^\nu \ell^\mu \ell^b \Sigma_{\nu b}\f{\p \Xi(-q)}{\p q^\mu}+i\zeta.q \ell^\mu \ell^b \f{\p^2 \Xi(-q)}{\p q^\mu \p q^b}+\mathcal{O}(\ell^4)\Bigg\rbrace\non\\
&&+\f{i\kappa}{q^2+m^2-i\epsilon}\Big\lbrace -2i\zeta.q \Xi(-q)+i\zeta^\nu \ell^b \Xi(-q)\Sigma_{\nu b}^T \Big\rbrace \ .
\ee
\endgroup
Now undoing the $\ell$ expansion for $\Xi(-q-\ell)$ the above expression can be re-written as 
\be 
&&\f{1}{q^2 +m^2-i\epsilon}\f{1}{(q+\ell)^2+m^2-i\epsilon}(\zeta^\mu \ell^\nu +\zeta^\nu \ell^\mu)\Big[ \Xi(-q)\Gamma^{(3)}_{\mu\nu}(q,-q-\ell,\ell)\Xi(-q-\ell)\Big]\non\\
&=&\kappa\Big\lbrace -2 \zeta.(q+\ell) -\zeta^\nu \ell^b \Sigma_{\nu b}\Big\rbrace \f{\Xi(-q-\ell)}{(q+\ell)^2+m^2-i\epsilon}+\f{\Xi(-q)}{q^2+m^2-i\epsilon}\ \kappa\Big\lbrace 2\zeta.q -\zeta^\nu \ell^b \Sigma_{\nu b}^T \Big\rbrace .\non\\ \label{KG_decomposition_graviton}
\ee
The above identity is diagrammatically represented in Fig.\ref{f:KG_graviton_figure}. It is worth noting that the blob vertices in this representation depend on the momenta of the massive particle and the attached K-graviton, as well as on the spin angular momenta of the massive particle, as indicated in Fig.\ref{f:KG_graviton_figure}.

In comparison to the QED Ward identity in \eqref{Ward_identity_photon}, the momenta dependence of the blob vertices weakens the power of KG-decomposition. For instance, if we wish to study the Ward identity for the four-point un-amputated Green's function associated with two massive particles, one graviton, and one K-graviton, analogous to Fig.\ref{f:KG_decomposition_2_photons}, it will result in new Feynman rules for the right blob vertex drawn in the second diagram after the equality of Fig.\ref{f:KG_decomposition_2_photons}. Specifically, the right blob vertex Feynman rule reads $\kappa\lbrace 2\zeta.(q+k) -\zeta^\nu \ell^b \Sigma_{\nu b}^T \rbrace$, which contains an additional term $2\kappa\zeta\cdot k$  relative to the three-point un-amputated Green's function right blob Feynman rule $\kappa\lbrace 2\zeta.q -\zeta^\nu \ell^b \Sigma_{\nu b}^T \rbrace$ in Fig.\ref{f:KG_graviton_figure}. In turn this implies that $\mathcal{A}^{(N+1)}$ contains an extra exponentiation term compared to the IR-exponentiation factor of $\mathcal{A}^{(N)}$. However this extra contribution arising from the momenta dependent right blob vertex rule contributes to IR-finite part in the loop integral, allowing us to follow the same strategy used in deriving the soft photon theorem. In \cite{1808.03288}, it is also observed that the right-hand side of Fig.\ref{f:KG_decomposition_photon_subdiagram} in presence of one external graviton, due to one K-graviton insertion, does not vanish, but instead leaves out some finite residual contribution. Fortunately, this residual terms also contributes to the IR-finite part in the loop integral when we evaluate $\mathcal{A}^{(N+1)}$, which we can systematically account for it as well. However, as mentioned in section-\ref{S:strategy}, we still need to regulate some additional IR divergence in $\mathcal{A}^{(N+1)}$ that arises from the Feynman diagram involving three graviton interaction vertices of which one graviton being the external graviton at the one-loop level. 
\paragraph{IR-finite amplitudes:}
The IR-finite amplitudes associated with the scattering of $N$ number of spinning massive particles, and associated with $N$ number of spinning massive particles plus one outgoing graviton are defined by
\be
&&\mathcal{A}^{(N)}\equiv \exp\lbrace K_{gr}\rbrace\ \mathcal{A}^{(N)}_{\text{IR-finite}} \label{eq:A_N_gr}\ ,\\
&&\mathcal{A}^{(N+1)}\overset{\text{reg}}{\equiv} \exp\lbrace K_{gr}\rbrace\ \mathcal{A}^{(N+1)}_{\text{IR-finite}}\ ,\label{eq:A_N+1_gr}
\ee 
where
\be
K_{gr}&= & -i\kappa^2\sum_{i=1}^{N}\  \sum_{\substack{j=1\\j\neq i}}^{N} \int \f{d^{4}\ell}{(2\pi)^{4}} \ \f{1}{\ell^{2}-i\epsilon}\ 
\f{1}{\ell.(\ell-2p_i)-i\epsilon}\ \f{1}{\ell.(\ell+2p_j)-i\epsilon}\nn\\
&&\times \Big[2(p_{i}.p_{j})^{2}-p_{i}^{2}p_{j}^{2}-\ell ^{2}(p_{i}.p_{j})+2(p_{i}.p_{j})(p_{i}.\ell )-2(p_{i}.p_{j})(p_{j}.\ell ) \Big]
\label{eq:K_gr}
\ee
is the Eikonal IR-divergent exponentiated factor.
In equations \eqref{eq:A_N_gr} and \eqref{eq:A_N+1_gr}, $\mathcal{A}^{(N)}_{\text{IR-finite}}$ and $\mathcal{A}^{(N+1)}_{\text{IR-finite}}$ represent the infrared finite components of the $N$-particle and $N$-particle-1-graviton amplitudes, respectively.\footnote{To determine $\mathcal{A}^{(N+1)}_{\text{IR-finite}}$, we need to use an explicit IR cut-off for the diagrams involving graviton self interaction vertices. The ``reg'' over the $\equiv$ sign in \eqref{eq:A_N+1_gr} corresponds to this particular IR regularization scheme. Effectively this IR-regularization scheme removes a factor of $\exp\lbrace K_{phase}\rbrace$ from $\mathcal{A}^{(N+1)}$ with an explicit expression for $K_{phase}$ being
\be
K_{phase}=  i\kappa^2\ \sum_{j=1}^{N}(p_{j}.k)^{2}\int_0^{R^{-1}}\f{d^{4}\ell}{(2\pi)^{4}}\ \f{1}{\ell^{2}-i\epsilon}\f{1}{k.\ell+i\epsilon}\f{1}{p_{j}.\ell -i\epsilon}\ .\label{eq:K_phase_div}
\ee
} These components are obtained by removing the exponentiated IR-divergent parts from the original divergent amplitudes defined through the relations \eqref{eq:A_N_gr} and \eqref{eq:A_N+1_gr}. Both $\mathcal{A}^{(N)}_{\text{IR-finite}}$ and $\mathcal{A}^{(N+1)}_{\text{IR-finite}}$ comprise contributions from the corresponding tree-level amplitudes and loop amplitudes up to all orders in perturbation theory. However, there is a condition: if both ends of a virtual graviton propagator are attached to different external massive spinning particle lines (which may already contain additional real or virtual graviton lines), then this graviton propagator should be replaced by a G-graviton propagator when we evaluate them for the IR-finite parts. Additionally the same set of diagrams need to be evaluated with K-graviton propagator as well and then have to subtract by a factor of $K_{gr}$ times the IR finite amplitude at one less loop level. On top of it to evaluate the full IR-finite part of $\mathcal{A}^{(N+1)}$, we need to IR regulate diagrams containing at least one graviton self interacting vertex involving the external graviton. Rigorous definitions of one-loop IR-finite amplitudes are provided below. 

Now analogous to \eqref{eq:AN_tree}, \eqref{eq:AN_finite} and \eqref{eq:AN+1_finite}, here we also split the IR-finite parts of the tree and one-loop amplitudes in the following way
\be
\mathcal{A}^{(N)}_{\text{IR-finite,0}}&\equiv & \Gamma^{(N)}\ ,\\
 \mathcal{A}^{(N+1)}_{\text{IR-finite,0}}&\equiv & \Gamma^{(N+1)}\ ,\\ 
\mathcal{A}^{(N)}_{\text{IR-finite,1}}&\equiv &\Big[\mathcal{A}^{(N)}_{G,1}+\mathcal{A}^{(N)}_{K-\text{finite},1}+\mathcal{A}^{(N)}_{\text{self,1}}+\mathcal{A}^{(N)}_{\text{non-div,1}}\Big]\ ,\label{eq:AN_finite_gr}\\
\mathcal{A}^{(N+1)}_{\text{IR-finite,1}}&\equiv &\Big[\mathcal{A}^{(N+1)}_{G,1}+\mathcal{A}^{(N+1)}_{K-\text{finite},1}+\mathcal{A}^{(N+1)}_\text{3-graviton-reg,1}+\mathcal{A}^{(N+1)}_{\text{self,1}}+\mathcal{A}^{(N+1)}_{\text{non-div,1}}\Big].\ \label{eq:AN+1_finite_gr}
\ee
Above different components of the IR-finite one-loop amplitudes are defined as
\begin{enumerate}
	\item $\mathcal{A}^{(N)}_{G,1}$ corresponds to the diagram in Fig.\ref{f:A^N_G} with the dashed line being  a G-graviton propagator.
	\item  $\mathcal{A}^{(N)}_{K-\text{finite},1}$ corresponds to the contribution from the diagram in Fig.\ref{f:A^N_G} , evaluated with K-graviton propagator representing the dashed line there and then subtracted the contribution $K_{gr}\times \Gamma^{(N)}$ from it at the integrand level.
	\item $\mathcal{A}^{(N)}_{\text{self,1}}$ represents the set of Feynman diagrams in Fig.\ref{f:A^N_self} evaluated with full graviton propagator representing the dashed lines.
	\item $\mathcal{A}^{(N)}_{\text{non-div,1}}$ represents the set of diagrams in Fig.\ref{f:A^N_non-div}, evaluated with full graviton propagator representing the dashed lines.
	\item $\mathcal{A}^{(N+1)}_{G,1}$ corresponds to the set of diagrams in Fig.\ref{f:A^N+1_G} where we need to evaluate the diagrams with G-graviton propagator representing the dashed virtual lines. The dashed external lines represent the on-shell graviton with momentum $k$.
	\item $\mathcal{A}^{(N+1)}_{K-\text{finite},1}$ corresponds to the contribution from the diagrams in Fig.\ref{f:A^N+1_G}, evaluated with K-graviton propagator representing the dashed line and then subtracted the contribution $K_{gr}\times \Gamma^{(N+1)}$ at the integrand level. The dashed external lines represent the on-shell graviton with momentum $k$.
	\item $\mathcal{A}^{(N+1)}_\text{3-graviton-reg,1}$ represents the sum of the contributions of Feynman diagrams in Fig.\ref{f:3-graviton_diagrams}, when we evaluate them using full graviton propagator and regulate the IR divergence considering detector resolution as the explicit IR cut-off. 
	\item $\mathcal{A}^{(N+1)}_{\text{self,1}}$ represents the set of Feynman diagrams in Fig.\ref{f:self_diagram_gr_N+1}, evaluated with full graviton propagator representing the dashed lines. The dashed external lines represent the on-shell graviton with momentum $k$.
	\item $\mathcal{A}^{(N+1)}_{\text{non-div,1}}$ represents the set of diagrams in Fig.\ref{f:A^N+1_non-div}, evaluated with full graviton propagator representing the dashed lines. The dashed external lines represent the on-shell graviton with momentum $k$.
\end{enumerate}

\subsection{Derivation of soft graviton theorem}
The goal here will be to derive the order $\omega^{-1}$ and $\ln\omega$ soft factors from the ratio of $\mathcal{A}^{(N+1)}_{\text{IR-finite}}$ and $\mathcal{A}^{(N)}_{\text{IR-finite}}$ when the external graviton energy is small i.e. $\omega<<|p_i^\mu |$.

\subsubsection{IR-finite one-loop $N$-particle amplitude}
Let us evaluate the sum of contributions from Fig.\ref{f:A^N_G} with insertion of G-graviton propagator and the finite part with the insertion of K-graviton propagator together. We can do this by examining the diagram in Fig.\ref{f:A^N_G}, where the dashed line represents a full graviton propagator. Then we need to subtract $K_{gr}\times \Gamma^{(N)}$ from the evaluated result. The mathematical expression reads
\begingroup
\allowdisplaybreaks
\be
\mathcal{A}^{(N)}_{G,1}+
\mathcal{A}^{(N)}_{K-\text{finite},1}
&=& \sum_{\substack{i,j=1\\i> j}}^{N} \int \f{d^4\ell}{(2\pi)^4}\  \f{1}{(p_i-\ell)^2+m_i^2-i\epsilon} \f{1}{(p_j+\ell)^2+m_j^2 -i\epsilon}\nn\\
&&\times\Big[ \epsilon_i^{T}(-p_i)\Gamma^{(3)}_{\mu\nu} (p_i,-p_i+\ell,-\ell)\Xi_i(-p_i+\ell)\Big]\Delta_F^{\mu\nu,\rho\sigma}(\ell)  \nn\\
&& \Big[ \epsilon_j^{T}(-p_j)\Gamma^{(3)}_{\rho\sigma} (p_j,-p_j-\ell,\ell)\Xi_j(-p_j-\ell)\Big] \Gamma^{(N)}_{(ij)}(p_i-\ell,p_j+\ell) \nn\\
&& -\ K_{gr}\times \Gamma^{(N)}\ .
\ee
\endgroup
We evaluate the above expression by using the identity derived in \eqref{Gamma3Xi_gravity}. The result in small $\ell$ expansion in the integrand turns out to be
\begingroup
\allowdisplaybreaks
\be
&&\mathcal{A}^{(N)}_{G,1}+\mathcal{A}^{(N)}_{K-\text{finite},1}\nn\\
&=& -i\kappa^2 \sum_{i=1}^{N}\sum_{\substack{j=1\\ j\neq i}}^{N}\epsilon_i^{T}\epsilon_j^{T}\int \f{d^4\ell}{(2\pi)^4}\f{1}{\ell^2-i\epsilon}\ \f{1}{\ell\cdot(\ell-2p_i)-i\epsilon}\ \f{1}{\ell\cdot (\ell+2p_j)-i\epsilon}\nn\\
&&\Bigg[\lbrace 2(p_i.p_j)^2-p_i^2 p_j^2\rbrace\left(-\ell^{\rho}\f{\p }{\p p_i^\rho}\Gamma^{(N)}_{(ij)}(p_i,p_j) +\ell^\rho\f{\p }{\p p_j^\rho}\Gamma^{(N)}_{(ij)}(p_i,p_j) \right)\nn\\
&&+\left\lbrace2(p_i.p_j)\left(p_j^\rho\Sigma_{i\rho\sigma}^T \ell^\sigma-p_i^\rho\Sigma_{j\rho\sigma}^T \ell^\sigma \right)+p_i^2 p_j^\rho\Sigma_{j\rho\sigma}^T \ell^\sigma-p_j^2 p_i^\rho\Sigma_{i\rho\sigma}^T \ell^\sigma\right\rbrace\Gamma^{(N)}_{(ij)}(p_i,p_j)\nn\\
&&+\mathcal{O}(\ell\ell)\Bigg]\label{AG+AK_finite}
\ee
\endgroup
Note that in the limit $\ell^\mu\rightarrow 0$ the integrand of the above expression behaves like $\int \f{d^4\ell}{|\ell|^3}$ at leading order, hence the contribution is IR-finite as promised. 

Diagrams in Fig.\ref{f:A^N_self} renormalizes the massive spinning particle propagators in presence of gravitational interaction, when the dashed lines represent graviton propagators. The contribution from the loop diagrams there are IR-finite. Hence following the analogous wave function renormalization condition \eqref{renormalization_condition}, the sum of the contribution vanishes i.e.
\be
\mathcal{A}^{(N)}_{\text{self,1}}=0\ .\label{eq:A_self^N_gr}
\ee
Now let us analyze the diagrams in Fig.\ref{f:A^N_non-div} with dashed lines being full graviton propagators are connected to some internal massive virtual lines or massive EFT vertices inside $\Gamma^{(N)}$. In the limit $\ell^\mu\rightarrow 0$ the integrand of the first diagram behaves like $\int \f{d^4\ell}{|\ell|^3}$ at leading order, and the second diagram behaves like $\int \f{d^4\ell}{|\ell|^2}$ at leading order, hence IR-finite. We do not need to evaluate them explicitly. Let the total contribution after removing the UV divergences by adding counter term diagrams reads
\be
\mathcal{A}^{(N)}_{\text{non-div,1}}\ .\label{eq:AN_non-div_gr}
\ee

Hence the total IR-finite contribution to $N$-particle amplitude follows from the definition \eqref{eq:AN_finite_gr} becomes
\begingroup
\allowdisplaybreaks
\be
\mathcal{A}^{(N)}_{\text{IR-finite,1}}&=& -i\kappa^2 \sum_{i=1}^{N}\sum_{\substack{j=1\\ j\neq i}}^{N}\epsilon_i^{T}\epsilon_j^{T}\int \f{d^4\ell}{(2\pi)^4}\f{1}{\ell^2-i\epsilon}\ \f{1}{\ell\cdot(\ell-2p_i)-i\epsilon}\ \f{1}{\ell\cdot (\ell+2p_j)-i\epsilon}\nn\\
&&\Bigg[\lbrace 2(p_i.p_j)^2-p_i^2 p_j^2\rbrace\left(-\ell^{\rho}\f{\p }{\p p_i^\rho}\Gamma^{(N)}_{(ij)}(p_i,p_j) +\ell^\rho\f{\p }{\p p_j^\rho}\Gamma^{(N)}_{(ij)}(p_i,p_j) \right)\nn\\
&&+\left\lbrace2(p_i.p_j)\left(p_j^\rho\Sigma_{i\rho\sigma}^T \ell^\sigma-p_i^\rho\Sigma_{j\rho\sigma}^T \ell^\sigma \right)+p_i^2 p_j^\rho\Sigma_{j\rho\sigma}^T \ell^\sigma-p_j^2 p_i^\rho\Sigma_{i\rho\sigma}^T \ell^\sigma\right\rbrace\Gamma^{(N)}_{(ij)}(p_i,p_j)\nn\\
&&+\mathcal{O}(\ell\ell)\Bigg] \ +\ \mathcal{A}^{(N)}_{\text{non-div,1}}.\label{eq:AN_ir-finite_gr}
\ee
\endgroup

\subsubsection{IR-finite one-loop $(N+1)$-particle amplitude in the soft limit}
Here we analyze all the Feynman diagrams contributing to \eqref{eq:AN+1_finite} in the soft limit i.e. $\omega\rightarrow 0$. We start by analyzing the diagrams in Fig.\ref{f:A^N+1_G} with the dashed lines being full graviton propagators, which evaluates the contribution of $\mathcal{A}^{(N+1)}_{G,1}+
\mathcal{A}^{(N+1)}_{K-\text{finite},1}+K_{gr}\Gamma^{(N+1)}$. By evaluating this sum with full graviton propagator, we avoid all the computational complicacies in the KG-decomposition in presence of external graviton as pointed out in  the paragraph below \eqref{KG_decomposition_graviton}. Then finally we subtract the contribution of $K_{gr}\Gamma^{(N+1)}$ from the sum to extract the $\omega^{-1}$ and $\ln\omega$ soft factors from $\mathcal{A}^{(N+1)}_{G,1}+
\mathcal{A}^{(N+1)}_{K-\text{finite},1}$ in the limit $\omega<<|p_i^\mu|$.

The first diagram in Fig.\ref{f:A^N+1_G} with full graviton propagator representing the internal dashed line takes the following form 
\begingroup
\allowdisplaybreaks
\be
B_{I}&\equiv &\ \sum_{i=1}^{N}\sum_{\substack{j=1\\j\neq i}}^{N}\f{\varepsilon^{\lambda\tau}(k)}{2p_i.k}\int \f{d^4\ell}{(2\pi)^4}\ \f{1}{(p_i+k-\ell)^2+m_i^2-i\epsilon}\f{1}{(p_j+\ell)^2+m_j^2 -i\epsilon} \non\\
&&\Big[ \epsilon_i^{T}(-p_i)\Gamma^{(3)}_{\lambda\tau} (p_i,-p_i-k,k)\Xi_i(-p_i-k)\Gamma^{(3)}_{\rho\sigma}(p_i+k,-p_i-k+\ell ,-\ell)\Xi_i(-p_i-k+\ell)\Big]\non\\
&&\times \Delta_F^{\rho\sigma,\mu\nu}(\ell) \Big[\epsilon_j^{T}(-p_j)\Gamma^{(3)}_{\mu\nu} (p_j,-p_j-\ell,\ell)\Xi_j(-p_j-\ell)\Big]\ \Gamma^{(N)}_{(ij)}(p_i+k-\ell,p_j+\ell)\ .
\ee
\endgroup
After using the identities in \eqref{Gamma3Xi_gravity} and \eqref{eq:Z1_gr} and simplifying the above expression reduces to
\begingroup
\allowdisplaybreaks
\be
B_{I}&= &i\kappa^3 \sum_{i=1}^{N}\sum_{\substack{j=1\\j\neq i}}^{N}\epsilon_i^{T}\epsilon_j^{T}\f{1}{p_i.k}\int \f{d^4\ell}{(2\pi)^4}\ \f{1}{\ell^2-i\epsilon}\f{1}{\ell.(\ell-2p_i-2k)+2p_i.k -i\epsilon}\f{1}{\ell.(\ell+2p_j)-i\epsilon}\nn\\
&&\Bigg[ \Bigg(-2(p_i.\varepsilon.p_i)\left\lbrace 2(p_i.p_j)^2-p_i^2p_j^2 +4p_i.p_j p_j.k-2p_i.p_j p_j.\ell +2p_i.p_j p_i.\ell\right\rbrace \nn\\
&& +4(p_i.\varepsilon.p_i) p_i.p_j p_i^\mu \Sigma_{j\mu\nu}^{T}\ell^\nu -2(p_i.\varepsilon.p_i) p_i^2 p_j^\mu \Sigma_{j\mu\nu}^T \ell^\nu+2\lbrace 2(p_i.p_j)^2-p_i^2p_j^2\rbrace p_{i\mu}\varepsilon^{\mu\rho}\Sigma_{i\rho \nu}^Tk^\nu\nn\\
&& -4(p_i.\varepsilon.p_i) p_i.p_j p_j^\mu \Sigma_{i\mu\nu}^{T}\ell^\nu +2(p_i.\varepsilon.p_i) p_j^2 p_i^\mu \Sigma_{j\mu\nu}^T \ell^\nu -2i p_i.k p_i.p_j p_{i\mu}\varepsilon^{\mu\nu}\f{\p \mathcal{K}_i(-p_i)}{\p p_i^\nu}\f{\p\Xi_i(-p_i)}{\p p_i^\sigma}p_j^\sigma\nn\\
&&+ ip_i.k p_j^2 p_{i\mu}\varepsilon^{\mu\nu}\f{\p \mathcal{K}_i(-p_i)}{\p p_i^\nu}\f{\p\Xi_i(-p_i)}{\p p_i^\sigma}p_i^\sigma\Bigg)\Gamma^{(N)}_{(ij)}(p_i,p_j)\ -2 (p_i.\varepsilon.p_i)\left\lbrace 2(p_i.p_j)^2-p_i^2p_j^2 \right\rbrace\nn\\
&&\times \left((k-\ell)^\mu \f{\p}{\p p_i^\mu}\Gamma^{(N)}_{(ij)}(p_i,p_j) +\ell^\mu \f{\p}{\p p_j^\mu}\Gamma^{(N)}_{(ij)}(p_i,p_j)\right)\ +\ \mathcal{O}(\ell\ell ,\ell k, kk)\Bigg]\ . \label{eq:B_I_gr}
\ee
\endgroup
The second diagram in Fig.\ref{f:A^N+1_G} with full graviton propagator representing the dashed internal line takes the following form 
\begingroup
\allowdisplaybreaks
\be
B_{II}&\equiv &\ \sum_{i=1}^{N}\sum_{\substack{j=1\\j\neq i}}^{N}\int \f{d^4\ell}{(2\pi)^4}\ \f{1}{(p_i+k-\ell)^2+m_i^2-i\epsilon}\f{1}{(p_j+\ell)^2+m_j^2 -i\epsilon} \Delta_F^{\rho\sigma,\lambda\tau}(\ell) \non\\
&&\times \Big[ \epsilon_i^{T}(-p_i)\varepsilon^{\mu\nu}(k)\Gamma^{(4)}_{\mu\nu, \rho\sigma} (p_i,-p_i-k+\ell,k,-\ell)\Xi_i(-p_i-k+\ell)\Big]\non\\
&&\times \  \Big[\epsilon_j^{T}(-p_j)\Gamma^{(3)}_{\lambda\tau} (p_j,-p_j-\ell,\ell)\Xi_j(-p_j-\ell)\Big]\ \Gamma^{(N)}_{(ij)}(p_i+k-\ell,p_j+\ell)\ .
\ee
\endgroup
After using the identities in \eqref{Gamma3Xi_gravity} and \eqref{eq:Z2_gr} and simplifying the above expression reduces to
\begingroup
\allowdisplaybreaks
\be
B_{II}& = &i\kappa^3 \sum_{i=1}^{N}\sum_{\substack{j=1\\j\neq i}}^{N}\epsilon_i^{T}\epsilon_j^{T}\int \f{d^4\ell}{(2\pi)^4}\ \f{1}{\ell^2-i\epsilon}\f{1}{\ell.(\ell-2p_i-2k)+2p_i.k -i\epsilon}\f{1}{\ell.(\ell+2p_j)-i\epsilon}\nn\\
&&\Bigg[ \Bigg\lbrace-4(p_i.\varepsilon.p_i)p_j^2+16 (p_i.\varepsilon.p_j)(p_i.p_j)+2ip_i.p_j p_{i\mu}\varepsilon^{\mu\nu}\f{\p\mathcal{K}_i(-p_i)}{\p p_i^\nu}p_j^\sigma \f{\p\Xi_i(-p_i)}{\p p_i^\sigma}\nn\\
&&+2ip_i.p_j p_j^\sigma\f{\p\mathcal{K}_i(-p_i)}{\p p_i^\sigma} p_{i\mu}\varepsilon^{\mu\nu}\f{\p\Xi_i(-p_i)}{\p p_i^\nu} -ip_j^2 p_{i\mu}\varepsilon^{\mu\nu}\f{\p\mathcal{K}_i(-p_i)}{\p p_i^\nu}p_i^\sigma \f{\p\Xi_i(-p_i)}{\p p_i^\sigma}\nn\\
&&-ip_j^2 p_i^\sigma\f{\p\mathcal{K}_i(-p_i)}{\p p_i^\sigma}p_{i\mu}\varepsilon^{\mu\nu}\f{\p\Xi_i(-p_i)}{\p p_i^\nu}\Bigg\rbrace \Gamma^{(N)}_{(ij)}(p_i,p_j) +\mathcal{O}(\ell, k)\Bigg]\ .\label{eq:B_II_gr}
\ee
\endgroup
The third diagram in Fig.\ref{f:A^N+1_G} with full graviton propagator representing the dashed internal line takes the following form 
\begingroup
\allowdisplaybreaks
\be
B_{III}&\equiv &\ \sum_{i=1}^{N}\sum_{\substack{j=1\\j\neq i}}^{N}\int \f{d^4\ell}{(2\pi)^4}\  \f{1}{(p_i-\ell)^2+m_i^2-i\epsilon}\f{1}{(p_i+k-\ell)^2+m_i^2-i\epsilon}\f{1}{(p_j+\ell)^2+m_j^2 -i\epsilon}\non\\
&&\Big[ \epsilon_i^{T}(-p_i)\Gamma^{(3)}_{\mu\nu} (p_i,-p_i+\ell,-\ell)\Xi_i(-p_i+\ell)\varepsilon^{\rho\sigma}(k)\Gamma^{(3)}_{\rho\sigma}(p_i-\ell,-p_i-k+\ell ,k)\Xi_i(-p_i-k+\ell)\Big]\non\\
&&\times \Delta_F^{\mu\nu,\lambda\tau}(\ell) \ \Big[\epsilon_j^{T}(-p_j)\Gamma^{(3)}_{\lambda\tau} (p_j,-p_j-\ell,\ell)\Xi_j(-p_j-\ell)\Big]\ \Gamma^{(N)}_{(ij)}(p_i+k-\ell,p_j+\ell)\ .
\ee
\endgroup
After using the identities in \eqref{Gamma3Xi_gravity} and \eqref{eq:Z1_gr} and simplifying the above expression reduces to
\begingroup
\allowdisplaybreaks
\be
B_{III}&= &i\kappa^3 \sum_{i=1}^{N}\sum_{\substack{j=1\\j\neq i}}^{N}\epsilon_i^{T}\epsilon_j^{T}\int \f{d^4\ell}{(2\pi)^4} \f{1}{\ell^2-i\epsilon}\f{1}{\ell.(\ell-2p_i) -i\epsilon}\f{1}{\ell.(\ell-2p_i-2k)+2p_i.k -i\epsilon}\nn\\
&&\times \f{1}{\ell.(\ell+2p_j)-i\epsilon}\Bigg[ \Bigg(-4(p_i.\varepsilon.p_i)\lbrace 2(p_i.p_j)^2 -p_i^2 p_j^2 \rbrace -8(p_i.\varepsilon.p_i) p_i.p_j p_i.\ell\nn\\
&& +8 (p_i.\varepsilon.\ell) \lbrace 2(p_i.p_j)^2 -p_i^2 p_j^2 \rbrace +8 (p_i.\varepsilon.p_i) p_i.p_j p_j.\ell +4 \lbrace 2(p_i.p_j)^2 -p_i^2 p_j^2 \rbrace p_{i\rho}\varepsilon^{\rho\sigma}\Sigma_{i\sigma \mu}^Tk^\mu\nn\\
&&-8p_i.p_j (p_i.\varepsilon.p_i) p_j^{\mu}\Sigma_{i\mu\nu}^T \ell^\nu +4p_j^2 (p_i.\varepsilon.p_i) p_i^{\mu}\Sigma_{i\mu\nu}^T \ell^\nu +8(p_i.\varepsilon.p_i) p_i.p_j p_i^{\mu}\Sigma_{j\mu\nu}^T \ell^\nu\nn\\
&& -4p_i^2 (p_i.\varepsilon.p_i) p_j^{\mu}\Sigma_{j\mu\nu}^T \ell^\nu +4i p_i.\ell p_i.p_j p_j^\mu \f{\p\mathcal{K}_i(-p_i)}{\p p_i^\mu}p_{i\rho}\varepsilon^{\rho\sigma }\f{\p \Xi_i(-p_i)}{\p p_i^\sigma}\nn\\
&&-2i p_i.\ell p_j^2 p_i^\mu \f{\p\mathcal{K}_i(-p_i)}{\p p_i^\mu}p_{i\rho}\varepsilon^{\rho\sigma }\f{\p \Xi_i(-p_i)}{\p p_i^\sigma}\Bigg)\Gamma^{(N)}_{(ij)}(p_i,p_j)\ -4 (p_i.\varepsilon.p_i)\left\lbrace 2(p_i.p_j)^2-p_i^2p_j^2 \right\rbrace\nn\\
&&\times \left((k-\ell)^\mu \f{\p}{\p p_i^\mu}\Gamma^{(N)}_{(ij)}(p_i,p_j) +\ell^\mu \f{\p}{\p p_j^\mu}\Gamma^{(N)}_{(ij)}(p_i,p_j)\right)\ +\ \mathcal{O}(\ell\ell ,\ell k, kk)\Bigg]\ .\label{eq:B_III_gr}
\ee
\endgroup
The fourth diagram in Fig.\ref{f:A^N+1_G} with full graviton propagator representing the dashed internal line takes the following form 
\begingroup
\allowdisplaybreaks
\be
B_{IV}&\equiv & \sum_{k=1}^N \f{1}{2p_k\cdot k}\epsilon_k^{T}(-p_k)\varepsilon^{\lambda\tau}(k)\Gamma^{(3)}_{\lambda\tau} (p_k,-p_k-k,k)\Xi_k(-p_k-k)\nn\\
&&\times \f{1}{2} \sum_{\substack{i=1\\ i\neq k}}^N\sum_{\substack{j=1\\j\neq i,k}}^{N} \int \f{d^4\ell}{(2\pi)^4}\  \f{1}{(p_i-\ell)^2+m_i^2-i\epsilon} \f{1}{(p_j+\ell)^2+m_j^2 -i\epsilon}\nn\\
&&\times\Big[ \epsilon_i^{T}(-p_i)\Gamma^{(3)}_{\mu\nu} (p_i,-p_i+\ell,-\ell)\Xi_i(-p_i+\ell)\Big]\Delta_F^{\mu\nu,\rho\sigma}(\ell)  \nn\\
&&\times \Big[ \epsilon_j^{T}(-p_j)\Gamma^{(3)}_{\rho\sigma} (p_j,-p_j-\ell,\ell)\Xi_j(-p_j-\ell)\Big] \Gamma^{(N)}_{(ijk)}(p_i-\ell,p_j+\ell ,p_k+k)\ . \label{eq:B_IV_gr}
\ee
\endgroup
The fifth diagram in Fig.\ref{f:A^N+1_G} with full graviton propagator representing the dashed internal line takes the following form 
\begingroup
\allowdisplaybreaks
\be
B_{V}&\equiv &  \f{1}{2}\sum_{\substack{i=1}}^N\sum_{\substack{j=1\\j\neq i}}^{N} \int \f{d^4\ell}{(2\pi)^4}\  \f{1}{(p_i-\ell)^2+m_i^2-i\epsilon} \f{1}{(p_j+\ell)^2+m_j^2 -i\epsilon}\nn\\
&&\times\Big[ \epsilon_i^{T}(-p_i)\Gamma^{(3)}_{\mu\nu} (p_i,-p_i+\ell,-\ell)\Xi_i(-p_i+\ell)\Big]\Delta_F^{\mu\nu,\rho\sigma}(\ell)\ \varepsilon^{\lambda\tau}(k)  \nn\\
&&\times \Big[ \epsilon_j^{T}(-p_j)\Gamma^{(3)}_{\rho\sigma} (p_j,-p_j-\ell,\ell)\Xi_j(-p_j-\ell)\Big] \widetilde{\Gamma}^{(N+1)}_{(ij), \lambda\tau}( p_i-\ell,p_j+\ell ,k)\ . \label{eq:B_V_gr}
\ee
\endgroup
In both the expressions of $B_{IV}$ and $B_{V}$ the integrands can be simplified using \eqref{Gamma3Xi_gravity} and the result of the following common loop integral reads
\begingroup
\allowdisplaybreaks
\be
&&\int \f{d^4\ell}{(2\pi)^4}\  \f{1}{(p_i-\ell)^2+m_i^2-i\epsilon} \f{1}{(p_j+\ell)^2+m_j^2 -i\epsilon}\Big[ \epsilon_i^{T}(-p_i)\Gamma^{(3)}_{\mu\nu} (p_i,-p_i+\ell,-\ell)\Xi_i(-p_i+\ell)\Big]\nn\\
&&\times \Delta_F^{\mu\nu,\rho\sigma}(\ell)\ \times \Big[ \epsilon_j^{T}(-p_j)\Gamma^{(3)}_{\rho\sigma} (p_j,-p_j-\ell,\ell)\Xi_j(-p_j-\ell)\Big] \nn\\
&=&-2i\kappa^2 \epsilon_i^{T}\epsilon_j^{T}\int \f{d^4\ell}{(2\pi)^4}\f{1}{\ell^2-i\epsilon}\ \f{1}{\ell\cdot(\ell-2p_i)-i\epsilon}\ \f{1}{\ell\cdot (\ell+2p_j)-i\epsilon}\Big[  2(p_{i}.p_{j})^{2}-p_{i}^{2}p_{j}^{2}\nn\\
&&+2(p_{i}.p_{j})(p_{i}.\ell )-2(p_{i}.p_{j})(p_{j}.\ell )  +2(p_i.p_j)\left(p_j^\rho\Sigma_{i\rho\sigma}^T \ell^\sigma-p_i^\rho\Sigma_{j\rho\sigma}^T \ell^\sigma \right)+p_i^2 p_j^\rho\Sigma_{j\rho\sigma}^T \ell^\sigma \nn\\
&&-p_j^2 p_i^\rho\Sigma_{i\rho\sigma}^T \ell^\sigma\ +\mathcal{O}(\ell\ell)\Big].
\ee
\endgroup
Hence the total IR-finite contribution from the sets of diagrams in Fig.\ref{f:A^N+1_G} reads
\be
&&\mathcal{A}^{(N+1)}_{G,1}+
\mathcal{A}^{(N+1)}_{K-\text{finite},1}\nn\\
&=& B_{I}+B_{II}+B_{III}+B_{IV}+B_{V}-K_{gr}\times \Gamma^{(N+1)}\ , \label{eq:A_finite_master}
\ee
where the $(N+1)$ particle tree level amplitude is given by
\be
\Gamma^{(N+1)} &=& \sum_{\substack{k=1}}^N \f{1}{2p_k\cdot k}\epsilon_k^{T}(-p_k)\varepsilon^{\lambda\tau}(k)\Gamma^{(3)}_{\lambda\tau} (p_k,-p_k-k,k)\Xi_k(-p_k-k) \Gamma^{(N)}_{(k)}(p_k+k)\nn\\
&& +\ \varepsilon^{\lambda\tau}(k)\ \widetilde{\Gamma}^{(N+1)}_{\lambda\tau}(k)\ .
\ee
In the soft limit the above tree level $(N+1)$ particle amplitude provides the tree level soft graviton theorem with soft factor given in \eqref{eq:S_tree_gr}.

Substituting the results of \eqref{eq:B_I_gr}, \eqref{eq:B_II_gr}, \eqref{eq:B_III_gr}, \eqref{eq:B_IV_gr}, \eqref{eq:B_V_gr} and \eqref{eq:K_gr} in \eqref{eq:A_finite_master}, it is easy to see that the final expression is IR-finite in the limit $\ell^\mu\rightarrow 0$. In explicit computation, the IR divergent contribution in the sum of $(B_{I}+B_{III})+B_{IV}+B_{V}$ cancels with the IR divergent contribution of $K_{gr}\times \Gamma^{(N+1)}$, and $B_{II}$ is IR finite at finite value of $k$. 

Now let us analyze the expression \eqref{eq:A_finite_master} in the soft limit i.e. $\omega<<|p_i^\mu|$. In this limit a part of the final expression \eqref{eq:A_finite_master} contributes to $\omega^{-1}$ for the full integration range of virtual graviton momenta, which reproduce the Weinberg's soft graviton factor relating two IR-finite one-loop amplitudes. For evaluating the rest of the IR-finite part of \eqref{eq:A_finite_master} we divide the virtual momenta integration range into three regions: $|\ell^\mu|\in [0,\omega], [\omega ,|p_i^\mu|]$ and $[|p_i^\mu|,\infty)$. It turns out that the integrand starts contributing at order $\omega^0$ in the region of integration $|\ell^\mu|\in [0,\omega]$ and $[|p_i^\mu|,\infty)$\footnote{In the integration region $|\ell^\mu|\in [|p_i^\mu|,\infty)$ the integrand is UV divergent, so needs to use UV regulator and add appropriate counter terms to extract finite result. } in the soft limit. Finally in the region of integration $|\ell^\mu|\in [\omega ,|p_i^\mu|]$ the integrand contributes at order $\ln\omega$ which is dominant compare to order $\omega^0$ in the soft expansion.  Hence the order $\omega^{-1}$ and $\ln\omega$ contribution from \eqref{eq:A_finite_master} turns out to be
\begingroup
\allowdisplaybreaks
\be
&&\mathcal{A}^{(N+1)}_{G,1}+
\mathcal{A}^{(N+1)}_{K-\text{finite},1}\nn\\
&=&  \kappa\sum_{k=1}^N \f{p_k. \varepsilon . p_k}{p_k. k} \times \left(\mathcal{A}^{(N)}_{G,1}+\mathcal{A}^{(N)}_{K-\text{finite},1}\right)\nn\\
&& -\f{i\kappa^3}{4} \sum_{i=1}^{N}\sum_{\substack{j=1\\j\neq i}}^{N}\epsilon_i^{T}\epsilon_j^{T}\int_{reg} \f{d^4\ell}{(2\pi)^4}\ \f{1}{\ell^2-i\epsilon} \f{1}{\ell.p_i +i\epsilon}\f{1}{\ell.p_j-i\epsilon}\nn\\
&&\times \Bigg[-8(p_i.\varepsilon.p_i) p_i.p_j \f{p_j.k}{p_i.k}  +\f{2\ell.k}{p_i.k}\f{1}{\ell.p_i +i\epsilon} (p_i.\varepsilon.p_i)\left\lbrace 2(p_i.p_j)^2-p_i^2p_j^2 \right\rbrace\nn\\
&&- \f{4 (p_i.\varepsilon.\ell)}{\ell.p_i+i\epsilon}   \lbrace 2(p_i.p_j)^2 -p_i^2 p_j^2 \rbrace -4(p_i.\varepsilon.p_i)p_j^2+16 (p_i.\varepsilon.p_j)(p_i.p_j)\Bigg] \Gamma^{(N)}_{(ij)}(p_i,p_j)\nn\\
&& +\mathcal{O}(\omega^0)\ .\label{eq:AK+AG_gr}
\ee
\endgroup
Above the expression of $\mathcal{A}^{(N)}_{G,1}+\mathcal{A}^{(N)}_{K-\text{finite},1}$ is given in \eqref{AG+AK_finite}. The ``$reg$'' in the subscript of loop momentum integration above refers to the integration range $|\ell^\mu|\in [\omega ,|p_i^\mu|],|p_j^\mu|]$. It is intriguing to observe that the components reliant on theory, such as $\mathcal{K}_i$, $\Xi_i$, and $\frac{\partial\Gamma^{(N)}}{\partial p_i}$, along with the dependence on spin angular momenta of the massive particles, vanish when the individual diagram contributions are summed in the $\ln\omega$ order soft factor contributing integrand. This theory and spin independence feature is not true for the order $\omega^0$ contribution from \eqref{eq:A_finite_master}, which we ignored here.
\begin{center}
\begin{figure}
	\includegraphics[scale=0.42]{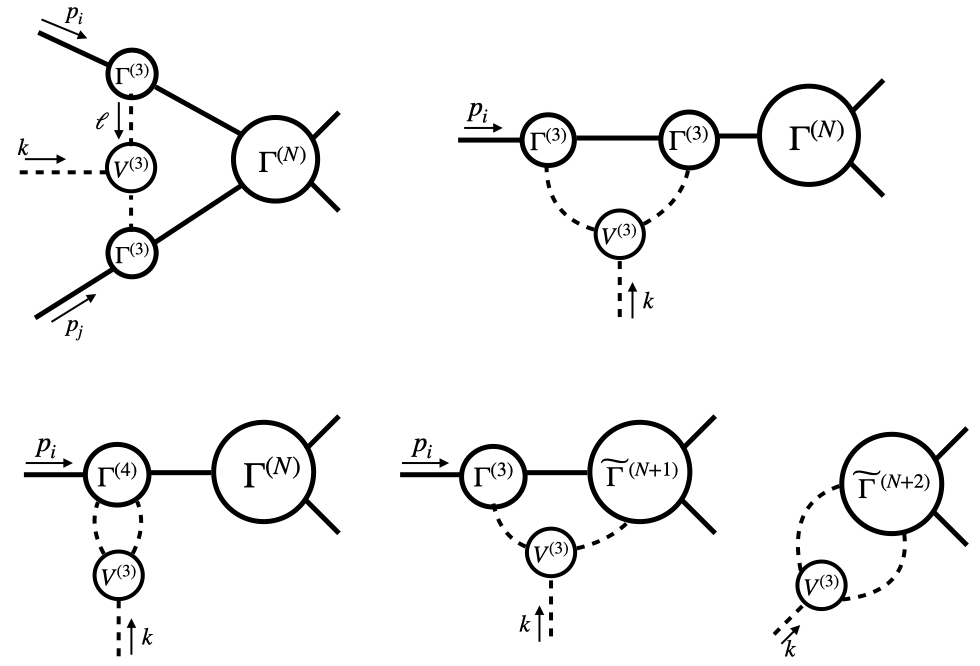}
	\caption{ Diagrams containing 3-graviton vertex contributing to $\mathcal{A}^{(N+1)}_\text{3-graviton-reg,1}$ after regulating the IR-divergence considering detector resolution as the explicit IR cut-off. The solid lines represent massive spinning particles and the dashed lines represent gravitons. }\label{f:3-graviton_diagrams}
\end{figure}
\end{center}

 Let us proceed to calculate the contribution of $\mathcal{A}^{(N+1)}_\text{3-graviton-reg,1}$. This term represents the sum of contributions from the Feynman diagrams shown in Fig.\ref{f:3-graviton_diagrams}.  We evaluate them using full graviton propagator and regulate the IR divergence of the virtual loop momentum integration by introducing an explicit IR cut-off $R^{-1}$, which serves as the resolution of the detector. The expression of the first diagram in Fig.\ref{f:3-graviton_diagrams} reads
 \begingroup
\allowdisplaybreaks
\be
C_I
&\equiv & \sum_{\substack{i,j=1\\i> j}}^{N} \int_{R^{-1}}^{\infty} \f{d^4\ell}{(2\pi)^4}\  \f{1}{(p_i-\ell)^2+m_i^2-i\epsilon} \f{1}{(p_j+\ell+k)^2+m_j^2 -i\epsilon}\nn\\
&&\times\Big[ \epsilon_i^{T}(-p_i)\Gamma^{(3)}_{ab} (p_i,-p_i+\ell,-\ell)\Xi_i(-p_i+\ell)\Big]  \nn\\
&&\times\Delta_F^{ab,\rho\sigma}(\ell)\varepsilon^{\mu\nu}(k)V^{(3)}_{\mu\nu, \rho\sigma,\tau\kappa}(k,\ell, -\ell-k) \Delta_F^{\tau\kappa,cd}(\ell)\nn\\
&& \Big[ \epsilon_j^{T}(-p_j)\Gamma^{(3)}_{cd} (p_j,-p_j-\ell-k,\ell+k)\Xi_j(-p_j-\ell-k)\Big] \Gamma^{(N)}_{(ij)}(p_i-\ell,p_j+\ell+k) \nn\\
 \ee
 \endgroup
Now using the identity in \eqref{Gamma3Xi_gravity} and substituting the 3-graviton vertex from \eqref{eq:V3_exp} with on-shell and transverse-traceless condition for the external graviton the above expression simplifies to
\begingroup
\allowdisplaybreaks
 \be
C_I
&=&-\f{i\kappa^3}{8} \sum_{\substack{i,j=1\\i\neq  j}}^{N} \epsilon_i^T \epsilon_j^{T}\int_{R^{-1}}^\infty \f{d^4\ell}{(2\pi)^4}\  \f{1}{(p_i-\ell)^2+m_i^2-i\epsilon} \f{1}{(p_j+\ell+k)^2+m_j^2 -i\epsilon}\f{1}{\ell^2-i\epsilon}\nn\\
&&\times \f{1}{(\ell+k)^2 -i\epsilon}\Big[4p_i^\rho p_i^\sigma -2p_i^2\eta^{\rho\sigma} -2p_i^\rho \ell^\sigma -2p_i^\sigma \ell^\rho +2p_i^\rho \Sigma^{T\sigma a}_{i}\ell_a+2p_i^\sigma \Sigma^{T\rho a}_{i}\ell_a\nn\\
&& -2\eta^{\rho\sigma}p_i^a\Sigma^T_{iac}\ell^c +\mathcal{O}(\ell\ell)  \Big]\times \varepsilon^{\mu\nu}(k)\Big[  -(k.\ell+\ell^2)\left( 
\eta_{\rho\sigma}\eta_{\mu\tau}\eta_{\nu\kappa}+\eta_{\tau\kappa}\eta_{\mu\rho}\eta_{\nu\sigma}\right)\nn\\
&&+\ 4\big(k.\ell+\ell^2\big)\eta_{\rho\nu}\eta_{\sigma\tau}\eta_{\kappa\mu}-\ 2\, \big(k_{\tau}\ell_{\kappa}\eta_{\mu\rho}\eta_{\nu\sigma}-\ell_{\mu}\ell_{\nu}\eta_{\rho\tau}\eta_{\sigma\kappa}-(k+\ell)_{\rho} k_{\sigma}\eta_{\mu\tau}\eta_{\nu\kappa}\big)\non\\
&&-2\, \big\lbrace(\ell_\mu\ell_\tau-k_{\tau}\ell_{\mu})\eta_{\nu\rho}\eta_{\sigma\kappa}+(2\ell_{\mu}k_{\rho}+\ell_\mu\ell_\rho )\eta_{\tau\nu}\eta_{\kappa\sigma}+(k_{\rho}k_{\tau} +\ell_\rho k_\tau -\ell_\tau k_\rho)\eta_{\mu\sigma}\eta_{\nu\kappa}\big\rbrace\nn\\
&&+\mathcal{O}(k^4, k^3\ell, k^2\ell^2, k\ell^3,\ell^4)\Big]\times  \Big[4p_j^\tau p_i^\kappa -2p_j^2\eta^{\tau\kappa} +2p_j^\tau (\ell+k)^\kappa +2p_j^\kappa (\ell+k)^\tau\nn\\
&& -2p_j^\tau \Sigma^{T\kappa b}_{j}(\ell+k)_b-2p_j^\kappa \Sigma^{T\tau b}_{j}(\ell+k)_b +2\eta^{\tau\kappa}p_j^b\Sigma^T_{jbd}(\ell+k)^d  +\mathcal{O}(\ell\ell,\ell k, kk)\Big] \nn\\
&&\times\, \Gamma^{(N)}_{(ij)}(p_i-\ell,p_j+\ell+k) \ .
 \ee
 \endgroup
The above expression contributes at order $\ln\omega$ in three regions of integration $|\ell^\mu|\in [R^{-1},\omega]$, $|(\ell+k)^\mu|\in [R^{-1},\omega]$, and  ``$reg$''$\equiv |\ell^\mu|\in [\omega, |p_i^\mu|]$ in the soft limit. The contribution from the regions $|\ell^\mu|\in [R^{-1},\omega]$ and $|(\ell+k)^\mu|\in [R^{-1},\omega]$ turn out to be same as $C_{I}$ is symmetric under the simultaneous exchange of $\ell\leftrightarrow (\ell+k)$ and $i\leftrightarrow j$. Hence we only evaluate the contribution in region $|\ell^\mu|\in [R^{-1},\omega]$ and multiply by a factor of $2$ to incorporate the contribution of other region.

In the region $|\ell^\mu|\in [R^{-1},\omega]$ we approximate the propagator denominators at leading order as
\be
&&\f{1}{(p_i-\ell)^2+m_i^2-i\epsilon}\simeq -\f{1}{2p_i.\ell+i\epsilon}\, ,\,  \f{1}{(p_j+\ell+k)^2+m_j^2 -i\epsilon}\simeq \f{1}{2p_j.k-i\epsilon}\, ,\nn\\
&&\f{1}{(\ell+k)^2 -i\epsilon} \simeq \f{1}{2\ell.k-i\epsilon}\, .
\ee
 Then only when we choose the order $\mathcal{O}(kk)$ terms from the $V^{(3)}$ vertex, after the loop momentum integration we can have $\mathcal{O}(\ln\omega)$ contribution. On the other hand in the region ``$reg$''$\equiv |\ell^\mu|\in [\omega, |p_i^\mu|]$ we approximate the propagator denominators at leading order as
\be
&&\f{1}{(p_i-\ell)^2+m_i^2-i\epsilon}\simeq -\f{1}{2p_i.\ell+i\epsilon}\, ,\,  \f{1}{(p_j+\ell+k)^2+m_j^2 -i\epsilon}\simeq \f{1}{2p_j.\ell-i\epsilon}\, ,\nn\\
&&\f{1}{(\ell+k)^2 -i\epsilon} \simeq \f{1}{\ell^2-i\epsilon}\, .
\ee
Then only when we choose the order $\mathcal{O}(\ell\ell)$ terms from the $V^{(3)}$ vertex, after the loop momentum integration we can have $\mathcal{O}(\ln\omega)$ contribution. 

Hence the full contribution at order $\ln\omega$ from all the three regions of integration becomes
\begingroup
\allowdisplaybreaks
\be
C_I
&=&\f{i\kappa^3}{4} \sum_{\substack{i,j=1\\i\neq  j}}^{N}\epsilon_i^T \epsilon_j^{T}\f{1}{p_j.k}\int_{R^{-1}}^\omega \f{d^4\ell}{(2\pi)^4}\  \f{1}{\ell^2-i\epsilon}\f{1}{p_i.\ell+i\epsilon}\f{1}{\ell.k-i\epsilon}\Big[ 4(p_i.k)^2 (p_j.\varepsilon.p_j)\nn\\
&&-4(p_i.k)(p_j.k)(p_i.\varepsilon.p_j)\Big]\ \Gamma^{(N)}_{(ij)}(p_i,p_j)\nn\\
&&+\f{i\kappa^3}{2} \sum_{\substack{i,j=1\\i\neq  j}}^{N} \epsilon_i^T \epsilon_j^{T}\int_{reg} \f{d^4\ell}{(2\pi)^4}\  \f{1}{\lbrace\ell^2-i\epsilon\rbrace^2}\f{1}{p_i.\ell+i\epsilon}\f{1}{p_j.\ell-i\epsilon}\Big[ -\ell^2 (p_i.\varepsilon.p_i)p_j^2\nn\\
&& -\ell^2 (p_j.\varepsilon.p_j)p_i^2 +4\ell^2 (p_i.p_j)(p_i.\varepsilon.p_j) +2(p_i.p_j)^2 (\ell.\varepsilon.\ell)-2(p_i.\varepsilon.\ell)(p_i.p_j)(p_j.\ell)\nn\\
&& +2p_j^2 (p_i.\varepsilon.\ell)(p_i.\ell)-p_i^2 p_j^2 (\ell.\varepsilon.\ell) -2(p_j.\varepsilon.\ell)(p_i.p_j)(p_i.\ell)+2p_i^2 (p_j.\varepsilon.\ell)(p_j.\ell)\Big]\nn\\
&&\ \times\, \Gamma^{(N)}_{(ij)}(p_i,p_j)\, +\, \mathcal{O}(\omega^0)\ , \label{eq:CI_2nd}
\ee
\endgroup
Note that the integrands above are independent of the spin angular momenta of external massive particles as well as do not depend on the theory dependent terms such as $\mathcal{K}_i, \Xi_i$ or non-minimal couplings. The first integrand above can be evaluated using the result of the integral (derived in \cite{1808.03288})
\be
&&\int_{R^{-1}}^\omega \f{d^4\ell}{(2\pi)^4}\  \f{1}{\ell^2-i\epsilon}\f{1}{p_i.\ell+i\epsilon}\f{1}{\ell.k-i\epsilon}\nn\\
&=& -\f{1}{4\pi}\f{1}{p_i.k}\ln(\omega R)\left[\delta_{\eta_i,-1}\ -\f{i}{2\pi}\ \ln\left(\f{p_i^2}{(p_i.\mathbf{n})^2}\right)\right]+\mathcal{O}(\omega^{-1})\ .
\ee
and momentum conservation relation
 \be
 \sum\limits_{\substack{j=1\\ j\neq i}}^N p_j^\mu=-p_i^\mu\ .\label{eq:mom_con}
 \ee
 Above $\eta_i$ convention is the same as described below \eqref{K_phase_reg}.
We won't evaluate the second integrand in \eqref{eq:CI_2nd} explicitly at this moment, but will simplify some of the terms which contains $(p_i.\ell)$ or $(p_j.\ell)$ in the numerator using the momentum conservation relation in \eqref{eq:mom_con}. Also the term containing $(\ell.\varepsilon.\ell)$ can be simplified by the following using integration by parts relation
\be
&&\int \f{d^4\ell}{(2\pi)^4}\f{\ell^\mu\ell^\nu}{(\ell^2 -i\epsilon)^2}\f{1}{p_i.\ell +i\epsilon}\f{1}{p_j.\ell -i\epsilon}\nn\\
&=&\f{1}{2}\int \f{d^4\ell}{(2\pi)^4}\f{1}{\ell^2-i\epsilon}\f{1}{p_i.\ell+i\epsilon}\f{1}{p_j.\ell -i\epsilon}\left(\eta^{\mu\nu}-\f{p_i^\mu\ell^\nu}{p_i.\ell+i\epsilon}-\f{p_j^\mu \ell^\nu}{p_j.\ell -i\epsilon}\right)\ .
\ee	
 Following all the steps outlined above and using exchange symmetry $(p_i,\ell)\leftrightarrow (p_j,-\ell)$ in the second integrand of \eqref{eq:CI_2nd}, the simplified expression of $C_I$ becomes
\begingroup
\allowdisplaybreaks
\be
C_I
&=&-\f{i\kappa^3}{4\pi} \ln\omega\ \sum_{\substack{j=1}}^{N}\epsilon_j^{T}\f{p_j.\varepsilon.p_j}{p_j.k}\times \sum_{\substack{i=1}}^N \epsilon_i^T  (p_i.k)\left[ \delta_{\eta_i,-1}\ -\f{i}{2\pi}\ \ln\left(\f{p_i^2}{(p_i.\mathbf{n})^2}\right)\right]  \Gamma^{(N)}_{(ij)}(p_i,p_j)  \nn\\
&&+\f{i\kappa^3}{2} \sum_{\substack{i,j=1\\i\neq  j}}^{N} \epsilon_i^T \epsilon_j^{T}\int_{reg} \f{d^4\ell}{(2\pi)^4}\  \f{1}{\ell^2-i\epsilon}\f{1}{p_i.\ell+i\epsilon}\f{1}{p_j.\ell-i\epsilon}\Big[ - (p_i.\varepsilon.p_i)p_j^2 - (p_j.\varepsilon.p_j)p_i^2\nn\\
&& +4 (p_i.p_j)(p_i.\varepsilon.p_j) -\f{1}{p_i.\ell +i\epsilon}\big\lbrace2(p_i.p_j)^2 -p_i^2 p_j^2 \big\rbrace (p_i.\varepsilon.\ell)\Big]\Gamma^{(N)}_{(ij)}(p_i,p_j)   + \mathcal{O}(\omega^0) .\label{eq:C_I_final}
\ee
\endgroup
Note that in the first line above $j=i$ sum is included while it was not present in the expression \eqref{eq:CI_2nd}. The inclusion of this term originates from the second term in the numerator of the first integral in \eqref{eq:CI_2nd} after using the momentum conservation relation \eqref{eq:mom_con}.

For completeness here we also briefly analyze the second diagram in Fig.\ref{f:3-graviton_diagrams}  which reads
\be
C_{II}&\equiv &\sum_{i=1}^N \f{1}{2p_i.k}\epsilon_i^T\int_{R^{-1}}^{\infty} \f{d^4\ell}{(2\pi)^4}\f{1}{(p_i-\ell)^2 +m_i^2-i\epsilon}  \ \Gamma^{(3)}_{ab}(p_i,-p_i+\ell,-\ell)\Xi_i(-p_i+\ell)\nn\\
&&\times \Gamma^{(3)}_{cd}(p_i-\ell, -p_i-k,\ell+k)\Xi_i(-p_i-k)\Delta_F^{ab,\rho\sigma}(\ell)\varepsilon^{\mu\nu}V^{(3)}_{\mu\nu,\rho\sigma,\tau\kappa}(k,\ell,-k-\ell)\nn\\
&&\times \Delta_F^{\tau\kappa,cd}(\ell+k)\Gamma^{(N)}_{(i)}(p_i+k)\ .
\ee
In the integration region  $|\ell^\mu|\in [R^{-1},\omega]$ or $|(\ell+k)^\mu|\in [R^{-1},\omega]$ the numerator of the potentially contributing $\ln\omega$ terms vanishes in the integrand. On the other hand, in the integration region ``$reg$'' using a set of integration by parts to cancel $(p_i.k)^{-1}$ factor, the terms potentially contributing at order $\ln\omega$ becomes
\be
C_{II}&=& -\f{i\kappa^3}{2}\sum_{i=1}^N\epsilon_i^{T}\int_{reg} \f{d^4\ell}{(2\pi)^4} \f{1}{\left(p_i.\ell +i\epsilon\right)^2}\f{1}{\ell^2 -i\epsilon} \Big[-2p_i^2 (p_i.\varepsilon.p_i)+\f{(p_i^2)^2}{p_i.\ell +i\epsilon} (p_i.\varepsilon.\ell)\Big]\nn\\
&&\times \Gamma^{(N)}_{(i)}(p_i)\ +\ \mathcal{O}(\omega^0)\ .
\ee
The above expression of $C_{II}$ can be simplified using the following identity
\be
\int_{reg} \f{d^4\ell}{(2\pi)^4} \f{\ell^\mu}{\left(p_i.\ell +i\epsilon\right)^3}\f{1}{\ell^2 -i\epsilon}=\f{1}{p_i^2} \int_{reg} \f{d^4\ell}{(2\pi)^4} \f{1}{\left(p_i.\ell +i\epsilon\right)^2}\f{1}{\ell^2 -i\epsilon}\ ,
\ee
and the simplified expression reads
\be
C_{II}&=& \f{i\kappa^3}{2}\sum_{i=1}^N\epsilon_i^{T}\int_{reg} \f{d^4\ell}{(2\pi)^4} \f{1}{\left(p_i.\ell +i\epsilon\right)^2}\f{1}{\ell^2 -i\epsilon} p_i^2 (p_i.\varepsilon.p_i)\times \Gamma^{(N)}_{(i)}(p_i) +\mathcal{O}(\omega^0)\ .\label{eq:CII}
\ee

The third, fourth and fifth diagrams in Fig.\ref{f:3-graviton_diagrams}  do not contribute to order $\ln\omega$ from any integration region. Hence the total order $\ln\omega$ contribution to $\mathcal{A}^{(N+1)}_\text{3-graviton-reg,1}$ after summing over \eqref{eq:C_I_final} and \eqref{eq:CII} turns out to be
\begingroup
\allowdisplaybreaks
\be
&&\mathcal{A}^{(N+1)}_\text{3-graviton-reg,1}\nn\\
&=&-\f{i\kappa^3}{4\pi} \ln\omega\ \sum_{\substack{j=1}}^{N}\epsilon_j^{T}\f{p_j.\varepsilon.p_j}{p_j.k}\times \sum_{\substack{i=1}}^N \epsilon_i^T  (p_i.k)\left[ \delta_{\eta_i,-1}\ -\f{i}{2\pi}\ \ln\left(\f{p_i^2}{(p_i.\mathbf{n})^2}\right)\right]  \Gamma^{(N)}_{(ij)}(p_i,p_j)  \nn\\
&&+\f{i\kappa^3}{2} \sum_{\substack{i,j=1\\i\neq  j}}^{N} \epsilon_i^T \epsilon_j^{T}\int_{reg} \f{d^4\ell}{(2\pi)^4}\  \f{1}{\ell^2-i\epsilon}\f{1}{p_i.\ell+i\epsilon}\f{1}{p_j.\ell-i\epsilon}\Bigg[ - (p_i.\varepsilon.p_i)p_j^2 - (p_j.\varepsilon.p_j)p_i^2\nn\\
&& +4 (p_i.p_j)(p_i.\varepsilon.p_j) -\f{1}{p_i.\ell +i\epsilon}\big\lbrace2(p_i.p_j)^2 -p_i^2 p_j^2 \big\rbrace (p_i.\varepsilon.\ell)\Bigg]\Gamma^{(N)}_{(ij)}(p_i,p_j)\nn\\
&&+\f{i\kappa^3}{2}\sum_{i=1}^N\epsilon_i^{T}\int_{reg} \f{d^4\ell}{(2\pi)^4} \f{1}{\left(p_i.\ell +i\epsilon\right)^2}\f{1}{\ell^2 -i\epsilon} p_i^2 (p_i.\varepsilon.p_i)\times \Gamma^{(N)}_{(i)}(p_i)  + \mathcal{O}(\omega^0)\ .\label{eq:A_3grav_N+1_final}
\ee
\endgroup
\begin{center}
\begin{figure}
	\includegraphics[scale=0.28]{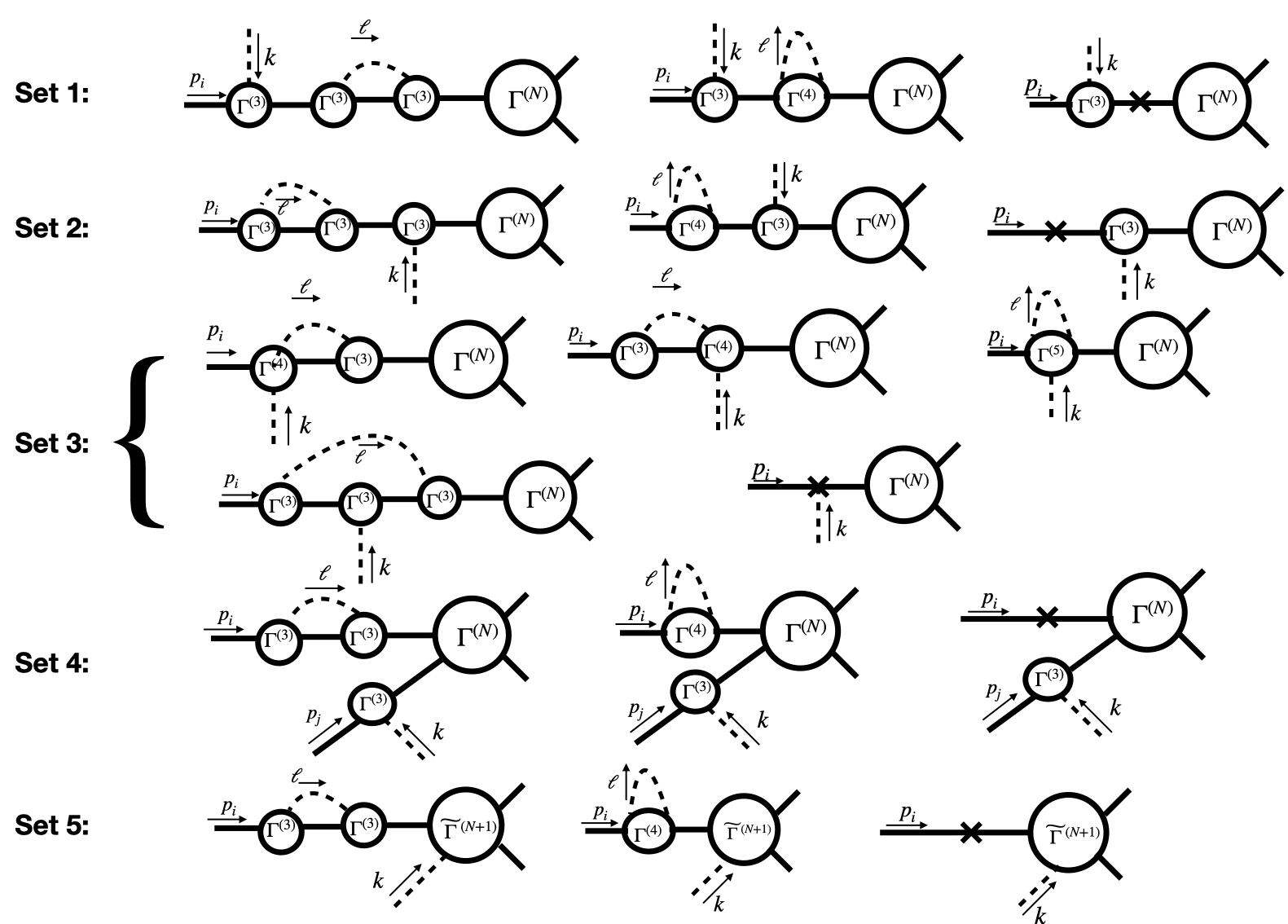}
	\caption{ Set of 1-loop diagrams contributing to $\mathcal{A}^{(N+1)}_{\text{self,1}}$, where the virtual dashed lines represent the full graviton propagator connecting two different points on the same massive spinning particle leg.  The cross appears in some diagrams above corresponds to counter term, which cancels the UV divergences in the renormalization prescription.}\label{f:self_diagram_gr_N+1}
\end{figure}
\end{center}

Let us now analyze the set of Feynman diagrams in Fig.\ref{f:self_diagram_gr_N+1} which contributes to $\mathcal{A}^{(N+1)}_{\text{self,1}}$. All the diagrams are IR-finite for finite $k$ and the UV divergences in the sum of contributions cancel by using on-shell renormalization condition with proper choice of counter terms. After renormalization a finite contribution remains, and it contributes at order $\ln\omega$ in the region ``$reg$''. In specific, the sum of the diagrams in Set 1, Set 2, Set 4 and Set 5 individually vanishes using the gravitational analogue of the wave functional renormalization condition  \eqref{renormalization_condition}. On the other hand the counter term choice of the last diagram in Set 3 cancels the sum of UV divergences appearing in the the first four diagrams in Set 3 and the second and third diagrams of Fig.\ref{f:3-graviton_diagrams}. After the cancellation of these UV divergences, only in the integration region ``$reg$'' we get the following expression, which can potentially contribute at order $\ln\omega$ in the soft limit
\be
\mathcal{A}^{(N+1)}_{\text{self,1}}&=& -\f{i\kappa^3}{2} \sum_{i=1}^N\epsilon_i^{T}\int_{reg} \f{d^4\ell}{(2\pi)^4} \f{1}{\left(p_i.\ell +i\epsilon\right)^2}\f{1}{\ell^2 -i\epsilon}\ p_i^2 (p_i.\varepsilon .p_i)\times \Gamma^{(N)}_{(i)}(p_i)\nn\\
&&+\mathcal{O}(\omega^0)\ . \label{eq:A_self_N+1_gr}
\ee

Before even evaluating $\mathcal{A}^{(N+1)}_{\text{non-div,1}}$ from the set of diagrams in Fig.\ref{f:A^N+1_non-div}, it becomes evident that these diagrams are IR-finite for finite $k$ when $\ell^\mu\rightarrow 0$.  The sum of first, fourth and sixth diagrams in Fig.\ref{f:A^N+1_non-div} contribute to leading soft graviton factor at order $\omega^{-1}$ multiplying $\mathcal{A}^{(N)}_{\text{non-div,1}}$. On the other hand in the integration region $\omega<<|\ell^\mu|<<|p_i^\mu |$, though individually the first and third diagrams in Fig.\ref{f:A^N+1_non-div} have the potential to contribute at order $\ln\omega$ as those behave like $\int_\omega\f{d^4\ell}{|\ell|^4}$, when we sum them up, in the numerator we get an extra factors of $\ell$ and/or  $k$. Hence they can only contribute from order $\omega^0$ or $\omega\ln\omega$. Hence, summing over all the contribution we get
\be
\mathcal{A}^{(N+1)}_{\text{non-div,1}}=\kappa \sum_{i=1}^{N}\f{p_i.\varepsilon.p_i}{p_i.k}\ \mathcal{A}^{(N)}_{\text{non-div,1}}\ +\mathcal{O}(\omega^0)\ .\label{N+1_non-div_contribution_gr}
\ee

Summing over the contribution of \eqref{eq:AK+AG_gr}, \eqref{eq:A_3grav_N+1_final}, \eqref{eq:A_self_N+1_gr} and \eqref{N+1_non-div_contribution_gr} we get
\begingroup
\allowdisplaybreaks
\be
\mathcal{A}^{(N+1)}_{\text{IR-finite,1}}&=&\kappa \sum_{i=1}^{N}\f{p_i.\varepsilon.p_i}{p_i.k}\ \times \mathcal{A}^{(N)}_{\text{IR-finite,1}}\nn\\
&&-\f{i\kappa^3}{4\pi} \ln\omega\ \sum_{\substack{j=1}}^{N}\f{p_j.\varepsilon.p_j}{p_j.k}\times \sum_{\substack{i=1}}^N   (p_i.k)\left[ \delta_{\eta_i,-1}\ -\f{i}{2\pi}\ \ln\left(\f{p_i^2}{(p_i.\mathbf{n})^2}\right)\right] \times\Gamma^{(N)}  \nn\\
&& -\f{i\kappa^3}{2} \sum_{i=1}^{N}\sum_{\substack{j=1\\j\neq i}}^{N}\int_{reg} \f{d^4\ell}{(2\pi)^4}\ \f{1}{\ell^2-i\epsilon} \f{1}{\ell.p_i +i\epsilon}\f{1}{\ell.p_j-i\epsilon}\nn\\
&&\times \Bigg[-4(p_i.\varepsilon.p_i) p_i.p_j \f{p_j.k}{p_i.k}  +\f{\ell.k}{p_i.k}\f{1}{\ell.p_i +i\epsilon} (p_i.\varepsilon.p_i)\left\lbrace 2(p_i.p_j)^2-p_i^2p_j^2 \right\rbrace\nn\\
&&- \f{ (p_i.\varepsilon.\ell)}{\ell.p_i+i\epsilon}   \lbrace 2(p_i.p_j)^2 -p_i^2 p_j^2 \rbrace +4 (p_i.\varepsilon.p_j)(p_i.p_j)\Bigg]\times \Gamma^{(N)}  + \mathcal{O}(\omega^0)\ , \label{eq:full_gr_amp}
\ee
\endgroup
where the expression of $\mathcal{A}^{(N)}_{\text{IR-finite,1}}$ is given in \eqref{eq:AN_ir-finite_gr}.

\paragraph{Soft graviton theorem result at one-loop:}
The expression for one-loop amplitude in the soft expansion derived in \eqref{eq:full_gr_amp} can be re-written in the following compact way
\begingroup
\allowdisplaybreaks
\be
\mathcal{A}^{(N+1)}_{\text{IR-finite,1}}&=& \kappa \sum_{i=1}^{N}\f{p_i.\varepsilon.p_i}{p_i.k}\ \times \mathcal{A}^{(N)}_{\text{IR-finite,1}}\nn\\
&&+  K_{phase}^{reg}\times \kappa \sum_{i=1}^{N}\f{p_i.\varepsilon.p_i}{p_i.k}\ \times \mathcal{A}^{(N)}_{\text{IR-finite,0}} \nn\\
&& +\kappa \sum_{i=1}^N\f{\varepsilon_{\mu\nu}p_i^\mu k_\rho }{p_i\cdot k} \left\lbrace p_i^\nu \f{\p K_{gr}^{reg}}{\p p_{i\rho}}-p_i^\rho \f{\p K_{gr}^{reg}}{\p p_{i\nu}}\right\rbrace \times \mathcal{A}^{(N)}_{\text{IR-finite,0}}   + \mathcal{O}(\omega^0)\ , \label{eq:oneloop-soft_gr-thm}
\ee
\endgroup
where the expressions of $K_{phase}^{reg}$ and $K_{gr}^{reg}$ are given in \eqref{K_phase_reg} and \eqref{K_gr_reg} respectively. This result agrees with the loop corrected subleading soft graviton theorem, originally derived in \cite{1808.03288} for minimally coupled scalar-gravity with scalar contact interaction. Here the re-derivation of this result from a scattering amplitude involving particles with arbitrary spins in a generic theory of quantum gravity confirms the universality of the $\ln\omega$ soft factor. The universal properties of tree-level soft graviton theorems in spacetime dimensions $D\geq 5$ have also been established from the soft expansion of string theory amplitudes in  \cite{ademollo,shapiro,
1406.4172,1406.5155,1411.6661,1502.05258,1505.05854,1507.08829,
1511.04921,1512.00803,1601.03457,1604.03355,1610.03481,1702.03934,1703.00024, Marotta:2019cip}. It would be interesting to explore whether the $\ln\omega$ soft factor can also be derived from one-loop amplitudes in string theory compactified into four spacetime dimensions.

The soft theorem result \eqref{eq:oneloop-soft_gr-thm} also verifies the well-known fact that Weinberg's leading soft graviton theorem remains unaltered by loop corrections. Therefore, even from the analysis presented above, we observe that Weinberg's soft theorem still holds, relating two one-loop IR-finite amplitudes.

\subsection{Discussion on generalization} \label{S:DiscussionII}
When some of the massive spinning particles carry electric charge, the order $\ln\omega$ soft graviton factor in \eqref{eq:oneloop-soft_gr-thm} undergoes correction due to electromagnetic interactions. The correction term has been derived in \cite{1808.03288} in a minimally coupled charged scalar theory. A straightforward generalization of our derivation, combined with the covariantization prescription outlined in \cite{1809.01675}, will be useful for establishing the universal nature of this correction. At one-loop order, it is also possible to derive the spin-dependent order $\omega\ln\omega$ soft graviton factor as provided in \eqref{S1_loop_gr}, once we understand how to account for the additional subtleties highlighted in the discussion below \eqref{eq:S_K_commute}. Furthermore, through an analysis of two-loop amplitudes in the soft limit, it is possible to derive the order $\omega(\ln\omega)^2$ soft graviton theorem as conjectured in \eqref{S2_loop_gr}, which is also expected to be universal. We intend to pursue these analyses in the future. 

Using the Feynman rules derived in subsection-\ref{S:Covariantization_gravity}, in combination with the identities in appendix-\ref{S:identities_gr}, it is possible to derive the tree-level simultaneous and consecutive double soft-graviton theorems up to sub-subleading orders in a generic theory of quantum gravity. This result does not exist in literature and will be interesting to explore in future.

\section{Summary and outlook}
\label{S:summary_outlook}
In this article, we have proved that the order $\ln\omega$ soft photon and graviton theorems are universal (theory independent) by working within a generic setup of arbitrary spinning particle scattering. Along the way, we have also provided the definitions of IR-finite amplitudes in the generic theory of QED and quantum gravity for scattering involving spinning particles. We used the Ward Identities along with the Grammer-Yennie decomposition to perform Eikonal exponentialization and extract the IR-finite amplitudes. At one-loop order, we provide integral expressions for the IR-finite amplitudes, which have been used to derive soft factors. While it would have been desirable to explicitly evaluate these one-loop IR-finite amplitudes and explore their crossing and unitarity properties in a specific theory, this remains an avenue for future investigation.\footnote{BS acknowledges Waël Aoun for pursuing his master thesis project on this topic using the IR-finite S-matrix formalism proposed in \cite{1911.06821} and making substantial progress.} We have also discussed that while the KG-decomposition is powerful for extracting IR-finite amplitudes in a generic theory of QED, it loses efficacy in the generic theory of quantum gravity. Furthermore, we have provided a set of soft photon and graviton theorems up to two-loop orders in section-\ref{S:intro_res}, with some of these theorems conjectured based on classical analysis. These set of new soft theorems can be derived by extending our analysis to higher loops as discussed in subsection-\ref{S:DiscussionI} and \ref{S:DiscussionII}.

In \cite{1801.07719, 1804.09193}, a relation between low-frequency electromagnetic/gravitational waveforms and the classical limit of soft photon/graviton factors has been derived. In frequency space, the low-frequency gravitational waveform in $D=4$ is given by \footnote{The explicit expression is given in equation (2.5) of \cite{2106.10741}, where we have adopted the convention that the energy of the outgoing particle is positive, which is contrary to the convention used in this paper. After Fourier transformation in frequency variable, it provides gravitational memory along with multiple tail contributions at late and early time for generic gravitational scattering event.}
\be
\varepsilon^{\mu\nu}(k)\widetilde{e}_{\mu\nu}(\omega, R\widehat{\bold{n}})&=& -i\ \f{2G}{R}\ \exp\left\lbrace i\omega \left( R+2G\ln R\sum\limits_{\substack{j=1\\ \eta_j=-1}}^N p_j\cdot \mathbf{n}\right)\right\rbrace \times \mathbb{S}^{gr}_{\text{classical}}\ , \label{eq:classical_quantum_relation}
\ee
where $\mathbb{S}^{gr}_{\text{classical}}$ represents the classical limit of quantum soft factor for single soft graviton emission with momentum $k^{\mu}=-\omega \mathbf{n}^\mu$. In the above expression $\widetilde{e}_{\mu\nu}$ is defined by
\be
\widetilde{e}_{\mu\nu}(\omega, \vec{x})&\equiv & \int_{-\infty}^\infty dt\ e^{i\omega t} \left(h_{\mu\nu}(t,\vec{x})-\f{1}{2}\eta_{\mu\nu}h^\rho_\rho(t,\vec{x})\right)\ .
\ee
Note that to derive the classical limit of the quantum ``soft factor'' from \eqref{Soft_theorem}, first we need to incorporate the sum over particles indexed by $i$. Subsequently, in their respective expressions, when the orbital momentum operator operates on $\mathcal{A}^{(N)}$, it must be substituted with the classical orbital momenta of the scattered objects. Furthermore, based on explicit classical computations of gravitational waveforms presented in \cite{1912.06413, 2008.04376, 2106.10741}, it has been conjectured that the validity of the relation \eqref{eq:classical_quantum_relation} holds true solely when, in the classical limit, the soft graviton factor is calculated using the retarded propagator for the graviton field instead of Feynman propagator. In practice this prescription suggests that in the integral representations of \eqref{K_gr_reg} and \eqref{K_phase_reg}, the term $(\ell^2-i\epsilon)^{-1}$ should be substituted with $-\left((\ell^0+i\epsilon)^2-\vec{\ell}^2\right)^{-1}$ in order to extract $\mathbb{S}^{gr}_{\text{classical}}$. With this substitution, $\mathbb{S}^{gr}_{\text{classical}}$ only receives contributions from the part of the loop integrals in which the virtual hard particle propagator goes on-shell (referred to as the potential region) and the graviton propagator with principal value. However, it does not receive contributions from the part of the loop integrals in which the virtual graviton goes on-shell (known as the radiation-reaction region). Unfortunately, we lack a fundamental understanding of why the classical limit of the quantum soft factor does not include contributions from radiation reactions.

For a $2\rightarrow 2$ scattering process with a large impact parameter or low momentum transfer, the contribution of radiation reaction to the quantum soft factor turns out to be suppressed compared to the contribution from the potential region (classical), as discussed in \cite{1808.03288}. This has also been established in \cite{2007.02077} using the KMOC formalism \cite{Kosower:2018adc, Cristofoli:2021vyo} after substituting the final momenta of the scattered particles in terms of initial momenta and the perturbatively computed momentum impulse in terms of initial scattering data and specified interaction. However, for hard scattering (small impact parameter scattering), the reason why the radiation reaction contribution to the quantum soft factor does not affect the classical waveform in the classical limit has not yet been resolved. We believe that a generalization of recent investigations into deriving classical gravitational waveforms from Eikonal exponentiation in \cite{DiVecchia:2023frv, Georgoudis:2023lgf, Elkhidir:2023dco, Brandhuber:2023hhy, Herderschee:2023fxh, Caron-Huot:2023vxl} could potentially resolve this puzzle even for hard scattering. It would be interesting to explore whether there are any observable consequences resulting from the contribution of radiation reaction in the quantum soft factor.

There have been many applications of soft theorems both in the context of scattering amplitudes and in relation to gravitational memory. For instance, the universal characteristic of Weinberg's soft theorem imposes an infinite hierarchy of constraints on the linear momentum impulse within the KMOC formalism, as derived in \cite{Bautista:2021llr}. It is also anticipated that the universality of the $\ln\omega$ soft theorem should impose an infinite hierarchy of constraints on the angular momentum impulse within the KMOC formalism.\footnote{BS acknowledges Alok Laddha for private communication.} As discussed in section-\ref{S:intro_res}, while the soft theorems alone cannot impose non-trivial constraints on the quantum theory in the UV, combining the results of the soft theorems with certain physical assumptions about scattering amplitudes, such as analyticity, unitarity and crossing grants them the ability to constrain the UV quantum theory. As an illustration of this concept, the article \cite{Karateev:2022jdb} derived non-perturbative bounds on the $a$-anomaly coefficient of the UV conformal field theory (CFT), the deformation of which leads to a massive QFT along the renormalization group (RG) flow. These bounds were established by incorporating the constraints from the double soft dilaton theorem within the framework of non-perturbative S-matrix bootstrap. For numerous astrophysical scattering events, the order of magnitude of gravitational tail memory follows from the $\ln\omega$ soft theorem has been estimated in \cite{1806.01872, 1912.06413}. In certain classical scattering scenarios, the gravitational waveform resulting from the $\ln\omega$ soft theorem has also been derived in \cite{Fernandes:2020tsq, Hait:2022ukn, Mohanty:2022abo}, carrying observable consequences in the present era of gravitational wave physics.

\acknowledgments
We would like to thank Waël Aoun, P.V. Athira, Thibault Damour, Hofie Hannesdottir,  Carlo Heissenberg, Alok Laddha, Joao Penedones, Ashoke Sen and Paolo Di Vecchia for useful discussion.
The work of HK is supported by NSF grant PHY 2210533. The work of BS is supported by the Simons Foundation grant 488649 (Simons Collaboration on the Nonperturbative Bootstrap) and by the Swiss National Science Foundation through the National Centre of Competence in Research SwissMAP. 

\appendix
\section{Intermediate steps in deriving soft photon theorem }
Let us evaluate the following expression appears in \eqref{A_I} after substituting the vertex \eqref{Gamma3_photon} and using the identities \eqref{KQ_relation}, \eqref{XiQ_relation} and \eqref{charge_eigen_value},
\begingroup
\allowdisplaybreaks
\be
\zeta_{1\mu}&\equiv & \epsilon_i^{T}(-p_i)\varepsilon^{\rho}(k)\Gamma^{(3)}_\rho (p_i,-p_i-k,k)\Xi_i(-p_i-k)\Gamma^{(3)}_\mu(p_i+k,-p_i-k+\ell ,-\ell)\Xi_i(-p_i-k+\ell)\non\\
&=&e_i^2\epsilon_i^{T}\Bigg[\varepsilon^{\rho}\f{\p \mathcal{K}_i(-p_i)}{\p p_i^\rho}\Xi_i(-p_i)+\varepsilon^\rho k^\sigma \f{\p \mathcal{K}_i(-p_i)}{\p p_i^\rho}\f{\p \Xi_i(-p_i)}{\p p_i^\sigma}+\f{1}{2}\varepsilon^\rho k^\sigma \f{\p^2 \mathcal{K}_i(-p_i)}{\p p_i^\rho \p p_i^\sigma}\Xi_i(-p_i)\non\\
&& +\f{2i}{e_i}\varepsilon^\rho k^\sigma \mathcal{B}^i_{\rho\sigma}(-p_i)\Xi_i(-p_i)+\mathcal{O}(kk)\Bigg]\times \Bigg[ -\f{\p \mathcal{K}_i(-p_i)}{\p p_i^\mu}\Xi_i(-p_i)-(k-\ell)^\kappa \f{\p \mathcal{K}_i(-p_i)}{\p p_i^\mu}\f{\p \Xi_i(-p_i)}{\p p_i^\kappa}\non\\
&& -\f{1}{2}(2k-\ell)^\kappa \f{\p^2 \mathcal{K}_i(-p_i) }{\p p_i^\mu \p p_i^\kappa}\Xi_i(-p_i)+\f{2i}{e_i}\ell^\kappa \mathcal{B}^i_{\mu\kappa}(-p_i)\Xi_i(-p_i)+\mathcal{O}(kk,k\ell,\ell\ell)\Bigg]\ .
\ee
\endgroup
Now using the identities \eqref{K2},\eqref{K3} and \eqref{on-shell_condition} we simplify the above expression and get
\begingroup
\allowdisplaybreaks
\be
\zeta_{1\mu}&=&e_i^2 \epsilon_i^{T}\Bigg[2i\varepsilon\cdot p_i +\f{1}{2}(\varepsilon^\rho k^\sigma -\varepsilon^\sigma k^\rho )\f{\p \mathcal{K}_i(-p_i)}{\p p_i^\rho}\f{\p \Xi_i(-p_i)}{\p p_i^\sigma}+\f{i}{e_i}(\varepsilon^\rho k^\sigma -\varepsilon^\sigma k^\rho ) \mathcal{B}^i_{\rho\sigma}(-p_i)\Xi_i(-p_i)\non\\
&&+ \mathcal{O}(kk)\Bigg]\times \Bigg[ -i(2p_i+2k-\ell)_\mu +\mathcal{K}_i(-p_i)\f{\p \Xi_i(-p_i)}{\p p_i^\mu}+\f{1}{2}(2k-\ell)^\kappa \mathcal{K}_i(-p_i)\f{\p^2 \Xi_i(-p_i)}{\p p_i^\mu \p p_i^\kappa}\non\\
&&+\f{1}{2}\ell^\kappa \f{\p \mathcal{K}_i(-p_i)}{\p p_i^\mu}\f{\p \Xi_i(-p_i)}{\p p_i^\kappa}+\f{1}{2}(2k-\ell)^\kappa \f{\p \mathcal{K}_i(-p_i)}{\p p_i^\kappa}\f{\p \Xi_i(-p_i)}{\p p_i^\mu}+\f{2i}{e_i}\ell^\kappa \mathcal{B}^i_{\mu\kappa}(-p_i)\Xi_i(-p_i)\non\\
&&+\mathcal{O}(kk,k\ell,\ell\ell)\Bigg]\non\\
&=&e_i^2 \epsilon_i^{T}\Bigg[ 2\varepsilon\cdot p_i(2p_i+2k-\ell)_\mu+i\varepsilon\cdot p_i \ell^\rho \Big\lbrace\f{\p \mathcal{K}_i(-p_i)}{\p p_i^\mu}\f{\p \Xi_i(-p_i)}{\p p_i^\rho}- \f{\p \mathcal{K}_i(-p_i)}{\p p_i^\rho}\f{\p \Xi_i(-p_i)}{\p p_i^\mu}\non\\
&&+\f{4i}{e_i}\mathcal{B}^i_{\mu\rho}(-p_i)\Xi_i(-p_i)\Big\rbrace -\ \f{i}{2}p_{i\mu}(\varepsilon^\rho k^\sigma -\varepsilon^\sigma k^\rho )\Big\lbrace \f{\p \mathcal{K}_i(-p_i)}{\p p_i^\rho}\f{\p \Xi_i(-p_i)}{\p p_i^\sigma}-\f{\p \mathcal{K}_i(-p_i)}{\p p_i^\sigma}\f{\p \Xi_i(-p_i)}{\p p_i^\rho}\non\\
&& +\f{4i}{e_i}\mathcal{B}^i_{\rho\sigma}(-p_i)\Xi_i(-p_i)\Big\rbrace+2ip_i.k \varepsilon^\rho \f{\p \mathcal{K}_i(-p_i)}{\p p_i^\rho}\f{\p \Xi_i(-p_i)}{\p p_i^\mu} +\mathcal{O}(kk,k\ell,\ell\ell)\Bigg]\ .
\ee
\endgroup
 Let us define a specific tensor structure which will appear together in all the computations in section-\ref{s:photon_one_loop_computation} , which is also the non-universal contribution to the tree-level subleading soft photon theorem as derived in \cite{1809.01675}
 \be
 \mathcal{N}^{i}_{\rho\sigma}(-p_i)\equiv -\f{i}{8}\Bigg[\f{\p \mathcal{K}_i(-p_i)}{\p p_i^\rho}\f{\p \Xi_i(-p_i)}{\p p_i^\sigma}-\f{\p \mathcal{K}_i(-p_i)}{\p p_i^\sigma}\f{\p \Xi_i(-p_i)}{\p p_i^\rho}+\f{4i}{e_i}\mathcal{B}^i_{\rho\sigma}(-p_i)\Xi_i(-p_i)\Bigg].\label{N_non-universal}
 \ee
With the above definition, the expression of $\zeta_{1\mu}$ can be written in the following compact form
\be
\zeta_{1\mu}&=& e_i^2 \epsilon_i^{T}\Bigg[ 2\varepsilon\cdot p_i(2p_i+2k-\ell)_\mu-8\varepsilon\cdot p_i \ell^\rho \mathcal{N}^i_{\mu\rho}(-p_i) +4\ p_{i\mu}(\varepsilon^\rho k^\sigma -\varepsilon^\sigma k^\rho ) \mathcal{N}^{i}_{\rho\sigma}(-p_i)\non\\
&&\ +2i\ p_i.k\ \varepsilon^\rho \f{\p \mathcal{K}_i(-p_i)}{\p p_i^\rho}\f{\p \Xi_i(-p_i)}{\p p_i^\mu} +\mathcal{O}(kk,k\ell,\ell\ell)\Bigg]\ .\label{zeta_1}
\ee
Let us evaluate the following expression appears in \eqref{A_II} after substituting the vertex \eqref{Gamma4_photon} and using the identities \eqref{KQ_relation}, \eqref{XiQ_relation} and \eqref{charge_eigen_value},
\be
\zeta_{2\mu}&\equiv & \epsilon_i^{T}(-p_i)\varepsilon^{\rho}(k)\Gamma^{(4)}_{\rho\mu} (p_i,-p_i-k+\ell,k,-\ell)\Xi_i(-p_i-k+\ell)\non\\
&=& ie_i^2\ \epsilon_i^T \varepsilon^\rho \f{\p^2 \mathcal{K}_i(-p_i)}{\p p_i^\mu \p p_i^\rho}\Xi_i(-p_i) +\mathcal{O}(\ell, k)\ .
\ee
Using the identity in \eqref{K3} and on-shell condition \eqref{on-shell_condition} the above expression reduces to
\be
\zeta_{2\mu}&=& -e_i^2\epsilon_i^{T}\Bigg[2\varepsilon_\mu +i\varepsilon^\rho \Big\lbrace \f{\p \mathcal{K}_i(-p_i)}{\p p_i^\mu}\f{\p \Xi_i(-p_i)}{\p p_i^\rho}+\f{\p \mathcal{K}_i(-p_i)}{\p p_i^\rho}\f{\p \Xi_i(-p_i)}{\p p_i^\mu}\Big\rbrace +\mathcal{O}(\ell, k)\Bigg].\label{zeta_2}
\ee
The expression below appears in \eqref{A_III} and can be evaluated analogous to the evaluation of $\zeta_{1\mu}$. The final result reads
\begingroup
\allowdisplaybreaks
\be
\zeta_{3\mu}&\equiv &\epsilon_i^{T}(-p_i)\varepsilon^{\rho}(k)\Gamma^{(3)}_\mu (p_i,-p_i+\ell,-\ell)\Xi_i(-p_i+\ell)\Gamma^{(3)}_\rho(p_i-\ell,-p_i-k+\ell ,k)\Xi_i(-p_i-k+\ell)\non\\
&=& e_i^2 \epsilon_i^{T}\Bigg[ 2\varepsilon\cdot p_i(2p_i-\ell)_\mu -4\varepsilon.\ell p_{i\mu}-8\varepsilon\cdot p_i \ell^\rho \mathcal{N}^i_{\mu\rho}(-p_i) +4\ p_{i\mu}(\varepsilon^\rho k^\sigma -\varepsilon^\sigma k^\rho ) \mathcal{N}^{i}_{\rho\sigma}(-p_i)\non\\
&&\ -2i\ p_i.\ell\ \varepsilon^\rho \f{\p \mathcal{K}_i(-p_i)}{\p p_i^\mu}\f{\p \Xi_i(-p_i)}{\p p_i^\rho} +\mathcal{O}(kk,k\ell,\ell\ell)\Bigg]\ .\label{zeta_3}
\ee
\endgroup
Let us evaluate the following expression appears in \eqref{A_G^N_result}, \eqref{A_I},\eqref{A_II} and \eqref{A_III}. After substituting the vertex from \eqref{Gamma3_photon} and using the identities \eqref{KQ_relation}, \eqref{XiQ_relation} and \eqref{charge_eigen_value} we get
\begingroup
\allowdisplaybreaks
\be
\zeta_{4\nu}&\equiv & \epsilon_j^{T}(-p_j)\Gamma^{(3)}_\nu (p_j,-p_j-\ell,\ell)\Xi_j(-p_j-\ell)\non\\
&=& -ie_j\ \epsilon_j^{T}\Bigg[ \f{\p \mathcal{K}_j(-p_j)}{\p p_j^\nu}\Xi_j(-p_j)+\ell^\sigma \f{\p \mathcal{K}_j(-p_j)}{\p p_j^\nu}\f{\p \Xi_j(-p_j)}{\p p_j^\sigma}+\f{1}{2}\ell^\sigma \f{\p^2 \mathcal{K}_j(-p_j)}{\p p_j^\nu \p p_j^\sigma}\Xi_j(-p_j)\non\\
&&\ +\f{2i}{e_j}\ell^\sigma\mathcal{B}^j_{\nu\sigma}(-p_j)\Xi_j(-p_j)+\mathcal{O}(\ell\ell)\Bigg]\ .
\ee
\endgroup
Using the identities \eqref{K2},\eqref{K3} and the on-shell condition \eqref{on-shell_condition}, the above expression reduces to
\be
\zeta_{4\nu}&=&e_j \epsilon_j^{T}\Bigg[ (2p_j+\ell)_\nu +4\ell^\sigma \mathcal{N}^j_{\nu\sigma}(-p_j)+\mathcal{O}(\ell\ell)\Bigg]\ .\label{zeta_4}
\ee
\section{Amputated Green's function involving single graviton}\label{appendix-Gamma-tilde}
Following the covariantization prescription described in section-\ref{S:Covariantization_gravity}, we compute the amputated Green's function involving $N$ number of massive spinning particles and one off-shell graviton, where the graviton is not attached to any external spinning particle leg. The resulting expression is given by
\be
h^{\mu\nu}(\ell)\widetilde{\Gamma}^{(N+1)\alpha_1\cdots \alpha_N}_{\mu\nu}(\ell)
&=&\kappa \ (2\pi)^4\delta^{(4)}(p_1+\cdots+p_N+\ell)\sum_{i=1}^{N}\Bigg[\delta_{\beta_i}^{\alpha_i}\ h(\ell)-\delta_{\beta_i}^{\alpha_i}\  h^{\mu\nu}(\ell)p_{i\mu}\f{\p }{\p p_i^{\nu}}\nn\\
&& +h^{\mu\nu}(\ell)(\Sigma_{i\nu b})_{\beta_i}^{\ \alpha_i}\ \ell^b \ \f{\p }{\p p_{i}^{\mu}} -\f{1}{2}\delta_{\beta_i}^{\alpha_i}h^{\mu\nu}(\ell)\Bigg\lbrace \ell^\rho p_{i\mu}\f{\p^2}{\p p_i^\rho \p p_i^\nu}+\ell^\rho p_{i\nu}\f{\p^2}{\p p_i^\rho \p p_i^\mu}\nn\\
&&-p_i.\ell \f{\p^2}{\p p_i^\mu \p p_i^\nu }\Bigg\rbrace+\mathcal{O}(\ell^2) \Bigg]\widehat{\Gamma}^{(N)\alpha_1\cdots \alpha_{i-1}\beta_i\alpha_{i+1}\cdots \alpha_N}\ .\label{Gamma_tilde_appendix}
\ee
Above $\widehat{\Gamma}^{(N)}$ is defined after stripping out the momentum conserving delta function $(2\pi)^4\delta^{(4)}(p_1+\cdots+p_N)$ from the expression of $\Gamma^{(N)}$, i.e. $\Gamma^{(N)}\equiv (2\pi)^4\delta^{(4)}(p_1+\cdots+p_N)\widehat{\Gamma}^{(N)}$. Now starting from the above covariantized expression, the goal is to express $\widetilde{\Gamma}^{(N+1)}_{\mu\nu}(\ell)$ in terms of some operator operating on $\Gamma^{(N)}$ up to linear order in $\ell$. To do that we mostly follow the analysis of \cite{1706.00759} with the only difference being $h^{\mu\nu}(\ell)$ is an off-shell graviton so we can not impose traceless or transverse condition. Instead in de Donder gauge, we use $\ell_\mu h^{\mu\nu}(\ell)=\f{1}{2}\ell^\nu h(\ell)$ in the intermediate stages of calculation.

Let us first analyze the first two terms within the square bracket in \eqref{Gamma_tilde_appendix} which is defined as
\be
J_1 &\equiv & \kappa \ (2\pi)^4\delta^{(4)}(p_1+\cdots+p_N+\ell)\Bigg[ h(\ell)-\ h^{\mu\nu}(\ell)\sum_{i=1}^{N} p_{i\mu}\f{\p }{\p p_i^{\nu}}\Bigg] \widehat{\Gamma}^{(N)\alpha_1\cdots \alpha_N}\non\\
&=&\kappa \ (2\pi)^4\delta^{(4)}(p_1+\cdots+p_N)\Bigg[ h(\ell)-\ h^{\mu\nu}(\ell)\sum_{i=1}^{N} p_{i\mu}\f{\p }{\p p_i^{\nu}}\Bigg] \widehat{\Gamma}^{(N)\alpha_1\cdots \alpha_N}\non\\
&&+\kappa \ (2\pi)^4\ \ell^\rho \Big\lbrace \f{\p }{\p \ell^\rho}\delta^{(4)}(p_1+\cdots+p_N+\ell)\Big\rbrace_{\ell=0} \Bigg[ h(\ell)-\ h^{\mu\nu}(\ell)\sum_{i=1}^{N} p_{i\mu}\f{\p }{\p p_i^{\nu}}\Bigg] \widehat{\Gamma}^{(N)\alpha_1\cdots \alpha_N}\non\\
&&+\mathcal{O}(\ell^2)\ .
\ee
Above we have just Taylor expanded the delta function and kept terms up to linear order in $\ell$. Now to evaluate the second line above we use the property involving momentum conserving delta function
\be
&&\sum_{i=1}^{N}p_{i\mu} \f{\p }{\p p_i^\nu}\delta^{(4)}(p_1+\cdots+p_N)\ =\sum_{i=1}^{N}p_{i\mu} \f{\p }{\p p_1^\nu}\delta^{(4)}(p_1+\cdots+p_N)\non\\
&=&\f{\p }{\p p_1^\nu}\sum_{i=1}^{N}p_{i\mu} \delta^{(4)}(p_1+\cdots+p_N)-\eta_{\mu\nu}\ \delta^{(4)}(p_1+\cdots+p_N)\non\\
&=& -\eta_{\mu\nu}\ \delta^{(4)}(p_1+\cdots+p_N)\ .
\ee
So using the above property and commuting the delta function through the momentum derivative the expression of $J_1$ becomes,
\be
J_1 &=& -\kappa \ (2\pi)^4\ h^{\mu\nu}(\ell)\sum_{i=1}^{N} p_{i\mu}\f{\p }{\p p_i^{\nu}}\Big\lbrace \delta^{(4)}(p_1+\cdots+p_N) \widehat{\Gamma}^{(N)\alpha_1\cdots \alpha_N}\Big\rbrace\non\\
&&+\kappa \ (2\pi)^4\ \ell^\rho \Big\lbrace \f{\p }{\p \ell^\rho}\delta^{(4)}(p_1+\cdots+p_N+\ell)\Big\rbrace_{\ell=0} \Bigg[ h(\ell)-\ h^{\mu\nu}(\ell)\sum_{i=1}^{N} p_{i\mu}\f{\p }{\p p_i^{\nu}}\Bigg] \widehat{\Gamma}^{(N)\alpha_1\cdots \alpha_N}\non\\
&&+\mathcal{O}(\ell^2)\ .\label{J1_alternative}
\ee
Last two terms within the square bracket in \eqref{Gamma_tilde_appendix} can be evaluated in the following way
\be
J_2&\equiv & \kappa \ (2\pi)^4\delta^{(4)}(p_1+\cdots+p_N+\ell)h^{\mu\nu}(\ell)\sum_{i=1}^{N}\Bigg[(\Sigma_{i\nu b})_{\beta_i}^{\ \alpha_i}\ \ell^b \ \f{\p }{\p p_{i}^{\mu}}-\f{1}{2}\delta_{\beta_i}^{\alpha_i}\Bigg\lbrace \ell^\rho p_{i\mu}\f{\p^2}{\p p_i^\rho \p p_i^\nu}\non\\
&&+\ell^\rho p_{i\nu}\f{\p^2}{\p p_i^\rho \p p_i^\mu}-p_i.\ell \f{\p^2}{\p p_i^\mu \p p_i^\nu }\Bigg\rbrace \Bigg]\widehat{\Gamma}^{(N)\alpha_1\cdots \alpha_{i-1}\beta_i\alpha_{i+1}\cdots \alpha_N}\non\\
&=&\kappa \ (2\pi)^4 h^{\mu\nu}(\ell)\sum_{i=1}^{N}\Bigg[(\Sigma_{i\nu b})_{\beta_i}^{\ \alpha_i}\ \ell^b \ \f{\p }{\p p_{i}^{\mu}}-\f{1}{2}\delta_{\beta_i}^{\alpha_i}\Bigg\lbrace \ell^\rho p_{i\mu}\f{\p^2}{\p p_i^\rho \p p_i^\nu}+\ell^\rho p_{i\nu}\f{\p^2}{\p p_i^\rho \p p_i^\mu}-p_i.\ell \f{\p^2}{\p p_i^\mu \p p_i^\nu }\Bigg\rbrace \Bigg]\non\\
&& \Big\lbrace \delta^{(4)}(p_1+\cdots+p_N+\ell) \widehat{\Gamma}^{(N)\alpha_1\cdots \alpha_{i-1}\beta_i\alpha_{i+1}\cdots \alpha_N}\Big\rbrace\ +\ \mathcal{L}\ , \label{J_2_final}
\ee
where
\be
\mathcal{L}&=& -\kappa \ (2\pi)^4\ \widehat{\Gamma}^{(N)\alpha_1\cdots \alpha_{i-1}\beta_i\alpha_{i+1}\cdots \alpha_N} h^{\mu\nu}(\ell)\sum_{i=1}^{N}\Bigg[(\Sigma_{i\nu b})_{\beta_i}^{\ \alpha_i}\ \ell^b \ \f{\p }{\p p_{i}^{\mu}}\delta^{(4)}(p_1+\cdots+p_N+\ell)\non\\
&&-\f{1}{2}\delta_{\beta_i}^{\alpha_i}\Bigg\lbrace \ell^\rho p_{i\mu}\f{\p^2}{\p p_i^\rho \p p_i^\nu}+\ell^\rho p_{i\nu}\f{\p^2}{\p p_i^\rho \p p_i^\mu}-p_i.\ell \f{\p^2}{\p p_i^\mu \p p_i^\nu }\Bigg\rbrace \delta^{(4)}(p_1+\cdots+p_N+\ell)\Bigg]\non\\
&&+\f{\kappa}{2}(2\pi)^4 h^{\mu\nu}(\ell)\sum_{i=1}^{N}\Big\lbrace \f{\p}{\p p_i^\rho}\delta^{(4)}(p_1+\cdots+p_N+\ell)\Big\rbrace \f{\p \widehat{\Gamma}^{\alpha_1\cdots \alpha_N}}{\p p_i^\sigma}\Big\lbrace \ell^\rho p_{i\mu}\delta_\nu^\sigma +\ell^\sigma p_{i\mu}\delta_\nu^\rho\non\\
&&+\ell^\rho p_{i\nu}\delta_\mu^\sigma +\ell^\sigma p_{i\nu}\delta_\mu^\rho -p_i.\ell \delta^\rho_\mu \delta^\sigma_\nu -p_i.\ell \delta^\sigma_\mu \delta^\rho_\nu \Big\rbrace\ .
\ee
To evaluate $\mathcal{L}$ we use the same trick as described earlier i.e. derivative w.r.t. $p_i$ on the delta function is same as derivative w.r.t. $\ell$ and then use the momentum conservation relation enforced by the delta function. After all these steps we get
\be
\mathcal{L}&=&-\kappa \ (2\pi)^4\ \widehat{\Gamma}^{(N)\alpha_1\cdots \alpha_{i-1}\beta_i\alpha_{i+1}\cdots \alpha_N}h^{\mu\nu}(\ell)\Bigg[\sum_{i=1}^{N}(\Sigma_{i\nu b})_{\beta_i}^{\ \alpha_i}\ \ell^b \ \f{\p }{\p \ell^{\mu}}\delta^{(4)}(p_1+\cdots+p_N+\ell)\non\\
&&+\f{1}{2}\delta_{\beta_i}^{\alpha_i}\Bigg\lbrace \Bigg(\ell^\rho \f{\p^2}{\p \ell^\rho \p \ell^\nu}\ell_\mu +\ell^\rho \f{\p^2}{\p \ell^\rho \p \ell^\mu}\ell_\nu -\ell^\rho \f{\p^2}{\p \ell^\mu \p \ell^\nu }\ell_\rho\Bigg)\delta^{(4)}(p_1+\cdots+p_N+\ell)\Bigg\rbrace \Bigg]\non\\
&&+\kappa(2\pi)^4\sum_{i=1}^{N}\Big\lbrace \f{\p}{\p \ell^\rho}\delta^{(4)}(p_1+\cdots+p_N+\ell)\Big\rbrace \f{\p \widehat{\Gamma}^{\alpha_1\cdots \alpha_N}}{\p p_i^\sigma}\Big\lbrace  p_{i\mu}\ell^\rho h^{\mu\sigma}(\ell)+p_{i\mu}\ell^\sigma h^{\mu\rho}(\ell) \non\\
&& -p_i.\ell h^{\rho\sigma}(\ell)\Big\rbrace\ . \label{L_expression}
\ee
To evaluate the first term within the square bracket above we use the conservation of total angular momenta follows from the Lorentz covariance of $\widehat{\Gamma}$
\be
\sum_{i=1}^{N}\Big[(\Sigma_{i\nu b})_{\beta_i}^{\ \alpha_i}\widehat{\Gamma}^{(N)\alpha_1\cdots \alpha_{i-1}\beta_i\alpha_{i+1}\cdots \alpha_N}-\Big\lbrace p_{i\nu}\f{\p}{\p p_i^b}-p_{i b}\f{\p }{\p p_i^\nu}\Big\rbrace \widehat{\Gamma}^{(N)\alpha_1\cdots  \alpha_N}\Big]\ =0\ .
\ee
Using the above relation in the first line of the expression of $\mathcal{L}$ in \eqref{L_expression}, operating the derivatives w.r.t. $\ell$ in the second term within the square bracket in \eqref{L_expression} and using de Donder gauge condition, we find 
\be
\mathcal{L}&=&-\f{\kappa}{2} \ (2\pi)^4  \widehat{\Gamma}^{\alpha_1\cdots \alpha_N}\Big\lbrace h(\ell)\ell^\nu \ell^\rho \f{\p^2}{\p \ell^\rho \p \ell^\nu}-\ell^2 h^{\mu\nu}(\ell)\f{\p^2}{\p \ell^\mu \p \ell^\nu }+2h(\ell)\ell^\rho \f{\p }{\p \ell^\rho}\Big\rbrace \delta^{(4)}(p_1+\cdots +p_N+\ell)\non\\
&&+\kappa(2\pi)^4\sum_{i=1}^{N}\Big\lbrace \f{\p}{\p \ell^\rho}\delta^{(4)}(p_1+\cdots+p_N+\ell)\Big\rbrace \f{\p \widehat{\Gamma}^{\alpha_1\cdots \alpha_N}}{\p p_i^\sigma}  p_{i\mu}\ell^\rho h^{\mu\sigma}(\ell)\ .\label{L_final}
\ee
Now after substituting the expression of $\mathcal{L}$ from \eqref{L_final} in \eqref{J_2_final}, then Taylor expanding the delta function and keeping terms up to linear in $\ell$ we get
\be
J_2&=&\ \kappa \ (2\pi)^4 h^{\mu\nu}(\ell)\sum_{i=1}^{N}\Bigg[(\Sigma_{i\nu b})_{\beta_i}^{\ \alpha_i}\ \ell^b \ \f{\p }{\p p_{i}^{\mu}}-\f{1}{2}\delta_{\beta_i}^{\alpha_i}\Bigg\lbrace \ell^\rho p_{i\mu}\f{\p^2}{\p p_i^\rho \p p_i^\nu}+\ell^\rho p_{i\nu}\f{\p^2}{\p p_i^\rho \p p_i^\mu}-p_i.\ell \f{\p^2}{\p p_i^\mu \p p_i^\nu }\Bigg\rbrace \Bigg]\non\\
&& \Big\lbrace \delta^{(4)}(p_1+\cdots+p_N) \widehat{\Gamma}^{(N)\alpha_1\cdots \alpha_{i-1}\beta_i\alpha_{i+1}\cdots \alpha_N}\Big\rbrace\ \non\\
&&-\kappa  \ (2\pi)^4  \widehat{\Gamma}^{\alpha_1\cdots \alpha_N}h(\ell)\ell^\rho \Big\lbrace\f{\p }{\p \ell^\rho} \delta^{(4)}(p_1+\cdots +p_N+\ell)\Big\rbrace_{\ell=0}\non\\
&&+\kappa(2\pi)^4\sum_{i=1}^{N}\Big\lbrace \f{\p}{\p \ell^\rho}\delta^{(4)}(p_1+\cdots+p_N+\ell)\Big\rbrace_{\ell=0} \f{\p \widehat{\Gamma}^{\alpha_1\cdots \alpha_N}}{\p p_i^\sigma}  p_{i\mu}\ell^\rho h^{\mu\sigma}(\ell)\ +\mathcal{O}(\ell^2)\ .\label{J2_alternative}
\ee
Finally after adding the expressions in \eqref{J1_alternative} and \eqref{J2_alternative} we get
\be
&&h^{\mu\nu}(\ell)\widetilde{\Gamma}^{(N+1)\alpha_1\cdots \alpha_N}_{\mu\nu}(\ell)\non\\
&=& -\kappa \ (2\pi)^4\ h^{\mu\nu}(\ell)\sum_{i=1}^{N} p_{i\mu}\f{\p }{\p p_i^{\nu}}\Big\lbrace \delta^{(4)}(p_1+\cdots+p_N) \widehat{\Gamma}^{(N)\alpha_1\cdots \alpha_N}\Big\rbrace\non\\
&&+\kappa \ (2\pi)^4 h^{\mu\nu}(\ell)\sum_{i=1}^{N}\Bigg[(\Sigma_{i\nu b})_{\beta_i}^{\ \alpha_i}\ \ell^b \ \f{\p }{\p p_{i}^{\mu}}-\f{1}{2}\delta_{\beta_i}^{\alpha_i}\Bigg\lbrace \ell^\rho p_{i\mu}\f{\p^2}{\p p_i^\rho \p p_i^\nu}+\ell^\rho p_{i\nu}\f{\p^2}{\p p_i^\rho \p p_i^\mu}-p_i.\ell \f{\p^2}{\p p_i^\mu \p p_i^\nu }\Bigg\rbrace \Bigg]\non\\
&& \Big\lbrace \delta^{(4)}(p_1+\cdots+p_N) \widehat{\Gamma}^{(N)\alpha_1\cdots \alpha_{i-1}\beta_i\alpha_{i+1}\cdots \alpha_N}\Big\rbrace\ +\mathcal{O}(\ell^2)\ .\label{tilde-gamma-final-appendix}
\ee
\section{Intermediate steps in deriving soft graviton theorem }\label{S:identities_gr}
We want to compute the following expression involving the vertex in \eqref{Gamma3_graviton} which takes the following form once we expanded in power of small $\ell$ and keep terms up to quadratic order
\begingroup
\allowdisplaybreaks
\be
&&\Gamma^{(3)}_{\mu\nu}(q,-q-\ell,\ell)\Xi(-q-\ell)\non\\
&=&\ i\kappa \Bigg[ \eta_{\mu\nu}\mathcal{K}(-q)\Xi(-q)+\eta_{\mu\nu}\mathcal{K}(-q)\ell^\rho \f{\p\Xi(-q)}{\p q^\rho}+\f{1}{2}\eta_{\mu\nu}\mathcal{K}(-q)\ell^\rho\ell^\sigma \f{\p^2\Xi(-q)}{\p q^\rho \p q^\sigma}\non\\
&&+\f{1}{2}\eta_{\mu\nu}\ell^\rho \f{\p \mathcal{K}(-q)}{\p q^\rho}\Xi(-q)+\f{1}{2}\eta_{\mu\nu}\ell^\rho\ell^\sigma \f{\p \mathcal{K}(-q)}{\p q^\rho}\f{\p \Xi(-q)}{\p q^\sigma}+\f{1}{4}\eta_{\mu\nu}\ell^\rho \ell^\sigma \f{\p^2 \mathcal{K}(-q)}{\p q^\rho \p q^\sigma}\Xi(-q)\non\\
&&- q_{(\mu} \f{\p \mathcal{K}(-q)}{\p q^{\nu )}}\Xi(-q)- q_{(\mu}\ell^\rho \f{\p \mathcal{K}(-q)}{\p q^{\nu )}}\f{\p\Xi(-q)}{\p q^\rho}- \f{1}{2}q_{(\mu}\ell^\rho \ell^\sigma\f{\p \mathcal{K}(-q)}{\p q^{\nu )}}\f{\p^2\Xi(-q)}{\p q^\rho\p q^\sigma}\non\\
&&-\f{1}{2}\ell_{(\mu} \f{\p \mathcal{K}(-q)}{\p q^{\nu)}}\Xi(-q)-\f{1}{2}\ell_{(\mu}\ell^\rho \f{\p \mathcal{K}(-q)}{\p q^{\nu)}}\f{\p \Xi(-q)}{\p q^\rho}-\f{1}{2}q_{(\mu}\ell^\rho \f{\p^2 \mathcal{K}(-q)}{\p q^{\nu)} \p q^\rho}\Xi(-q)\non\\
&&-\f{1}{2}q_{(\mu}\ell^\rho \ell^\sigma\f{\p^2 \mathcal{K}(-q)}{\p q^{\nu)} \p q^\rho}\f{\p \Xi(-q)}{\p q^\sigma} -\f{1}{4}q_{(\mu}\ell^\rho \ell^\sigma \f{\p^3 \mathcal{K}(-q)}{\p q^{\nu)} \p q^\rho \p q^\sigma}\Xi(-q)\non\\
&&-\f{1}{2}\ell_{(\mu} \ell^\rho \f{\p^2 \mathcal{K}(-q)}{\p q^{\nu)} \p q^\rho}\Xi(-q)-\f{1}{2}\ell^b\f{\p \mathcal{K}(-q)}{\p q^{(\mu}}\Sigma_{\nu )b}\Xi(-q)-\f{1}{2}\ell^b\ell^\rho\f{\p \mathcal{K}(-q)}{\p q^{(\mu}}\Sigma_{\nu )b}\f{\p \Xi(-q)}{\p q^\rho} \non\\
&&+\f{1}{2}\ell^b \ \Sigma^T_{(\nu b} \f{\p \mathcal{K}(-q)}{\p q^{\mu)}}\Xi(-q)+\f{1}{2}\ell^b \ell^\rho\ \Sigma^T_{(\nu b} \f{\p \mathcal{K}(-q)}{\p q^{\mu)}}\f{\p \Xi(-q)}{\p q^\rho} -\f{1}{4}\ell^b \ell^\rho\ \f{\p^2 \mathcal{K}(-q)}{\p q^{(\mu} \p q^\rho}\Sigma_{\nu) b}\Xi(-q)\non\\
&& +\f{1}{4}\ell^b \ell^\rho \Sigma^T_{(\nu b}\f{\p^2 \mathcal{K}(-q)}{\p q^{\mu)} \p q^\rho}\Xi(-q)+\f{1}{4}\Bigg\lbrace \ell_{\mu} \ell^\rho \f{\p^2 \mathcal{K}(-q)}{\p q^\rho \p q^{\nu}}+\ell_{\nu} \ell^\rho \f{\p^2 \mathcal{K}(-q)}{\p q^\rho \p q^{\mu}}-\ell^2   \f{\p^2 \mathcal{K}(-q)}{\p q^\mu \p q^\nu}\Bigg\rbrace \Xi(-q) \non\\
&&+\f{1}{12}\ell^\sigma\Bigg\lbrace q_{\mu} \ell^\rho \f{\p^3 \mathcal{K}(-q)}{\p q^\rho \p q^{\nu} \p q^\sigma}+q_{\nu} \ell^\rho \f{\p^3 \mathcal{K}(-q)}{\p q^\rho \p q^{\mu} \p q^\sigma}-q.\ell   \f{\p^3 \mathcal{K}(-q)}{\p q^\mu \p q^\nu \p q^\sigma}\Bigg\rbrace \Xi(-q)\non\\
&& -\ell^\rho \ell^\sigma \Big\lbrace \mathcal{G}_{(\mu\rho\sigma \nu)}(-q)+\mathcal{G}_{\sigma(\nu\mu)\rho}(-q)-\mathcal{G}_{(\mu\rho\nu)\sigma }(-q) -\mathcal{G}_{\sigma(\nu\rho\mu)}(-q)\Big\rbrace \Xi(-q)+\mathcal{O}(\ell^3)\Bigg].
\ee
\endgroup
Now to simplify the above expression we need to move the momenta derivatives from $\mathcal{K}$ to $\Xi$ as much possible and also move the spin operator to the extreme right using the identities \eqref{K1}-\eqref{SigmaXi}. Following these steps the result up to order $\mathcal{O}(\ell^2)$ reads
\begingroup
\allowdisplaybreaks
\be
&&\Gamma^{(3)}_{\mu\nu}(q,-q-\ell,\ell)\Xi(-q-\ell)\non\\
&=&\ i\kappa \Bigg[ i\eta_{\mu\nu}(q^2+m^2+q.\ell)-2iq_\mu q_\nu  -2iq_{(\mu}\ell_{\nu)} +2i\ell^b q_{(\mu}\Sigma^T_{\nu)b}+i\ell^b \ell_{(\mu}\Sigma^T_{\nu)b}\non\\
&&  +q_{(\mu}\mathcal{K}(-q)\f{\p \Xi(-q)}{\p q^{\nu)}}+\f{1}{4}\eta_{\mu\nu}\mathcal{K}(-q)\ell^\rho\ell^\sigma \f{\p^2\Xi(-q)}{\p q^\rho \p q^\sigma}\non +\f{1}{2}q.\ell \mathcal{K}(-q)\f{\p^2\Xi(-q)}{\p q^\mu \p q^\nu} \non\\
&&+\ell_{(\mu} \mathcal{K}(-q)\f{\p \Xi(-q)}{\p q^{\nu)}}-\f{1}{2}\ell^b \mathcal{K}(-q)\f{\p\Xi(-q)}{\p q^{(\mu}}\Sigma^T_{\nu)b} -\f{1}{4}\ell^\rho \ell^b \mathcal{K}(-q)\f{\p^2 \Xi(-q)}{\p q^{(\mu}\p q^\rho}\Sigma^T_{\nu)b}\non\\
&& -\f{1}{2}\ell^b \  \mathcal{K}(-q)\f{\p \Xi(-q)}{\p q^{(\mu}}\Sigma^T_{\nu) b}  +\f{1}{4}\ell^b \ell^\rho \mathcal{K}(-q)\Sigma_{(\nu b}\f{\p^2 \Xi(-q)}{\p q^{\mu)} \p q^\rho}+\f{1}{4}\ell^2 \mathcal{K}(-q)\f{\p^2 \Xi(-q)}{\p q^\mu \p q^\nu} \non\\
&&+\f{1}{12}\ell^\sigma\Bigg\lbrace q_{(\mu} \ell^\rho\mathcal{K}(-q) \f{\p^3 \Xi(-q)}{\p q^\rho \p q^{\nu)} \p q^\sigma}+q.\ell \mathcal{K}(-q)  \f{\p^3 \Xi(-q)}{\p q^\mu \p q^\nu \p q^\sigma}\Bigg\rbrace \non\\
&& +\f{1}{3}q.\ell \ell^\rho \f{\p \mathcal{K}(-q)}{\p q^\rho} \f{\p^2 \Xi(-q)}{\p q^\mu \p q^\nu} +\f{1}{3}\ell^b \ell^\rho q_{(\mu}\f{\p\mathcal{K}(-q)}{\p q^{\nu)}}\f{\p^2\Xi(-q)}{\p q^\rho \p q^b}-\f{1}{3}q.\ell \ell^\rho \f{\p \mathcal{K}(-q)}{\p q^{(\mu}}\f{\p^2\Xi(-q)}{\p q^{\nu)}\p q^\rho}\non\\
&&-\f{1}{3}\ell^b \ell^\rho q_{(\mu}\f{\p \mathcal{K}(-q)}{\p q^\rho}\f{\p^2 \Xi(-q)}{\p q^{\nu)}\p q^b}   +\f{1}{6}q.\ell \ell^\rho \f{\p^2 \mathcal{K}(-q)}{\p q^{(\mu}\p q^\rho}\f{\p \Xi(-q)}{\p q^{\nu)}}  -\f{1}{6}\ell^b\ell^\rho q_{(\mu}\f{\p^2 \mathcal{K}(-q)}{\p q^\rho \p q^b}\f{\p \Xi(-q)}{\p q^{\nu)}}\non\\
&&+\f{1}{6}q_{(\mu}\ell^\rho \ell^\sigma \f{\p^2\mathcal{K}(-q)}{\p q^\rho \p q^{\nu)}}\f{\p \Xi(-q)}{\p q^\sigma}-\f{1}{6}q.\ell \ell^\rho \f{\p^2 \mathcal{K}(-q)}{\p q^\mu \p q^\nu}\f{\p \Xi(-q)}{\p q^\rho}\non\\
&&-\f{1}{2}\ell^b \ell^\rho\  \f{\p \mathcal{K}(-q)}{\p q^{\rho}}\f{\p \Xi(-q)}{\p q^{(\mu}}\Sigma^T_{\nu) b} +\f{1}{2}\ell^b\ell^\rho\f{\p \mathcal{K}(-q)}{\p q^{(\mu}}\f{\p \Xi(-q)}{\p q^\rho}\Sigma^T_{\nu )b}\non\\
&& -\ell^\rho \ell^\sigma \Big\lbrace \mathcal{G}_{(\mu\rho\sigma \nu)}(-q)+\mathcal{G}_{\sigma(\nu\mu)\rho}(-q)-\mathcal{G}_{(\mu\rho\nu)\sigma }(-q) -\mathcal{G}_{\sigma(\nu\rho\mu)}(-q)\Big\rbrace \Xi(-q)+\mathcal{O}(\ell^3)\Bigg].\label{Gamma3Xi_gravity_intermediate}
\ee
\endgroup
The symmetrization in the exchange between $\mu$ and $\nu$ can be omitted since any contraction involving the above expression in any Feynman diagram calculation will always exhibit symmetry under the exchange of $\mu$ and $\nu$. Also the above expression can be written in a compact way by introducing the following tensor structure
\be
\Delta_{\mu\rho\nu\sigma}(-q)&\equiv & \f{1}{3}q_\mu \f{\p\mathcal{K}(-q)}{\p q^\nu}\f{\p^2\Xi(-q)}{\p q^\rho \p q^\sigma}-\f{1}{6}q_\rho \f{\p^2\mathcal{K}(-q)}{\p q^\mu \p q^\nu}\f{\p \Xi(-q)}{\p q^\sigma}+\f{1}{4}\f{\p \mathcal{K}(-q)}{\p q^\mu}\f{\p\Xi(-q)}{\p q^\rho}\Sigma_{\nu\sigma}^{T}\non\\
&&\ +\ \mathcal{G}_{\mu\rho\nu\sigma}(-q)\Xi(-q)\ .\label{Delta}
\ee
Hence removing the $\mu\leftrightarrow\nu$ symmetrization, the expression in \eqref{Gamma3Xi_gravity_intermediate} can be compactly written as
\begingroup
\allowdisplaybreaks
\be 
&&\Gamma^{(3)}_{\mu\nu}(q,-q-\ell,\ell)\Xi(-q-\ell)\non\\
&=&\ i\kappa \Bigg[ i\eta_{\mu\nu}(q^2+m^2+q.\ell)-2iq_\mu q_\nu  -2iq_{\mu}\ell_{\nu} +2i\ell^b q_{\mu}\Sigma^T_{\nu b}+i\ell^b \ell_{\mu}\Sigma^T_{\nu b}\non\\
&&  +q_{\mu}\mathcal{K}(-q)\f{\p \Xi(-q)}{\p q^{\nu}}+\f{1}{4}\eta_{\mu\nu}\mathcal{K}(-q)\ell^\rho\ell^\sigma \f{\p^2\Xi(-q)}{\p q^\rho \p q^\sigma} +\f{1}{2}q.\ell \mathcal{K}(-q)\f{\p^2\Xi(-q)}{\p q^\mu \p q^\nu} \non\\
&&+\ell_{\mu} \mathcal{K}(-q)\f{\p \Xi(-q)}{\p q^{\nu}}-\ell^b \mathcal{K}(-q)\f{\p\Xi(-q)}{\p q^{\mu}}\Sigma^T_{\nu b} -\f{1}{4}\ell^\rho \ell^b \mathcal{K}(-q)\f{\p^2 \Xi(-q)}{\p q^{\mu}\p q^\rho}\Sigma^T_{\nu b}\non\\
&&   +\f{1}{4}\ell^b \ell^\rho \mathcal{K}(-q)\Sigma_{\nu b}\f{\p^2 \Xi(-q)}{\p q^{\mu} \p q^\rho}+\f{1}{4}\ell^2 \mathcal{K}(-q)\f{\p^2 \Xi(-q)}{\p q^\mu \p q^\nu} \non\\
&&+\f{1}{12}\ell^\sigma\Bigg\lbrace q_{\mu} \ell^\rho\mathcal{K}(-q) \f{\p^3 \Xi(-q)}{\p q^\rho \p q^{\nu} \p q^\sigma}+q.\ell \mathcal{K}(-q)  \f{\p^3 \Xi(-q)}{\p q^\mu \p q^\nu \p q^\sigma}\Bigg\rbrace \non\\
&&+\ell^\rho\ell^\sigma \Big\lbrace \Delta_{\mu\rho\nu\sigma}(-q)+\Delta_{\rho\mu\sigma\nu}(-q)-\Delta_{\rho\mu\nu\sigma}(-q)-\Delta_{\mu\rho\sigma\nu}(-q)\Big\rbrace +\mathcal{O}(\ell^3)\Bigg]. \label{Gamma3Xi_gravity}
\ee
\endgroup
Now we want to compute the following expression involving two vertices of kind \eqref{Gamma3_graviton} which takes the following form once we expanded in power of small $\ell_1,\ell_2$ and keep terms up to quadratic order
\begingroup
\allowdisplaybreaks
\be 
&&\Gamma^{(3)}_{\mu\nu}(q,-q-\ell_1,\ell_1)\Xi(-q-\ell_1)\Gamma^{(3)}_{\rho\sigma}(q+\ell_1,-q-\ell_1-\ell_2,\ell_2)\Xi(-q-\ell_1-\ell_2)\non\\
&=& -\kappa^2 \Bigg[ i\eta_{\mu\nu}(q^2+m^2+q.\ell_1)-2iq_\mu q_\nu  -2iq_{\mu}\ell_{1\nu} +2i\ell_1^b q_{\mu}\Sigma^T_{\nu b}+i\ell_1^b \ell_{1\mu}\Sigma^T_{\nu b}\non\\
&&  +q_{\mu}\mathcal{K}(-q)\f{\p \Xi(-q)}{\p q^{\nu}}+\f{1}{4}\eta_{\mu\nu}\mathcal{K}(-q)\ell_1^\lambda\ell_1^b \f{\p^2\Xi(-q)}{\p q^\lambda \p q^b} +\f{1}{2}q.\ell_1 \mathcal{K}(-q)\f{\p^2\Xi(-q)}{\p q^\mu \p q^\nu} \non\\
&&+\ell_{1\mu} \mathcal{K}(-q)\f{\p \Xi(-q)}{\p q^{\nu}}-\ell_1^b \mathcal{K}(-q)\f{\p\Xi(-q)}{\p q^{\mu}}\Sigma^T_{\nu b} -\f{1}{4}\ell_1^\lambda \ell_1^b \mathcal{K}(-q)\f{\p^2 \Xi(-q)}{\p q^{\mu}\p q^\lambda}\Sigma^T_{\nu b}\non\\
&&   +\f{1}{4}\ell_1^b \ell_1^\lambda \mathcal{K}(-q)\Sigma_{\nu b}\f{\p^2 \Xi(-q)}{\p q^{\mu} \p q^\lambda}+\f{1}{4}\ell_1^2 \mathcal{K}(-q)\f{\p^2 \Xi(-q)}{\p q^\mu \p q^\nu} \non\\
&&+\f{1}{12}\ell_1^b\Bigg\lbrace q_{\mu} \ell_1^\lambda\mathcal{K}(-q) \f{\p^3 \Xi(-q)}{\p q^\lambda \p q^{\nu} \p q^b}+q.\ell_1 \mathcal{K}(-q)  \f{\p^3 \Xi(-q)}{\p q^\mu \p q^\nu \p q^b}\Bigg\rbrace \non\\
&&+\ell_1^\lambda\ell_1^b \Big\lbrace \Delta_{\mu\lambda\nu b}(-q)+\Delta_{\lambda\mu b\nu}(-q)-\Delta_{\lambda\mu\nu b}(-q)-\Delta_{\mu\lambda b\nu}(-q)\Big\rbrace\Bigg]\non\\
&&\times \Bigg[ i\eta_{\rho\sigma}(q^2+m^2+2q.\ell_1+q.\ell_2+\ell_1^2+\ell_1.\ell_2)-2i(q+\ell_1)_\rho (q+\ell_1)_\sigma  -2i(q+\ell_1)_{\rho}\ell_{2\sigma}\non\\
&& +2i\ell_2^a (q+\ell_1)_{\rho}\Sigma^T_{\sigma a}+i\ell_2^a \ell_{2\rho}\Sigma^T_{\sigma a}\non\\
&&  +(q+\ell_1)_{\rho}\mathcal{K}(-q)\f{\p \Xi(-q)}{\p q^{\sigma}}+(q+\ell_1)_\rho \ell_1^\tau \f{\p\mathcal{K}(-q)}{\p q^\tau}\f{\p\Xi(-q)}{\p q^\sigma}+(q+\ell_1)_\rho \ell_1^\tau \mathcal{K}(-q) \f{\p^2\Xi(-q)}{\p q^\sigma \p q^\tau}\non\\
&&+\f{1}{2}q_\rho \ell_1^\tau \ell_1^\kappa\Bigg\lbrace \f{\p^2 \mathcal{K}(-q)}{\p q^\tau \p q^\kappa}\f{\p\Xi(-q)}{\p q^\sigma}+2\f{\p \mathcal{K}(-q)}{\p q^\tau }\f{\p^2\Xi(-q)}{\p q^\sigma \p q^\kappa}+ \mathcal{K}(-q)\f{\p^3\Xi(-q)}{\p q^\sigma \p q^\kappa \p q^\tau}\Bigg\rbrace \non\\
&&+\f{1}{4}\eta_{\rho\sigma}\mathcal{K}(-q)\ell_2^\kappa\ell_2^\tau \f{\p^2\Xi(-q)}{\p q^\kappa \p q^\tau} +\f{1}{2}(q+\ell_1).\ell_2 \mathcal{K}(-q)\f{\p^2\Xi(-q)}{\p q^\rho \p q^\sigma} \non\\
&&+\f{1}{2} q.\ell_2 \ell_1^\tau \f{\p\mathcal{K}(-q)}{\p q^\tau}\f{\p^2\Xi(-q)}{\p q^\rho \p q^\sigma}+\f{1}{2}q.\ell_2 \ell_1^\tau\mathcal{K}(-q)\f{\p^3\Xi(-q)}{\p q^\rho \p q^\sigma \p q^\tau}\non\\
&&+\ell_{2\rho} \mathcal{K}(-q)\f{\p \Xi(-q)}{\p q^{\sigma}}+\ell_{2\rho} \ell_1^\tau \f{\p \mathcal{K}(-q)}{\p q^\tau}\f{\p \Xi(-q)}{\p q^{\sigma}}+\ell_{2\rho}\ell_1^\tau \mathcal{K}(-q)\f{\p^2 \Xi(-q)}{\p q^{\sigma} \p q^\tau}\non\\
&&-\ell_2^a \mathcal{K}(-q)\f{\p\Xi(-q)}{\p q^{\rho}}\Sigma^T_{\sigma a}-\ell_2^a \ell_1^\tau \f{\p \mathcal{K}(-q)}{\p q^\tau}\f{\p\Xi(-q)}{\p q^{\rho}}\Sigma^T_{\sigma a}-\ell_2^a \ell_1^\tau \mathcal{K}(-q)\f{\p^2\Xi(-q)}{\p q^{\rho}\p q^\tau}\Sigma^T_{\sigma a}\non\\
&& -\f{1}{4}\ell_2^\tau \ell_2^a \mathcal{K}(-q)\f{\p^2 \Xi(-q)}{\p q^{\rho}\p q^\tau}\Sigma^T_{\sigma a}  +\f{1}{4}\ell_2^a \ell_2^\tau \mathcal{K}(-q)\Sigma_{\sigma a}\f{\p^2 \Xi(-q)}{\p q^{\rho} \p q^\tau}+\f{1}{4}\ell_2^2 \mathcal{K}(-q)\f{\p^2 \Xi(-q)}{\p q^\rho \p q^\sigma} \non\\
&&+\f{1}{12}\ell_2^\tau \Bigg\lbrace q_{\rho} \ell_2^\kappa\mathcal{K}(-q) \f{\p^3 \Xi(-q)}{\p q^\kappa \p q^{\sigma} \p q^\tau}+q.\ell_2 \mathcal{K}(-q)  \f{\p^3 \Xi(-q)}{\p q^\rho \p q^\sigma \p q^\tau}\Bigg\rbrace \non\\
&&+\ell_2^\kappa\ell_2^\tau \Big\lbrace \Delta_{\rho\kappa\sigma\tau}(-q)+\Delta_{\kappa\rho\tau\sigma}(-q)-\Delta_{\kappa\rho\sigma\tau}(-q)-\Delta_{\rho\kappa\tau\sigma}(-q)\Big\rbrace\Bigg]\ .
\ee
\endgroup
In principle the above expression can be evaluated using the identities \eqref{K1}-\eqref{SigmaXi}, but it is tedious and we don't need the full contribution. Instead we only evaluate the above expression by contracting $\epsilon(-q)^{T}$ from the left considering the particle with momentum $q$ being on-shell i.e. $q^2+m^2=0$ and $\epsilon(-q)^{T}\mathcal{K}(-q)=0$. Implementing these conditions and using the identities \eqref{K1}-\eqref{SigmaXi}, the above expression simplifies to
\begingroup
\allowdisplaybreaks
\be 
&&Z_{1,\mu\nu\rho\sigma}\nn\\
&\equiv & \epsilon(-q)^T\Gamma^{(3)}_{\mu\nu}(q,-q-\ell_1,\ell_1)\Xi(-q-\ell_1)\Gamma^{(3)}_{\rho\sigma}(q+\ell_1,-q-\ell_1-\ell_2,\ell_2)\Xi(-q-\ell_1-\ell_2)\non\\
&=&  -\kappa^2 \epsilon^{T}\Bigg[ -\eta_{\mu\nu}\eta_{\rho\sigma}q.\ell_1(2q.\ell_1+q.\ell_2)+2q_\mu q_\nu \eta_{\rho\sigma}(2q.\ell_1+q.\ell_2+\ell_1^2+\ell_1.\ell_2)\non\\
&&+2q_\mu \ell_{1\nu}\eta_{\rho\sigma}(2q.\ell_1+q.\ell_2)-2\eta_{\rho\sigma}q_\mu \ell_1^b\Sigma_{\nu b}^{T}(2q.\ell_1+q.\ell_2)+2\eta_{\mu\nu}q.\ell_1 (q_\rho q_\sigma +q_\rho \ell_{1\sigma}+q_\sigma \ell_{1\rho})\non\\
&&-4q_\mu q_\nu (q+\ell_1)_\rho (q+\ell_1)_\sigma -4q_\mu \ell_{1\nu}(q_\rho q_\sigma +q_\rho \ell_{1\sigma}+q_\sigma \ell_{1\rho})+4q_\mu \ell_1^b \Sigma_{\nu b}^T(q_\rho q_\sigma +q_\rho \ell_{1\sigma}+q_\sigma \ell_{1\rho})\non\\
&&+2\ell_1^b \ell_{1\mu}\Sigma_{\nu b}^T q_\rho q_\sigma +2\eta_{\mu\nu}q.\ell_1 q_\rho \ell_{2\sigma}-4q_\mu q_\nu (q+\ell_1)_\rho \ell_{2\sigma}-4q_\mu \ell_{1\nu}q_\rho \ell_{2\sigma}+4q_\mu \ell_1^b \Sigma_{\nu b}^T q_\rho \ell_{2\sigma}\non\\
&&-2\eta_{\mu\nu}q.\ell_1 \ell_2^a q_\rho \Sigma_{\sigma a}^T +4q_\mu q_\nu \ell_2^a (q+\ell_1)_\rho \Sigma_{\sigma a}^T +4q_\mu \ell_{1\nu}\ell_2^a q_\rho \Sigma_{\sigma a}^T -4q_\mu \ell_1^b \Sigma_{\nu b}^T \ell_2^a q_\rho \Sigma_{\sigma a}^T\non\\
&&+2q_\mu q_\nu \ell_2^a \ell_{2\rho}\Sigma_{\sigma a}^T -2iq.\ell_1 q_\mu  q_\rho  \f{\p \mathcal{K}(-q)}{\p q^\nu}\f{\p\Xi(-q)}{\p q^\sigma}\non\\
&&-2iq_\rho q_\sigma \ell_1^\lambda\ell_1^b \Big\lbrace \Delta_{\mu\lambda\nu b}(-q)+\Delta_{\lambda\mu b\nu}(-q)-\Delta_{\lambda\mu\nu b}(-q)-\Delta_{\mu\lambda b\nu}(-q)\Big\rbrace\non\\
&&-2i q_\mu q_\nu \ell_2^\kappa\ell_2^\tau \Big\lbrace \Delta_{\rho\kappa\sigma\tau}(-q)+\Delta_{\kappa\rho\tau\sigma}(-q)-\Delta_{\kappa\rho\sigma\tau}(-q)-\Delta_{\rho\kappa\tau\sigma}(-q)\Big\rbrace\non\\
&&+\ell_1^\lambda\ell_1^b \Bigg\lbrace -\f{1}{6}q_\mu q_\rho \f{\p\mathcal{K}(-q)}{\p q^\nu}\f{\p\Xi(-q)}{\p q^\lambda}\f{\p\mathcal{K}(-q)}{\p q^b}\f{\p\Xi(-q)}{\p q^\sigma}-\f{1}{6}q_\mu q_\rho \f{\p\mathcal{K}(-q)}{\p q^\lambda}\f{\p\Xi(-q)}{\p q^\nu}\f{\p\mathcal{K}(-q)}{\p q^b}\f{\p\Xi(-q)}{\p q^\sigma}\non\\
&& -\f{1}{6}q_\lambda q_\rho \f{\p\mathcal{K}(-q)}{\p q^\mu}\f{\p\Xi(-q)}{\p q^b}\f{\p\mathcal{K}(-q)}{\p q^\nu}\f{\p\Xi(-q)}{\p q^\sigma}-\f{1}{6}q_\lambda q_\rho \f{\p\mathcal{K}(-q)}{\p q^b}\f{\p\Xi(-q)}{\p q^\mu}\f{\p\mathcal{K}(-q)}{\p q^\nu}\f{\p\Xi(-q)}{\p q^\sigma}\non\\
&&+\f{1}{3}q_\mu q_\rho \f{\p\mathcal{K}(-q)}{\p q^b}\f{\p\Xi(-q)}{\p q^\lambda}\f{\p\mathcal{K}(-q)}{\p q^\nu}\f{\p\Xi(-q)}{\p q^\sigma}+\f{1}{3}q_\lambda q_\rho \f{\p\mathcal{K}(-q)}{\p q^\mu}\f{\p\Xi(-q)}{\p q^b}\f{\p\mathcal{K}(-q)}{\p q^\nu}\f{\p\Xi(-q)}{\p q^\sigma}\Bigg\rbrace\non\\
&& -i\ell_{1\nu}\ell_1^b q_\mu q_\rho \f{\p\mathcal{K}(-q)}{\p q^b}\f{\p\Xi(-q)}{\p q^\sigma}+2i \eta_{\nu\sigma}q_\rho q.\ell_1 \ell_1^b \f{\p\mathcal{K}(-q)}{\p q^\mu}\f{\p\Xi(-q)}{\p q^b}-2i\ell_{1\sigma}q_\rho q.\ell_1 \f{\p\mathcal{K}(-q)}{\p q^\mu}\f{\p\Xi(-q)}{\p q^\nu} \non\\
&& -i(q.\ell_1)^2 q_\rho \f{\p^2 \mathcal{K}(-q)}{\p q^\mu \p q^\nu}\f{\p\Xi(-q)}{\p q^\sigma} +iq.\ell_1 \ell_1^b q_\nu q_\rho \f{\p^2 \mathcal{K}(-q)}{\p q^\mu \p q^b}\f{\p\Xi(-q)}{\p q^\sigma}+2i q_\rho q.\ell_1 \ell_1^b \f{\p\mathcal{K}(-q)}{\p q^\mu}\f{\p\Xi(-q)}{\p q^\sigma}\Sigma_{\nu b}^T \non\\
&&+2i q_\rho q.\ell_1 \ell_1^b q_\nu \f{\p\mathcal{K}(-q)}{\p q^\mu}\f{\p^2 \Xi(-q)}{\p q^b \p q^\sigma}-2i q_\rho (q.\ell_1)^2  \f{\p\mathcal{K}(-q)}{\p q^\mu}\f{\p^2 \Xi(-q)}{\p q^\nu \p q^\sigma}\non\\
&& -2iq.\ell_1 q_\mu  \ell_{1\rho}  \f{\p \mathcal{K}(-q)}{\p q^\nu}\f{\p\Xi(-q)}{\p q^\sigma}+i\ell_1^b \ell_{1\mu} q_\rho q_\nu \f{\p \mathcal{K}(-q)}{\p q^b}\f{\p\Xi(-q)}{\p q^\sigma}-2iq.\ell_1 \ell_{1\mu} q_\rho  \f{\p \mathcal{K}(-q)}{\p q^\nu}\f{\p\Xi(-q)}{\p q^\sigma}\non\\
&&+2i\eta_{\mu\nu}q.\ell_1 q_\rho \ell_1^\tau \f{\p \mathcal{K}(-q)}{\p q^\tau}\f{\p\Xi(-q)}{\p q^\sigma}-iq.\ell_1 q_\mu q_\rho \ell_1^\tau \f{\p^2 \mathcal{K}(-q)}{\p q^\nu \p q^\tau}\f{\p\Xi(-q)}{\p q^\sigma} -i\ell_1^2 q_\mu q_\rho \f{\p\mathcal{K}(-q)}{\p q^\nu}\f{\p\Xi(-q)}{\p q^\sigma}\non\\
&&-2iq.\ell_1 q_\mu  q_\rho \ell_1^\tau   \f{\p \mathcal{K}(-q)}{\p q^\nu}\f{\p^2\Xi(-q)}{\p q^\sigma \p q^\tau}+i\ell_1^b q_{\mu} q.\ell_2 q_\nu \f{\p \mathcal{K}(-q)}{\p q^b}\f{\p^2\Xi(-q)}{\p q^\rho \p q^\sigma}-iq.\ell_1 q_{\mu} q.\ell_2  \f{\p \mathcal{K}(-q)}{\p q^\nu}\f{\p^2\Xi(-q)}{\p q^\rho \p q^\sigma}\non\\
&&-iq_\mu q_\nu q.\ell_2 \ell_1^\tau \f{\p\mathcal{K}(-q)}{\p q^\tau}\f{\p^2\Xi(-q)}{\p q^\rho \p q^\sigma}+2i\ell_1^b q_{\mu}\ell_{2\rho} q_\nu  \f{\p \mathcal{K}(-q)}{\p q^b}\f{\p \Xi(-q)}{\p q^{\sigma}}-2iq.\ell_1 q_{\mu}\ell_{2\rho}  \f{\p \mathcal{K}(-q)}{\p q^\nu}\f{\p \Xi(-q)}{\p q^{\sigma}}\non\\
&&-2iq_\mu q_\nu \ell_{2\rho} \ell_1^\tau \f{\p \mathcal{K}(-q)}{\p q^\tau}\f{\p \Xi(-q)}{\p q^{\sigma}}-2i\ell_1^b q_{\mu}\ell_2^a q_\nu \f{\p \mathcal{K}(-q)}{\p q^b}\f{\p\Xi(-q)}{\p q^{\rho}}\Sigma^T_{\sigma a}+2iq.\ell_1 q_{\mu}\ell_2^a  \f{\p \mathcal{K}(-q)}{\p q^\nu}\f{\p\Xi(-q)}{\p q^{\rho}}\Sigma^T_{\sigma a}\non\\
&&+2iq_\mu q_\nu \ell_2^a \ell_1^\tau \f{\p \mathcal{K}(-q)}{\p q^\tau}\f{\p\Xi(-q)}{\p q^{\rho}}\Sigma^T_{\sigma a}+\mathcal{O}(\ell_{1,2}^3)\Bigg]\ . \label{eq:Z1_gr}
\ee
\endgroup
To write the above expression in a compact form we used our compact structure $\Delta$ defined in \eqref{Delta}. 

We also need to compute the following expression involving the vertex in \eqref{Gamma4_graviton} which takes the following form once we expanded in power of small $\ell_1,\ell_2$ and keep terms up to linear order
\begingroup
\allowdisplaybreaks
\be
&&\Gamma^{(4)}_{\mu\nu ,\rho\sigma}(q,-q-\ell_1-\ell_2,\ell_1,\ell_2)\Xi(-q-\ell_1-\ell_2)\non\\
&=&i\kappa^2 \Bigg[ (\eta_{\mu\nu}\eta_{\rho\sigma}-2\eta_{\mu\rho}\eta_{\nu\sigma})\Bigg\lbrace\mathcal{K}(-q)\Xi(-q)+(\ell_1+\ell_2)^\kappa \mathcal{K}(-q)\f{\p \Xi(-q)}{\p q^\kappa}+\f{1}{2}(\ell_1+\ell_2)^\kappa \f{\p \mathcal{K}(-q)}{\p q^\kappa}\Xi(-q)\Bigg\rbrace \non\\
&& -\eta_{\mu\nu}\Bigg\lbrace q_\rho \f{\p \mathcal{K}(-q)}{\p q^\sigma}\Xi(-q)+q_\rho (\ell_1+\ell_2)^\kappa\f{\p \mathcal{K}(-q)}{\p q^\sigma}\f{\p \Xi(-q)}{\p q^\kappa}+\f{1}{2}(\ell_1+\ell_2)_\rho \f{\p \mathcal{K}(-q)}{\p q^\sigma}\Xi(-q)\non\\
&&+\f{1}{2}q_\rho (\ell_1+\ell_2)^\kappa \f{\p^2 \mathcal{K}(-q)}{\p q^\sigma \p q^\kappa}\Xi(-q)+\f{1}{2}\ell_2^b \f{\p \mathcal{K}(-q)}{\p q^\rho}\Sigma_{\sigma b}\Xi(-q) -\f{1}{2}\ell_2^b \Sigma^T_{\sigma b}\f{\p \mathcal{K}(-q)}{\p q^\rho}\Xi(-q)\Bigg\rbrace\non\\
&&-\eta_{\rho\sigma}\Bigg\lbrace q_\mu \f{\p \mathcal{K}(-q)}{\p q^\nu}\Xi(-q)+q_\mu (\ell_1+\ell_2)^\kappa \f{\p \mathcal{K}(-q)}{\p q^\nu}\f{\p \Xi(-q)}{\p q^\kappa}+\f{1}{2}(\ell_1+\ell_2)_\mu \f{\p \mathcal{K}(-q)}{\p q^\nu}\Xi(-q)\non\\
&&+\f{1}{2}q_\mu (\ell_1+\ell_2)^\kappa \f{\p^2 \mathcal{K}(-q)}{\p q^\nu \p q^\kappa}\Xi(-q)+\f{1}{2}\ell_1^b \f{\p \mathcal{K}(-q)}{\p q^\mu}\Sigma_{\nu b}\Xi(-q) -\f{1}{2}\ell_1^b \Sigma^T_{\nu b}\f{\p \mathcal{K}(-q)}{\p q^\mu}\Xi(-q)\Bigg\rbrace\non\\
&&+\f{3}{2}\eta_{\mu\rho}\Bigg\lbrace q_\sigma \f{\p \mathcal{K}(-q)}{\p q^\nu}\Xi(-q)+q_\sigma (\ell_1+\ell_2)^\kappa\f{\p \mathcal{K}(-q)}{\p q^\nu}\f{\p\Xi(-q)}{\p q^\kappa}+\f{1}{2}(\ell_1+\ell_2)_\sigma \f{\p \mathcal{K}(-q)}{\p q^\nu}\Xi(-q)\non\\
&&+\f{1}{2}q_\sigma (\ell_1+\ell_2)^\kappa \f{\p^2 \mathcal{K}(-q)}{\p q^\kappa \p q^\nu}\Xi(-q)+q_\nu \f{\p \mathcal{K}(-q)}{\p q^\sigma}\Xi(-q)+q_\nu (\ell_1+\ell_2)^\kappa\f{\p \mathcal{K}(-q)}{\p q^\sigma}\f{\p \Xi(-q)}{\p q^\kappa} \non\\
&&+\f{1}{2}(\ell_1+\ell_2)_\nu \f{\p \mathcal{K}(-q)}{\p q^\sigma}\Xi(-q)+\f{1}{2}q_\nu (\ell_1+\ell_2)^\kappa \f{\p^2 \mathcal{K}(-q)}{\p q^\kappa \p q^\sigma}\Xi(-q)\Bigg\rbrace +q_\mu q_\rho \f{\p^2 \mathcal{K}(-q)}{\p q^\nu \p q^\sigma}\Xi(-q)\non\\
&& +q_\mu q_\rho (\ell_1+\ell_2)^\kappa\f{\p^2 \mathcal{K}(-q)}{\p q^\nu \p q^\sigma}\f{\p \Xi(-q)}{\p q^\kappa}+\f{1}{2}\Big\lbrace q_\mu (\ell_1 +\ell_2)_\rho +q_\rho (\ell_1+\ell_2)_\mu \Big\rbrace \f{\p^2 \mathcal{K}(-q)}{\p q^\nu \p q^\sigma}\Xi(-q)\non\\
&&+\f{1}{2}q_\mu q_\rho (\ell_1 +\ell_2)^\kappa\f{\p^3 \mathcal{K}(-q)}{\p q^\nu \p q^\sigma \p q^\kappa}\Xi(-q) -\f{1}{4}\eta_{\nu\sigma}(\ell_2-\ell_1)^\kappa \Bigg\lbrace \f{\p \mathcal{K}(-q)}{\p q^\kappa}\Sigma_{\rho\mu}-\Sigma_{\rho\mu}^{T}\f{\p \mathcal{K}(-q)}{\p q^\kappa}\Bigg\rbrace\Xi(-q) \non\\
&&+\f{1}{2}\eta_{\mu\rho}\Bigg\lbrace \ell_2^b \f{\p \mathcal{K}(-q)}{\p q^\nu}\Sigma_{\sigma b}-\ell_2^b \Sigma^T_{\sigma b}\f{\p \mathcal{K}(-q)}{\p q^\nu} +\ell_1^b \f{\p \mathcal{K}(-q)}{\p q^\sigma}\Sigma_{\nu b} -\ell_1^b \Sigma^T_{\nu b}\f{\p \mathcal{K}(-q)}{\p q^\sigma}\Bigg\rbrace\Xi(-q)\non\\
&&+\f{1}{4}q_\mu \ell_2^b \Bigg\lbrace \f{\p^2 \mathcal{K}(-q)}{\p q^\nu \p q^\rho}\Sigma_{\sigma b} -\Sigma_{\sigma b}^T \f{\p^2 \mathcal{K}(-q)}{\p q^\nu \p q^\rho}\Bigg\rbrace\Xi(-q) +\f{1}{4}q_\rho \ell_1^b \Bigg\lbrace \f{\p^2 \mathcal{K}(-q)}{\p q^\sigma \p q^\mu}\Sigma_{\nu b}-\Sigma_{\nu b}^T \f{\p^2 \mathcal{K}(-q)}{\p q^\sigma \p q^\mu}\Bigg\rbrace\Xi(-q) \non\\
&&-\f{1}{2}\eta_{\nu\sigma}\ell_2^a \Bigg\lbrace \f{\p \mathcal{K}(-q)}{\p q^\rho}\Sigma_{a\mu} -\Sigma_{a\mu}^T \f{\p \mathcal{K}(-q)}{\p q^\rho}\Bigg\rbrace\Xi(-q) -\f{1}{2}\eta_{\nu\sigma}\ell_1^a \Bigg\lbrace \f{\p \mathcal{K}(-q)}{\p q^\mu}\Sigma_{a\rho} -\Sigma_{a\rho}^T \f{\p \mathcal{K}(-q)}{\p q^\mu}\Bigg\rbrace\Xi(-q)\non\\
&& -\f{1}{2}\ell_{2\nu}\Bigg\lbrace \f{\p \mathcal{K}(-q)}{\p q^\sigma}\Sigma_{\mu\rho}-\Sigma_{\mu\rho}^T \f{\p \mathcal{K}(-q)}{\p q^\sigma}\Bigg\rbrace\Xi(-q) -\f{1}{2}\ell_{1\sigma}\Bigg\lbrace \f{\p \mathcal{K}(-q)}{\p q^\nu}\Sigma_{\rho\mu}-\Sigma_{\rho\mu}^T \f{\p \mathcal{K}(-q)}{\p q^\nu}\Bigg\rbrace \Xi(-q)\non\\
&&+\mathcal{O}(\ell_1^2, \ell_1\ell_2,\ell_2^2)\Bigg]\ .
\ee
\endgroup
Now we contract the above expression with $\epsilon(-q)^{T}$ from the left considering the particle with momentum $q$ being on-shell i.e. $q^2+m^2=0$ and $\epsilon(-q)^{T}\mathcal{K}(-q)=0$. Then using the identities \eqref{K1}-\eqref{SigmaXi}, the above expression up to order $\mathcal{O}(\ell_1,\ell_2)$ simplifies to
\begingroup
\allowdisplaybreaks
\be
&&Z_{2,\mu\nu\rho\sigma}\nn\\
&\equiv & \epsilon(-q)^T\ \Gamma^{(4)}_{\mu\nu ,\rho\sigma}(q,-q-\ell_1-\ell_2,\ell_1,\ell_2)\Xi(-q-\ell_1-\ell_2)\non\\
&=&i\kappa^2 \epsilon^{T}\Bigg[ (\eta_{\mu\nu}\eta_{\rho\sigma}-2\eta_{\mu\rho}\eta_{\nu\sigma})\Big\lbrace i(q^2+m^2)+iq.(\ell_1 +\ell_2)\Big\rbrace    -\eta_{\mu\nu}\Big\lbrace 2iq_\rho q_\sigma +i(\ell_1+\ell_2)_\rho q_\sigma\nn\\
&& +i q_\rho (\ell_1+\ell_2)_\sigma  \Big\rbrace  -\eta_{\rho\sigma} \Big\lbrace 2iq_\mu q_\nu +i(\ell_1+\ell_2)_\mu q_\nu +i q_\mu (\ell_1+\ell_2)_\nu  \Big\rbrace +3\eta_{\mu\rho }\Big\lbrace 2iq_\nu q_\sigma\nn\\
&& +iq_\nu (\ell_1+\ell_2)_\sigma +iq_\sigma (\ell_1+\ell_2)_\nu\Big\rbrace +\eta_{\nu\sigma}\Big\lbrace 2iq_\mu q_\nu+i q_\mu (\ell_1 +\ell_2)_\rho +iq_\rho (\ell_1+\ell_2)_\mu \Big\rbrace \non\\
&&-i\eta_{\nu\rho}q_\mu \ell_2^b \Sigma_{\sigma b}^T-i\eta_{\sigma\mu}q_\rho \ell_1^b \Sigma_{\nu b}^T +2i\eta_{\mu\nu} \ell_2^b q_\rho\Sigma^T_{\sigma b}+2i\eta_{\rho\sigma}\ell_1^b q_\mu\Sigma^T_{\nu b} \non\\
&&+2i\eta_{\nu\sigma}q_\rho \ell_2^a \Sigma_{a\mu}^T +2i\eta_{\nu\sigma}q_\mu \ell_1^a \Sigma_{a\rho}^T +2i\ell_{2\nu}q_\sigma \Sigma_{\mu\rho}^T +2i\ell_{1\sigma}q_\nu \Sigma_{\rho\mu}^T\non\\
&& -2i\eta_{\mu\rho}q_\nu \ell_2^b \Sigma_{\sigma b}^T -2i\eta_{\mu\rho}q_\sigma\ell_1^b \Sigma_{\nu b}^T +i\eta_{\nu\sigma} q.(\ell_2-\ell_1)\Sigma^T_{\rho\mu}\non\\
&& -\eta_{\mu\nu}\Bigg\lbrace \f{1}{2}q_\rho \ell_1^\kappa\f{\p \mathcal{K}(-q)}{\p q^\sigma}\f{\p \Xi(-q)}{\p q^\kappa}-\f{1}{2}q_\rho \ell_1^\kappa\f{\p \mathcal{K}(-q)}{\p q^\kappa}\f{\p \Xi(-q)}{\p q^\sigma} \Bigg\rbrace\non\\
&& -\eta_{\rho\sigma}\Bigg\lbrace \f{1}{2}q_\mu \ell_2^\kappa\f{\p \mathcal{K}(-q)}{\p q^\nu}\f{\p \Xi(-q)}{\p q^\kappa}-\f{1}{2}q_\mu \ell_2^\kappa\f{\p \mathcal{K}(-q)}{\p q^\kappa}\f{\p \Xi(-q)}{\p q^\nu} \Bigg\rbrace\non\\
&& +\f{3}{2}\eta_{\mu\rho}\Bigg\lbrace -\f{1}{2}q_\sigma (\ell_1+\ell_2)^\kappa \f{\p \mathcal{K}(-q)}{\p q^\kappa}\f{\p \Xi(-q)}{\p q^\nu}  +\f{1}{2}q_\sigma (\ell_1+\ell_2)^\kappa \f{\p \mathcal{K}(-q)}{\p q^\nu}\f{\p \Xi(-q)}{\p q^\kappa}\non\\
&& -\f{1}{2}q_\nu (\ell_1+\ell_2)^\kappa \f{\p \mathcal{K}(-q)}{\p q^\kappa}\f{\p \Xi(-q)}{\p q^\sigma}  +\f{1}{2}q_\nu (\ell_1+\ell_2)^\kappa \f{\p \mathcal{K}(-q)}{\p q^\sigma}\f{\p \Xi(-q)}{\p q^\kappa}\Bigg\rbrace\non\\
&& -q_\mu q_\rho \f{\p \mathcal{K}(-q)}{\p q^\nu}\f{\p \Xi(-q)}{\p q^\sigma}-q_\mu q_\rho \f{\p \mathcal{K}(-q)}{\p q^\sigma}\f{\p \Xi(-q)}{\p q^\nu}  +q_\mu q_\rho (\ell_1+\ell_2)^\kappa\f{\p^2 \mathcal{K}(-q)}{\p q^\nu \p q^\sigma}\f{\p \Xi(-q)}{\p q^\kappa}\non\\
&&-\f{1}{2}\Big\lbrace q_\mu (\ell_1 +\ell_2)_\rho +q_\rho (\ell_1+\ell_2)_\mu \Big\rbrace \Bigg\lbrace  \f{\p \mathcal{K}(-q)}{\p q^\nu}\f{\p \Xi(-q)}{\p q^\sigma}+ \f{\p \mathcal{K}(-q)}{\p q^\sigma}\f{\p \Xi(-q)}{\p q^\nu}\Bigg\rbrace \non\\
&&-\f{1}{2}q_\mu q_\rho (\ell_1 +\ell_2)^\kappa \Bigg\lbrace\f{\p^2 \mathcal{K}(-q)}{\p q^\nu \p q^\sigma }\f{\p \Xi(-q)}{\p q^\kappa}+\f{\p^2 \mathcal{K}(-q)}{\p q^\nu \p q^\kappa }\f{\p \Xi(-q)}{\p q^\sigma}+\f{\p^2 \mathcal{K}(-q)}{\p q^\kappa \p q^\sigma }\f{\p \Xi(-q)}{\p q^\nu}\non\\
&& + \f{\p \mathcal{K}(-q)}{\p q^\nu }\f{\p^2 \Xi(-q)}{\p q^\kappa \p q^\sigma}+\f{\p \mathcal{K}(-q)}{ \p q^\kappa }\f{\p^2 \Xi(-q)}{\p q^\sigma \p q^\nu}+\f{\p \mathcal{K}(-q)}{ \p q^\sigma }\f{\p^2 \Xi(-q)}{\p q^\nu \p q^\kappa}\Bigg\rbrace\non\\
&&+\f{1}{2}\eta_{\mu\rho}\ell_2^b\Bigg\lbrace  -q_\sigma \f{\p \mathcal{K}(-q)}{\p q^\nu}\f{\p \Xi(-q)}{\p q^b}+q_\sigma \f{\p \mathcal{K}(-q)}{\p q^b}\f{\p \Xi(-q)}{\p q^\nu}+q_b \f{\p \mathcal{K}(-q)}{\p q^\nu}\f{\p \Xi(-q)}{\p q^\sigma}-q_b \f{\p \mathcal{K}(-q)}{\p q^\sigma}\f{\p \Xi(-q)}{\p q^\nu}\Bigg\rbrace \non\\
&& +\f{1}{2}\eta_{\mu\rho}\ell_1^b \Bigg\lbrace   -q_\nu \f{\p \mathcal{K}(-q)}{\p q^\sigma}\f{\p \Xi(-q)}{\p q^b}+q_\nu \f{\p \mathcal{K}(-q)}{\p q^b}\f{\p \Xi(-q)}{\p q^\sigma}+q_b \f{\p \mathcal{K}(-q)}{\p q^\sigma}\f{\p \Xi(-q)}{\p q^\nu}-q_b \f{\p \mathcal{K}(-q)}{\p q^\nu}\f{\p \Xi(-q)}{\p q^\sigma} \Bigg\rbrace \non\\
&&+\f{1}{4}q_\mu \ell_2^b \Bigg\lbrace 2\f{\p \mathcal{K}(-q)}{\p q^\nu}\f{\p\Xi(-q)}{\p q^\rho}\Sigma_{\sigma b}^T+2\f{\p \mathcal{K}(-q)}{\p q^\rho}\f{\p\Xi(-q)}{\p q^\nu}\Sigma_{\sigma b}^T -q_\sigma \f{\p^2 \mathcal{K}(-q)}{\p q^\nu \p q^\rho}\f{\p \Xi(-q)}{\p q^b} \nn\\
&&+q_\sigma \f{\p^2 \mathcal{K}(-q)}{\p q^\nu \p q^b}\f{\p \Xi(-q)}{\p q^\rho}+q_\sigma \f{\p^2 \mathcal{K}(-q)}{\p q^b \p q^\rho}\f{\p \Xi(-q)}{\p q^\nu}-q_b \f{\p^2 \mathcal{K}(-q)}{\p q^\sigma \p q^\rho}\f{\p \Xi(-q)}{\p q^\nu}+\eta_{\sigma\nu}\f{\p\mathcal{K}(-q)}{\p q^b}\f{\p\Xi(-q)}{\p q^\rho}\nn\\
&& -\eta_{b\nu}\f{\p\mathcal{K}(-q)}{\p q^\sigma}\f{\p\Xi(-q)}{\p q^\rho} +\eta_{\sigma\rho}\f{\p\mathcal{K}(-q)}{\p q^b}\f{\p\Xi(-q)}{\p q^\nu}-\eta_{b\rho}\f{\p\mathcal{K}(-q)}{\p q^\sigma}\f{\p\Xi(-q)}{\p q^\nu}+\eta_{\sigma\rho}\f{\p\mathcal{K}(-q)}{\p q^\nu}\f{\p\Xi(-q)}{\p q^b}\nn\\
&&-\eta_{b\rho}\f{\p\mathcal{K}(-q)}{\p q^\nu}\f{\p\Xi(-q)}{\p q^\sigma}+\eta_{\sigma\nu}\f{\p\mathcal{K}(-q)}{\p q^\rho}\f{\p\Xi(-q)}{\p q^b} -\eta_{b\nu}\f{\p\mathcal{K}(-q)}{\p q^\rho}\f{\p\Xi(-q)}{\p q^\sigma}+q_\sigma \f{\p \mathcal{K}(-q)}{\p q^b}\f{\p^2\Xi(-q)}{\p q^\nu \p q^\rho}\nn\\
&& -q_b \f{\p \mathcal{K}(-q)}{\p q^\sigma}\f{\p^2\Xi(-q)}{\p q^\nu \p q^\rho}  +q_\sigma \f{\p \mathcal{K}(-q)}{\p q^\nu}\f{\p^2\Xi(-q)}{\p q^b \p q^\rho} -q_b \f{\p \mathcal{K}(-q)}{\p q^\nu}\f{\p^2\Xi(-q)}{\p q^\sigma \p q^\rho} +q_\sigma \f{\p \mathcal{K}(-q)}{\p q^\rho}\f{\p^2\Xi(-q)}{\p q^\nu \p q^b}\nn\\
&& -q_b \f{\p \mathcal{K}(-q)}{\p q^\rho}\f{\p^2\Xi(-q)}{\p q^\nu \p q^\sigma}\Bigg\rbrace \non\\
&& +\f{1}{4}q_\rho \ell_1^b \Bigg\lbrace 2\f{\p \mathcal{K}(-q)}{\p q^\sigma}\f{\p\Xi(-q)}{\p q^\mu}\Sigma_{\nu b}^T+2\f{\p \mathcal{K}(-q)}{\p q^\mu}\f{\p\Xi(-q)}{\p q^\sigma}\Sigma_{\nu b}^T -q_\nu \f{\p^2 \mathcal{K}(-q)}{\p q^\sigma \p q^\mu}\f{\p \Xi(-q)}{\p q^b}\nn\\
&& +q_\nu \f{\p^2 \mathcal{K}(-q)}{\p q^\sigma \p q^b}\f{\p \Xi(-q)}{\p q^\mu}+q_\nu \f{\p^2 \mathcal{K}(-q)}{\p q^b \p q^\mu}\f{\p \Xi(-q)}{\p q^\sigma}-q_b \f{\p^2 \mathcal{K}(-q)}{\p q^\nu \p q^\mu}\f{\p \Xi(-q)}{\p q^\sigma}+\eta_{\sigma\nu}\f{\p\mathcal{K}(-q)}{\p q^b}\f{\p\Xi(-q)}{\p q^\mu} \nn\\
&&-\eta_{b\sigma}\f{\p\mathcal{K}(-q)}{\p q^\nu}\f{\p\Xi(-q)}{\p q^\mu} +\eta_{\nu\mu}\f{\p\mathcal{K}(-q)}{\p q^b}\f{\p\Xi(-q)}{\p q^\sigma}-\eta_{b\mu}\f{\p\mathcal{K}(-q)}{\p q^\nu}\f{\p\Xi(-q)}{\p q^\sigma}   +\eta_{\nu\mu}\f{\p\mathcal{K}(-q)}{\p q^\sigma}\f{\p\Xi(-q)}{\p q^b}\nn\\
&&-\eta_{b\mu}\f{\p\mathcal{K}(-q)}{\p q^\sigma}\f{\p\Xi(-q)}{\p q^\nu}+\eta_{\sigma\nu}\f{\p\mathcal{K}(-q)}{\p q^\mu}\f{\p\Xi(-q)}{\p q^b}  -\eta_{b\sigma}\f{\p\mathcal{K}(-q)}{\p q^\mu}\f{\p\Xi(-q)}{\p q^\nu}+q_\nu \f{\p \mathcal{K}(-q)}{\p q^b}\f{\p^2\Xi(-q)}{\p q^\sigma \p q^\mu} \nn\\
&&-q_b \f{\p \mathcal{K}(-q)}{\p q^\nu}\f{\p^2\Xi(-q)}{\p q^\sigma \p q^\mu}  +q_\nu \f{\p \mathcal{K}(-q)}{\p q^\sigma}\f{\p^2\Xi(-q)}{\p q^b \p q^\mu} -q_b \f{\p \mathcal{K}(-q)}{\p q^\sigma}\f{\p^2\Xi(-q)}{\p q^\nu \p q^\mu} +q_\nu \f{\p \mathcal{K}(-q)}{\p q^\mu}\f{\p^2\Xi(-q)}{\p q^\sigma \p q^b}\nn\\
&& -q_b \f{\p \mathcal{K}(-q)}{\p q^\mu}\f{\p^2\Xi(-q)}{\p q^\sigma \p q^\nu}\Bigg\rbrace \non\\
&& -\f{1}{4}\eta_{\nu\sigma}(\ell_2+\ell_1)^\kappa \Bigg\lbrace  -q_\rho \f{\p \mathcal{K}(-q)}{\p q^\kappa}\f{\p \Xi(-q)}{\p q^\mu}+q_\rho \f{\p \mathcal{K}(-q)}{\p q^\mu}\f{\p \Xi(-q)}{\p q^\kappa} -q_\mu \f{\p \mathcal{K}(-q)}{\p q^\kappa}\f{\p \Xi(-q)}{\p q^\rho}\nn\\
&&+q_\mu \f{\p \mathcal{K}(-q)}{\p q^\rho}\f{\p \Xi(-q)}{\p q^\kappa}\Bigg\rbrace  -\f{1}{2}\eta_{\nu\sigma}q.(\ell_1-\ell_2) \Bigg\lbrace   - \f{\p \mathcal{K}(-q)}{\p q^\mu}\f{\p \Xi(-q)}{\p q^\rho}+ \f{\p \mathcal{K}(-q)}{\p q^\rho}\f{\p \Xi(-q)}{\p q^\mu} \Bigg\rbrace \non\\
&& -\f{1}{2}\ell_{2\nu}\Bigg\lbrace  q_\rho \f{\p \mathcal{K}(-q)}{\p q^\sigma}\f{\p \Xi(-q)}{\p q^\mu}-q_\rho \f{\p \mathcal{K}(-q)}{\p q^\mu}\f{\p \Xi(-q)}{\p q^\sigma} \Bigg\rbrace  -\f{1}{2}\ell_{1\sigma}\Bigg\lbrace   q_\mu \f{\p \mathcal{K}(-q)}{\p q^\nu}\f{\p \Xi(-q)}{\p q^\rho}\nn\\
&&-q_\mu \f{\p \mathcal{K}(-q)}{\p q^\rho}\f{\p \Xi(-q)}{\p q^\nu} \Bigg\rbrace +\mathcal{O}(\ell_1^2, \ell_1\ell_2,\ell_2^2)\Bigg]\ . \label{eq:Z2_gr}
\ee
\endgroup

Above we removed the terms which are anti-symmetric under $\mu\leftrightarrow\nu$ and/or $\rho\leftrightarrow\sigma$ exchanges.

\bibliography{classicalref.bib}

\providecommand{\href}[2]{#2}\begingroup\raggedright\begin{thebibliography}{10}

\bibitem{1808.03288}
B.~Sahoo and A.~Sen, ``{Classical and Quantum Results on Logarithmic Terms in
  the Soft Theorem in Four Dimensions},''
  \href{http://dx.doi.org/10.1007/JHEP02(2019)086}{{\em JHEP} {\bfseries 02}
  (2019) 086},
\href{http://arxiv.org/abs/1808.03288}{{\ttfamily arXiv:1808.03288 [hep-th]}}.

\bibitem{1703.00024}
A.~Sen, ``{Subleading Soft Graviton Theorem for Loop Amplitudes},''
  \href{http://dx.doi.org/10.1007/JHEP11(2017)123}{{\em JHEP} {\bfseries 11}
  (2017) 123},
\href{http://arxiv.org/abs/1703.00024}{{\ttfamily arXiv:1703.00024 [hep-th]}}.

\bibitem{1706.00759}
A.~Laddha and A.~Sen, ``{Sub-subleading Soft Graviton Theorem in Generic
  Theories of Quantum Gravity},''
  \href{http://dx.doi.org/10.1007/JHEP10(2017)065}{{\em JHEP} {\bfseries 10}
  (2017) 065},
\href{http://arxiv.org/abs/1706.00759}{{\ttfamily arXiv:1706.00759 [hep-th]}}.

\bibitem{1707.06803}
S.~Chakrabarti, S.~P. Kashyap, B.~Sahoo, A.~Sen, and M.~Verma, ``{Subleading
  Soft Theorem for Multiple Soft Gravitons},''
  \href{http://dx.doi.org/10.1007/JHEP12(2017)150}{{\em JHEP} {\bfseries 12}
  (2017) 150},
\href{http://arxiv.org/abs/1707.06803}{{\ttfamily arXiv:1707.06803 [hep-th]}}.

\bibitem{1809.01675}
S.~Atul~Bhatkar and B.~Sahoo, ``{Subleading Soft Theorem for arbitrary number
  of external soft photons and gravitons},''
  \href{http://dx.doi.org/10.1007/JHEP01(2019)153}{{\em JHEP} {\bfseries 01}
  (2019) 153},
\href{http://arxiv.org/abs/1809.01675}{{\ttfamily arXiv:1809.01675 [hep-th]}}.

\bibitem{1802.03148}
Z.-Z. Li, H.-H. Lin, and S.-Q. Zhang, ``{Infinite Soft Theorems from Gauge
  Symmetry},'' \href{http://dx.doi.org/10.1103/PhysRevD.98.045004}{{\em Phys.
  Rev.} {\bfseries D98} no.~4, (2018) 045004},
\href{http://arxiv.org/abs/1802.03148}{{\ttfamily arXiv:1802.03148 [hep-th]}}.

\bibitem{1703.05448}
A.~Strominger, ``{Lectures on the Infrared Structure of Gravity and Gauge
  Theory},''
\href{http://arxiv.org/abs/1703.05448}{{\ttfamily arXiv:1703.05448 [hep-th]}}.

\bibitem{1611.07534}
H.~Elvang, C.~R.~T. Jones, and S.~G. Naculich, ``{Soft Photon and Graviton
  Theorems in Effective Field Theory},''
  \href{http://dx.doi.org/10.1103/PhysRevLett.118.231601}{{\em Phys. Rev.
  Lett.} {\bfseries 118} no.~23, (2017) 231601},
  \href{http://arxiv.org/abs/1611.07534}{{\ttfamily arXiv:1611.07534
  [hep-th]}}.

\bibitem{1405.1410}
S.~He, Y.-t. Huang, and C.~Wen, ``{Loop Corrections to Soft Theorems in Gauge
  Theories and Gravity},''
  \href{http://dx.doi.org/10.1007/JHEP12(2014)115}{{\em JHEP} {\bfseries 12}
  (2014) 115},
\href{http://arxiv.org/abs/1405.1410}{{\ttfamily arXiv:1405.1410 [hep-th]}}.

\bibitem{1405.1015}
Z.~Bern, S.~Davies, and J.~Nohle, ``{On Loop Corrections to Subleading Soft
  Behavior of Gluons and Gravitons},''
  \href{http://dx.doi.org/10.1103/PhysRevD.90.085015}{{\em Phys. Rev.}
  {\bfseries D90} no.~8, (2014) 085015},
\href{http://arxiv.org/abs/1405.1015}{{\ttfamily arXiv:1405.1015 [hep-th]}}.

\bibitem{1901.10986}
A.~Addazi, M.~Bianchi, and G.~Veneziano, ``{Soft gravitational radiation from
  ultra-relativistic collisions at sub- and sub-sub-leading order},''
  \href{http://dx.doi.org/10.1007/JHEP05(2019)050}{{\em JHEP} {\bfseries 05}
  (2019) 050},
\href{http://arxiv.org/abs/1901.10986}{{\ttfamily arXiv:1901.10986 [hep-th]}}.

\bibitem{1812.08137}
M.~Ciafaloni, D.~Colferai, and G.~Veneziano, ``{Infrared features of
  gravitational scattering and radiation in the eikonal approach},''
  \href{http://dx.doi.org/10.1103/PhysRevD.99.066008}{{\em Phys. Rev.}
  {\bfseries D99} no.~6, (2019) 066008},
\href{http://arxiv.org/abs/1812.08137}{{\ttfamily arXiv:1812.08137 [hep-th]}}.

\bibitem{1801.07719}
A.~Laddha and A.~Sen, ``{Gravity Waves from Soft Theorem in General
  Dimensions},'' \href{http://dx.doi.org/10.1007/JHEP09(2018)105}{{\em JHEP}
  {\bfseries 09} (2018) 105},
\href{http://arxiv.org/abs/1801.07719}{{\ttfamily arXiv:1801.07719 [hep-th]}}.

\bibitem{1804.09193}
A.~Laddha and A.~Sen, ``{Logarithmic Terms in the Soft Expansion in Four
  Dimensions},'' \href{http://dx.doi.org/10.1007/JHEP10(2018)056}{{\em JHEP}
  {\bfseries 10} (2018) 056},
\href{http://arxiv.org/abs/1804.09193}{{\ttfamily arXiv:1804.09193 [hep-th]}}.

\bibitem{1906.08288}
A.~Laddha and A.~Sen, ``{Classical proof of the classical soft graviton theorem
  in D>4},'' \href{http://dx.doi.org/10.1103/PhysRevD.101.084011}{{\em Phys.
  Rev. D} {\bfseries 101} no.~8, (2020) 084011},
  \href{http://arxiv.org/abs/1906.08288}{{\ttfamily arXiv:1906.08288 [gr-qc]}}.

\bibitem{1912.06413}
A.~P. Saha, B.~Sahoo, and A.~Sen, ``{Proof of the classical soft graviton
  theorem in $D$ = 4},'' \href{http://dx.doi.org/10.1007/JHEP06(2020)153}{{\em
  JHEP} {\bfseries 06} (2020) 153},
  \href{http://arxiv.org/abs/1912.06413}{{\ttfamily arXiv:1912.06413
  [hep-th]}}.

\bibitem{1411.5745}
A.~Strominger and A.~Zhiboedov, ``{Gravitational Memory, BMS Supertranslations
  and Soft Theorems},'' \href{http://dx.doi.org/10.1007/JHEP01(2016)086}{{\em
  JHEP} {\bfseries 01} (2016) 086},
\href{http://arxiv.org/abs/1411.5745}{{\ttfamily arXiv:1411.5745 [hep-th]}}.

\bibitem{1712.01204}
M.~Pate, A.-M. Raclariu, and A.~Strominger, ``{Gravitational Memory in Higher
  Dimensions},'' \href{http://dx.doi.org/10.1007/JHEP06(2018)138}{{\em JHEP}
  {\bfseries 06} (2018) 138},
\href{http://arxiv.org/abs/1712.01204}{{\ttfamily arXiv:1712.01204 [hep-th]}}.

\bibitem{1502.06120}
S.~Pasterski, A.~Strominger, and A.~Zhiboedov, ``{New Gravitational
  Memories},'' \href{http://dx.doi.org/10.1007/JHEP12(2016)053}{{\em JHEP}
  {\bfseries 12} (2016) 053},
\href{http://arxiv.org/abs/1502.06120}{{\ttfamily arXiv:1502.06120 [hep-th]}}.

\bibitem{1806.01872}
A.~Laddha and A.~Sen, ``{Observational Signature of the Logarithmic Terms in
  the Soft Graviton Theorem},''
  \href{http://dx.doi.org/10.1103/PhysRevD.100.024009}{{\em Phys. Rev.}
  {\bfseries D100} no.~2, (2019) 024009},
\href{http://arxiv.org/abs/1806.01872}{{\ttfamily arXiv:1806.01872 [hep-th]}}.

\bibitem{2008.04376}
B.~Sahoo, ``{Classical Sub-subleading Soft Photon and Soft Graviton Theorems in
  Four Spacetime Dimensions},''
  \href{http://dx.doi.org/10.1007/JHEP12(2020)070}{{\em JHEP} {\bfseries 12}
  (2020) 070}, \href{http://arxiv.org/abs/2008.04376}{{\ttfamily
  arXiv:2008.04376 [hep-th]}}.

\bibitem{2105.08739}
B.~Sahoo and A.~Sen, ``{Classical soft graviton theorem rewritten},''
  \href{http://dx.doi.org/10.1007/JHEP01(2022)077}{{\em JHEP} {\bfseries 01}
  (2022) 077}, \href{http://arxiv.org/abs/2105.08739}{{\ttfamily
  arXiv:2105.08739 [hep-th]}}.

\bibitem{2106.10741}
D.~Ghosh and B.~Sahoo, ``{Spin-dependent gravitational tail memory in $D=4$},''
  \href{http://dx.doi.org/10.1103/PhysRevD.105.025024}{{\em Phys. Rev. D}
  {\bfseries 105} no.~2, (2022) 025024},
  \href{http://arxiv.org/abs/2106.10741}{{\ttfamily arXiv:2106.10741
  [hep-th]}}.

\bibitem{weinberg1}
S.~Weinberg, ``{Photons and Gravitons in $S$-Matrix Theory: Derivation of
  Charge Conservation and Equality of Gravitational and Inertial Mass},''
\href{http://dx.doi.org/10.1103/PhysRev.135.B1049}{{\em Phys. Rev.} {\bfseries
  135} (1964) B1049--B1056}.

\bibitem{weinberg2}
S.~Weinberg, ``{Infrared photons and gravitons},''
\href{http://dx.doi.org/10.1103/PhysRev.140.B516}{{\em Phys. Rev.} {\bfseries
  140} (1965) B516--B524}.

\bibitem{jackiw1}
D.~J. Gross and R.~Jackiw, ``{Low-Energy Theorem for Graviton Scattering},''
\href{http://dx.doi.org/10.1103/PhysRev.166.1287}{{\em Phys. Rev.} {\bfseries
  166} (1968) 1287--1292}.

\bibitem{jackiw2}
R.~Jackiw, ``{Low-Energy Theorems for Massless Bosons: Photons and
  Gravitons},''
\href{http://dx.doi.org/10.1103/PhysRev.168.1623}{{\em Phys. Rev.} {\bfseries
  168} (1968) 1623--1633}.

\bibitem{1103.2981}
C.~D. White, ``{Factorization Properties of Soft Graviton Amplitudes},''
  \href{http://dx.doi.org/10.1007/JHEP05(2011)060}{{\em JHEP} {\bfseries 05}
  (2011) 060},
\href{http://arxiv.org/abs/1103.2981}{{\ttfamily arXiv:1103.2981 [hep-th]}}.

\bibitem{1401.7026}
T.~He, V.~Lysov, P.~Mitra, and A.~Strominger, ``{BMS supertranslations and
  Weinberg’s soft graviton theorem},''
  \href{http://dx.doi.org/10.1007/JHEP05(2015)151}{{\em JHEP} {\bfseries 05}
  (2015) 151},
\href{http://arxiv.org/abs/1401.7026}{{\ttfamily arXiv:1401.7026 [hep-th]}}.

\bibitem{1404.4091}
F.~Cachazo and A.~Strominger, ``{Evidence for a New Soft Graviton Theorem},''
\href{http://arxiv.org/abs/1404.4091}{{\ttfamily arXiv:1404.4091 [hep-th]}}.

\bibitem{1405.3533}
N.~Afkhami-Jeddi, ``{Soft Graviton Theorem in Arbitrary Dimensions},''
\href{http://arxiv.org/abs/1405.3533}{{\ttfamily arXiv:1405.3533 [hep-th]}}.

\bibitem{1406.6987}
Z.~Bern, S.~Davies, P.~Di~Vecchia, and J.~Nohle, ``{Low-Energy Behavior of
  Gluons and Gravitons from Gauge Invariance},''
  \href{http://dx.doi.org/10.1103/PhysRevD.90.084035}{{\em Phys. Rev.}
  {\bfseries D90} no.~8, (2014) 084035},
\href{http://arxiv.org/abs/1406.6987}{{\ttfamily arXiv:1406.6987 [hep-th]}}.

\bibitem{1406.6574}
J.~Broedel, M.~de~Leeuw, J.~Plefka, and M.~Rosso, ``{Constraining subleading
  soft gluon and graviton theorems},''
  \href{http://dx.doi.org/10.1103/PhysRevD.90.065024}{{\em Phys. Rev.}
  {\bfseries D90} no.~6, (2014) 065024},
\href{http://arxiv.org/abs/1406.6574}{{\ttfamily arXiv:1406.6574 [hep-th]}}.

\bibitem{1801.05528}
Y.~Hamada and G.~Shiu, ``{Infinite Set of Soft Theorems in Gauge-Gravity
  Theories as Ward-Takahashi Identities},''
  \href{http://dx.doi.org/10.1103/PhysRevLett.120.201601}{{\em Phys. Rev.
  Lett.} {\bfseries 120} no.~20, (2018) 201601},
\href{http://arxiv.org/abs/1801.05528}{{\ttfamily arXiv:1801.05528 [hep-th]}}.

\bibitem{1503.04816}
F.~Cachazo, S.~He, and E.~Y. Yuan, ``{New Double Soft Emission Theorems},''
  \href{http://dx.doi.org/10.1103/PhysRevD.92.065030}{{\em Phys. Rev.}
  {\bfseries D92} no.~6, (2015) 065030},
\href{http://arxiv.org/abs/1503.04816}{{\ttfamily arXiv:1503.04816 [hep-th]}}.

\bibitem{1504.05558}
T.~Klose, T.~McLoughlin, D.~Nandan, J.~Plefka, and G.~Travaglini,
  ``{Double-Soft Limits of Gluons and Gravitons},''
  \href{http://dx.doi.org/10.1007/JHEP07(2015)135}{{\em JHEP} {\bfseries 07}
  (2015) 135}, \href{http://arxiv.org/abs/1504.05558}{{\ttfamily
  arXiv:1504.05558 [hep-th]}}.

\bibitem{1504.05559}
A.~Volovich, C.~Wen, and M.~Zlotnikov, ``{Double Soft Theorems in Gauge and
  String Theories},'' \href{http://dx.doi.org/10.1007/JHEP07(2015)095}{{\em
  JHEP} {\bfseries 07} (2015) 095},
\href{http://arxiv.org/abs/1504.05559}{{\ttfamily arXiv:1504.05559 [hep-th]}}.

\bibitem{1507.00938}
P.~Di~Vecchia, R.~Marotta, and M.~Mojaza, ``{Double-soft behavior for scalars
  and gluons from string theory},''
  \href{http://dx.doi.org/10.1007/JHEP12(2015)150}{{\em JHEP} {\bfseries 12}
  (2015) 150},
\href{http://arxiv.org/abs/1507.00938}{{\ttfamily arXiv:1507.00938 [hep-th]}}.

\bibitem{1604.02834}
S.~He, Z.~Liu, and J.-B. Wu, ``{Scattering Equations, Twistor-string Formulas
  and Double-soft Limits in Four Dimensions},''
  \href{http://dx.doi.org/10.1007/JHEP07(2016)060}{{\em JHEP} {\bfseries 07}
  (2016) 060},
\href{http://arxiv.org/abs/1604.02834}{{\ttfamily arXiv:1604.02834 [hep-th]}}.

\bibitem{1607.02700}
A.~P. Saha, ``{Double Soft Theorem for Perturbative Gravity},''
  \href{http://dx.doi.org/10.1007/JHEP09(2016)165}{{\em JHEP} {\bfseries 09}
  (2016) 165},
\href{http://arxiv.org/abs/1607.02700}{{\ttfamily arXiv:1607.02700 [hep-th]}}.

\bibitem{1702.02350}
A.~P. Saha, ``{Double soft limit of the graviton amplitude from the
  Cachazo-He-Yuan formalism},''
  \href{http://dx.doi.org/10.1103/PhysRevD.96.045002}{{\em Phys. Rev.}
  {\bfseries D96} no.~4, (2017) 045002},
\href{http://arxiv.org/abs/1702.02350}{{\ttfamily arXiv:1702.02350 [hep-th]}}.

\bibitem{1705.06175}
P.~Di~Vecchia, R.~Marotta, and M.~Mojaza, ``{Double-soft behavior of the
  dilaton of spontaneously broken conformal invariance},''
  \href{http://dx.doi.org/10.1007/JHEP09(2017)001}{{\em JHEP} {\bfseries 09}
  (2017) 001},
\href{http://arxiv.org/abs/1705.06175}{{\ttfamily arXiv:1705.06175 [hep-th]}}.

\bibitem{1709.07883}
S.~Chakrabarti, S.~P. Kashyap, B.~Sahoo, A.~Sen, and M.~Verma, ``{Testing
  Subleading Multiple Soft Graviton Theorem for CHY Prescription},''
  \href{http://dx.doi.org/10.1007/JHEP01(2018)090}{{\em JHEP} {\bfseries 01}
  (2018) 090}, \href{http://arxiv.org/abs/1709.07883}{{\ttfamily
  arXiv:1709.07883 [hep-th]}}.

\bibitem{DiVecchia:2023frv}
P.~Di~Vecchia, C.~Heissenberg, R.~Russo, and G.~Veneziano, ``{The gravitational
  eikonal: from particle, string and brane collisions to black-hole
  encounters},'' \href{http://arxiv.org/abs/2306.16488}{{\ttfamily
  arXiv:2306.16488 [hep-th]}}.

\bibitem{Jakobsen:2021lvp}
G.~U. Jakobsen, G.~Mogull, J.~Plefka, and J.~Steinhoff, ``{Gravitational
  Bremsstrahlung and Hidden Supersymmetry of Spinning Bodies},''
  \href{http://dx.doi.org/10.1103/PhysRevLett.128.011101}{{\em Phys. Rev.
  Lett.} {\bfseries 128} no.~1, (2022) 011101},
  \href{http://arxiv.org/abs/2106.10256}{{\ttfamily arXiv:2106.10256
  [hep-th]}}.

\bibitem{Aoude:2023dui}
R.~Aoude, K.~Haddad, C.~Heissenberg, and A.~Helset, ``{Leading-order
  gravitational radiation to all spin orders},''
  \href{http://arxiv.org/abs/2310.05832}{{\ttfamily arXiv:2310.05832
  [hep-th]}}.

\bibitem{Gell-Mann}
M.~Gell-Mann and M.~L. Goldberger, ``{Scattering of low-energy photons by
  particles of spin 1/2},''
\href{http://dx.doi.org/10.1103/PhysRev.96.1433}{{\em Phys. Rev.} {\bfseries
  96} (1954) 1433--1438}.

\bibitem{Low1}
F.~E. Low, ``{Scattering of light of very low frequency by systems of spin
  1/2},''
\href{http://dx.doi.org/10.1103/PhysRev.96.1428}{{\em Phys. Rev.} {\bfseries
  96} (1954) 1428--1432}.

\bibitem{low}
F.~E. Low, ``{Bremsstrahlung of very low-energy quanta in elementary particle
  collisions},''
\href{http://dx.doi.org/10.1103/PhysRev.110.974}{{\em Phys. Rev.} {\bfseries
  110} (1958) 974--977}.

\bibitem{saito}
S.~Saito, ``{Low-energy theorem for Compton scattering},''
\href{http://dx.doi.org/10.1103/PhysRev.184.1894}{{\em Phys. Rev.} {\bfseries
  184} (1969) 1894--1902}.

\bibitem{burnett}
T.~H. Burnett and N.~M. Kroll, ``{Extension of the low soft photon theorem},''
\href{http://dx.doi.org/10.1103/PhysRevLett.20.86}{{\em Phys. Rev. Lett.}
  {\bfseries 20} (1968) 86}.

\bibitem{bell}
J.~S. Bell and R.~Van~Royen, ``{On the low-burnett-kroll theorem for
  soft-photon emission},''
\href{http://dx.doi.org/10.1007/BF02823297}{{\em Nuovo Cim.} {\bfseries A60}
  (1969) 62--68}.

\bibitem{duca}
V.~Del~Duca, ``{High-energy Bremsstrahlung Theorems for Soft Photons},''
\href{http://dx.doi.org/10.1016/0550-3213(90)90392-Q}{{\em Nucl. Phys.}
  {\bfseries B345} (1990) 369--388}.

\bibitem{grammer}
G.~Grammer, Jr. and D.~R. Yennie, ``{Improved treatment for the infrared
  divergence problem in quantum electrodynamics},''
\href{http://dx.doi.org/10.1103/PhysRevD.8.4332}{{\em Phys. Rev.} {\bfseries
  D8} (1973) 4332--4344}.

\bibitem{Sahoo:2020csy}
B.~Sahoo, {\em {CLASSICAL AND QUANTUM SUBLEADING SOFT THEOREM IN FOUR SPACETIME
  DIMENSIONS}}.
\newblock PhD thesis, HBNI, Mumbai, 2020.

\bibitem{Faddeev-Kulish}
P.~P. Kulish and L.~D. Faddeev, ``{Asymptotic conditions and infrared
  divergences in quantum electrodynamics},''
  \href{http://dx.doi.org/10.1007/BF01066485}{{\em Theor. Math. Phys.}
  {\bfseries 4} (1970) 745}.
[Teor. Mat. Fiz.4,153(1970)].

\bibitem{senqcd}
A.~Sen, ``{Asymptotic Behavior of the Sudakov Form-Factor in QCD},''
\href{http://dx.doi.org/10.1103/PhysRevD.24.3281}{{\em Phys. Rev.} {\bfseries
  D24} (1981) 3281}.

\bibitem{Himwich:2020rro}
E.~Himwich, S.~A. Narayanan, M.~Pate, N.~Paul, and A.~Strominger, ``{The Soft
  $\mathcal{S}$-Matrix in Gravity},''
  \href{http://dx.doi.org/10.1007/JHEP09(2020)129}{{\em JHEP} {\bfseries 09}
  (2020) 129}, \href{http://arxiv.org/abs/2005.13433}{{\ttfamily
  arXiv:2005.13433 [hep-th]}}.

\bibitem{Naculich:2011ry}
S.~G. Naculich and H.~J. Schnitzer, ``{Eikonal methods applied to gravitational
  scattering amplitudes},''
  \href{http://dx.doi.org/10.1007/JHEP05(2011)087}{{\em JHEP} {\bfseries 05}
  (2011) 087}, \href{http://arxiv.org/abs/1101.1524}{{\ttfamily arXiv:1101.1524
  [hep-th]}}.

\bibitem{Peskin:1995ev}
M.~E. Peskin and D.~V. Schroeder, {\em {An Introduction to quantum field
  theory}}.
\newblock Addison-Wesley, Reading, USA, 1995.

\bibitem{9411092}
S.~Y. Choi, J.~S. Shim, and H.~S. Song, ``{Factorization and polarization in
  linearized gravity},'' \href{http://dx.doi.org/10.1103/PhysRevD.51.2751}{{\em
  Phys. Rev.} {\bfseries D51} (1995) 2751--2769},
\href{http://arxiv.org/abs/hep-th/9411092}{{\ttfamily arXiv:hep-th/9411092
  [hep-th]}}.

\bibitem{Gervais:2017zky}
H.~Gervais, ``{Soft Graviton Emission at High and Low Energies in Yukawa and
  Scalar Theories},'' \href{http://dx.doi.org/10.1103/PhysRevD.96.065007}{{\em
  Phys. Rev. D} {\bfseries 96} no.~6, (2017) 065007},
  \href{http://arxiv.org/abs/1706.03453}{{\ttfamily arXiv:1706.03453
  [hep-th]}}.

\bibitem{ademollo}
M.~Ademollo, A.~D'Adda, R.~D'Auria, F.~Gliozzi, E.~Napolitano, S.~Sciuto, and
  P.~Di~Vecchia, ``{Soft Dilations and Scale Renormalization in Dual
  Theories},''
\href{http://dx.doi.org/10.1016/0550-3213(75)90491-5}{{\em Nucl. Phys.}
  {\bfseries B94} (1975) 221--259}.

\bibitem{shapiro}
J.~A. Shapiro, ``{On the Renormalization of Dual Models},''
\href{http://dx.doi.org/10.1103/PhysRevD.11.2937}{{\em Phys. Rev.} {\bfseries
  D11} (1975) 2937}.

\bibitem{1406.4172}
B.~U.~W. Schwab, ``{Subleading Soft Factor for String Disk Amplitudes},''
  \href{http://dx.doi.org/10.1007/JHEP08(2014)062}{{\em JHEP} {\bfseries 08}
  (2014) 062},
\href{http://arxiv.org/abs/1406.4172}{{\ttfamily arXiv:1406.4172 [hep-th]}}.

\bibitem{1406.5155}
M.~Bianchi, S.~He, Y.-t. Huang, and C.~Wen, ``{More on Soft Theorems: Trees,
  Loops and Strings},''
  \href{http://dx.doi.org/10.1103/PhysRevD.92.065022}{{\em Phys. Rev.}
  {\bfseries D92} no.~6, (2015) 065022},
\href{http://arxiv.org/abs/1406.5155}{{\ttfamily arXiv:1406.5155 [hep-th]}}.

\bibitem{1411.6661}
B.~U.~W. Schwab, ``{A Note on Soft Factors for Closed String Scattering},''
  \href{http://dx.doi.org/10.1007/JHEP03(2015)140}{{\em JHEP} {\bfseries 03}
  (2015) 140},
\href{http://arxiv.org/abs/1411.6661}{{\ttfamily arXiv:1411.6661 [hep-th]}}.

\bibitem{1502.05258}
P.~Di~Vecchia, R.~Marotta, and M.~Mojaza, ``{Soft theorem for the graviton,
  dilaton and the Kalb-Ramond field in the bosonic string},''
  \href{http://dx.doi.org/10.1007/JHEP05(2015)137}{{\em JHEP} {\bfseries 05}
  (2015) 137},
\href{http://arxiv.org/abs/1502.05258}{{\ttfamily arXiv:1502.05258 [hep-th]}}.

\bibitem{1505.05854}
M.~Bianchi and A.~L. Guerrieri, ``{On the soft limit of open string disk
  amplitudes with massive states},''
  \href{http://dx.doi.org/10.1007/JHEP09(2015)164}{{\em JHEP} {\bfseries 09}
  (2015) 164},
\href{http://arxiv.org/abs/1505.05854}{{\ttfamily arXiv:1505.05854 [hep-th]}}.

\bibitem{1507.08829}
A.~L. Guerrieri, ``{Soft behavior of string amplitudes with external massive
  states},'' \href{http://dx.doi.org/10.1393/ncc/i2016-16221-2}{{\em Nuovo
  Cim.} {\bfseries C39} no.~1, (2016) 221},
\href{http://arxiv.org/abs/1507.08829}{{\ttfamily arXiv:1507.08829 [hep-th]}}.

\bibitem{1511.04921}
P.~Di~Vecchia, R.~Marotta, and M.~Mojaza, ``{Soft Theorems from String
  Theory},'' \href{http://dx.doi.org/10.1002/prop.201500068}{{\em Fortsch.
  Phys.} {\bfseries 64} (2016) 389--393},
\href{http://arxiv.org/abs/1511.04921}{{\ttfamily arXiv:1511.04921 [hep-th]}}.

\bibitem{1512.00803}
M.~Bianchi and A.~L. Guerrieri, ``{On the soft limit of closed string
  amplitudes with massive states},''
  \href{http://dx.doi.org/10.1016/j.nuclphysb.2016.02.005}{{\em Nucl. Phys.}
  {\bfseries B905} (2016) 188--216},
\href{http://arxiv.org/abs/1512.00803}{{\ttfamily arXiv:1512.00803 [hep-th]}}.

\bibitem{1601.03457}
M.~Bianchi and A.~L. Guerrieri,
  \href{http://dx.doi.org/10.1142/9789813226609_0555}{``{On the soft limit of
  tree-level string amplitudes},''} in {\em {Proceedings, 14th Marcel Grossmann
  Meeting on Recent Developments in Theoretical and Experimental General
  Relativity, Astrophysics, and Relativistic Field Theories (MG14) (In 4
  Volumes): Rome, Italy, July 12-18, 2015}}, vol.~4, pp.~4157--4163.
\newblock 2017.
\newblock
\href{http://arxiv.org/abs/1601.03457}{{\ttfamily arXiv:1601.03457 [hep-th]}}.
\newblock

\bibitem{1604.03355}
P.~Di~Vecchia, R.~Marotta, and M.~Mojaza, ``{Subsubleading soft theorems of
  gravitons and dilatons in the bosonic string},''
  \href{http://dx.doi.org/10.1007/JHEP06(2016)054}{{\em JHEP} {\bfseries 06}
  (2016) 054},
\href{http://arxiv.org/abs/1604.03355}{{\ttfamily arXiv:1604.03355 [hep-th]}}.

\bibitem{1610.03481}
P.~Di~Vecchia, R.~Marotta, and M.~Mojaza, ``{Soft behavior of a closed massless
  state in superstring and universality in the soft behavior of the dilaton},''
  \href{http://dx.doi.org/10.1007/JHEP12(2016)020}{{\em JHEP} {\bfseries 12}
  (2016) 020},
\href{http://arxiv.org/abs/1610.03481}{{\ttfamily arXiv:1610.03481 [hep-th]}}.

\bibitem{1702.03934}
A.~Sen, ``{Soft Theorems in Superstring Theory},''
  \href{http://dx.doi.org/10.1007/JHEP06(2017)113}{{\em JHEP} {\bfseries 06}
  (2017) 113},
\href{http://arxiv.org/abs/1702.03934}{{\ttfamily arXiv:1702.03934 [hep-th]}}.

\bibitem{Marotta:2019cip}
R.~Marotta and M.~Verma, ``{Soft Theorems from Compactification},''
  \href{http://dx.doi.org/10.1007/JHEP02(2020)008}{{\em JHEP} {\bfseries 02}
  (2020) 008}, \href{http://arxiv.org/abs/1911.05099}{{\ttfamily
  arXiv:1911.05099 [hep-th]}}.

\bibitem{1911.06821}
H.~Hannesdottir and M.~D. Schwartz, ``{An $S$-Matrix for Massless Particles},''
\href{http://arxiv.org/abs/1911.06821}{{\ttfamily arXiv:1911.06821 [hep-th]}}.

\bibitem{2007.02077}
M.~A, D.~Ghosh, A.~Laddha, and A.~P., ``{Soft Radiation from Scattering
  Amplitudes Revisited},'' \href{http://arxiv.org/abs/2007.02077}{{\ttfamily
  arXiv:2007.02077 [hep-th]}}.

\bibitem{Kosower:2018adc}
D.~A. Kosower, B.~Maybee, and D.~O'Connell, ``{Amplitudes, Observables, and
  Classical Scattering},''
  \href{http://dx.doi.org/10.1007/JHEP02(2019)137}{{\em JHEP} {\bfseries 02}
  (2019) 137}, \href{http://arxiv.org/abs/1811.10950}{{\ttfamily
  arXiv:1811.10950 [hep-th]}}.

\bibitem{Cristofoli:2021vyo}
A.~Cristofoli, R.~Gonzo, D.~A. Kosower, and D.~O'Connell, ``{Waveforms from
  amplitudes},'' \href{http://dx.doi.org/10.1103/PhysRevD.106.056007}{{\em
  Phys. Rev. D} {\bfseries 106} no.~5, (2022) 056007},
  \href{http://arxiv.org/abs/2107.10193}{{\ttfamily arXiv:2107.10193
  [hep-th]}}.

\bibitem{Georgoudis:2023lgf}
A.~Georgoudis, C.~Heissenberg, and I.~Vazquez-Holm, ``{Inelastic exponentiation
  and classical gravitational scattering at one loop},''
  \href{http://dx.doi.org/10.1007/JHEP06(2023)126}{{\em JHEP} {\bfseries 06}
  (2023) 126}, \href{http://arxiv.org/abs/2303.07006}{{\ttfamily
  arXiv:2303.07006 [hep-th]}}.

\bibitem{Elkhidir:2023dco}
A.~Elkhidir, D.~O'Connell, M.~Sergola, and I.~A. Vazquez-Holm, ``{Radiation and
  Reaction at One Loop},'' \href{http://arxiv.org/abs/2303.06211}{{\ttfamily
  arXiv:2303.06211 [hep-th]}}.

\bibitem{Brandhuber:2023hhy}
A.~Brandhuber, G.~R. Brown, G.~Chen, S.~De~Angelis, J.~Gowdy, and
  G.~Travaglini, ``{One-loop gravitational bremsstrahlung and waveforms from a
  heavy-mass effective field theory},''
  \href{http://dx.doi.org/10.1007/JHEP06(2023)048}{{\em JHEP} {\bfseries 06}
  (2023) 048}, \href{http://arxiv.org/abs/2303.06111}{{\ttfamily
  arXiv:2303.06111 [hep-th]}}.

\bibitem{Herderschee:2023fxh}
A.~Herderschee, R.~Roiban, and F.~Teng, ``{The sub-leading scattering waveform
  from amplitudes},'' \href{http://dx.doi.org/10.1007/JHEP06(2023)004}{{\em
  JHEP} {\bfseries 06} (2023) 004},
  \href{http://arxiv.org/abs/2303.06112}{{\ttfamily arXiv:2303.06112
  [hep-th]}}.

\bibitem{Caron-Huot:2023vxl}
S.~Caron-Huot, M.~Giroux, H.~S. Hannesdottir, and S.~Mizera, ``{What can be
  measured asymptotically?},''
  \href{http://arxiv.org/abs/2308.02125}{{\ttfamily arXiv:2308.02125
  [hep-th]}}.

\bibitem{Bautista:2021llr}
Y.~F. Bautista and A.~Laddha, ``{Soft constraints on KMOC formalism},''
  \href{http://dx.doi.org/10.1007/JHEP12(2022)018}{{\em JHEP} {\bfseries 12}
  (2022) 018}, \href{http://arxiv.org/abs/2111.11642}{{\ttfamily
  arXiv:2111.11642 [hep-th]}}.

\bibitem{Karateev:2022jdb}
D.~Karateev, J.~Marucha, J.~a. Penedones, and B.~Sahoo, ``{Bootstrapping the
  a-anomaly in 4d QFTs},''
  \href{http://dx.doi.org/10.1007/JHEP12(2022)136}{{\em JHEP} {\bfseries 12}
  (2022) 136}, \href{http://arxiv.org/abs/2204.01786}{{\ttfamily
  arXiv:2204.01786 [hep-th]}}.

\bibitem{Fernandes:2020tsq}
K.~Fernandes and A.~Mitra, ``{Soft factors from classical scattering on the
  Reissner-Nordstr\"om spacetime},''
  \href{http://dx.doi.org/10.1103/PhysRevD.102.105015}{{\em Phys. Rev. D}
  {\bfseries 102} no.~10, (2020) 105015},
  \href{http://arxiv.org/abs/2005.03613}{{\ttfamily arXiv:2005.03613
  [hep-th]}}.

\bibitem{Hait:2022ukn}
A.~Hait, S.~Mohanty, and S.~Prakash, ``{Frequency space derivation of linear
  and non-linear memory gravitational wave signals from eccentric binary
  orbits},'' \href{http://arxiv.org/abs/2211.13120}{{\ttfamily arXiv:2211.13120
  [gr-qc]}}.

\bibitem{Mohanty:2022abo}
S.~Mohanty, ``{Gravitational Waves from a Quantum Field Theory Perspective},''
  \href{http://dx.doi.org/10.1007/978-3-031-23770-6}{{\em Lect.Notes Phys.}
  {\bfseries 1013} (2023) }.

\end{thebibliography}\endgroup
\bibliographystyle{utphys}

\end{document}